\documentclass[prd,showpacs, %twocolumn,
nofootinbib, preprintnumbers,floatfix]{revtex4}%

\usepackage{amssymb,hyperref, amsmath}
\usepackage[dvips]{color}
\usepackage[dvips]{graphicx}
\usepackage{epsfig}

\begin{document}
\newcommand{\ignore}[1]{}
\bibliographystyle{apsrev}

\unitlength=1.2mm \preprint{NT@UW-10-13,JLAB-THY-10-1173}
\title{Lattice Calculations of Nucleon Electromagnetic Form Factors\\ at Large Momentum Transfer}

\author{Huey-Wen Lin}
%\email{hwlin@phys.washington.edu}
\affiliation{Department of Physics, University of Washington, Seattle, WA 98195}
\author{Saul D. Cohen}
\affiliation{Center for Computational Science, Boston University, Boston, MA 02215}

\author{Robert G. Edwards, Kostas Orginos, David G. Richards}
\affiliation{Thomas Jefferson National Accelerator Facility, Newport News, VA 23606}

%\date{\today}
\date{May 3, 2010}
\begin{abstract}
In this work, we report a novel technique in lattice QCD for studying the high momentum-transfer region of nucleon form factors. These calculations could give important theoretical input to experiments, such as those of JLab's 12-GeV program and studies of nucleon deformation. There is an extensive history of form-factor calculations on the lattice, primarily with ground states for both the initial and final state. However, determining form factors at large momentum transfer ($Q^2$) has been difficult due to large statistical and systematic errors in this regime. We study the nucleon form factors using three pion masses with both quenched and 2+1-flavor anisotropic lattice configurations with $Q^2$ as large as $6\mbox{ GeV}^2$.
These form factors are further processed to obtain transverse charge and magnetization densities across 2-dimensional impact-parameter space. Our approach can be applied to isotropic lattices and lattices with smaller lattice spacing to calculate even larger-$Q^2$ form factors.
\end{abstract}

\maketitle

\section{Introduction}\label{sec:intro}

The structure of hadrons is revealed by their interactions with various probes in scattering experiments. For electromagnetic interactions that do not change the particle content, we describe them in terms of the elastic electromagnetic form factors, whose dependence on transfer momentum ($Q^2$) provides information about hadronic structure at different scales. Many experimental studies of these nucleon form factors have been conducted. Recently, a Jefferson Lab experiment using both a polarized target and longitudinally polarized beam (so called double-polarization) revealed a non-trivial momentum dependence for the ratio $G_E^p/G_M^p$. This contradicted previous results using the Rosenbluth separation method, which suggested $\mu_p G_E^p/G_M^p \approx 1$. The apparent contradiction highlights the possibility of a systematic correction associated with two-photon exchange, affecting the Rosenbluth separation method more significantly than double-polarization. (For details and further references, see the recent review articles: Refs.~\cite{Arrington:2007ux,Perdrisat:2006hj,Arrington:2006zm}.)

Measuring the form factors at higher $Q^2$ will help us to understand hadrons and challenge models based on quantum chromodynamics (QCD).
Future experimental facilities, such as the 12-GeV upgrade at Jefferson Lab, will provide precision data at large values of $Q^2$.
Experimentally, it is easier to study proton form factors at larger $Q^2$, but rather challenging to study the neutron sector due to its neutral net charge and the lack of a free-neutron target. Current data for the neutron $G_E$ form factor only reaches to $\approx 4\mbox{ GeV}^2$ transfer momentum.
However, the 12-GeV upgrade at Jefferson Lab will provide precision form factors to around 10--$18\mbox{ GeV}^2$ for both the proton and neutron.
Theoretically, perturbative QCD should converge better at larger $Q^2$. However, it fails to describe recent BaBar results for $\gamma^*\gamma \rightarrow \pi^0$ over a wide range of transfer momenta
$4\mbox{ GeV}^2 < Q^2 < 40\mbox{ GeV}^2$\cite{:2009mc,Radyushkin:2009zg}. This leaves an opportunity for nonperturbative QCD approaches to extend their region of applicability to these high-$Q^2$ regions. Lattice QCD is a perfect candidate for the job.

Studies in the nonperturbative regime of QCD theory have been difficult without resorting to model-dependent calculations or making approximations due to the strong coupling at long distances. However, by discretizing space-time into a four-dimensional lattice with a fixed lattice spacing and volume, we are able to compute the path integral (in terms of discretized versions of the QCD Lagrangian and operators) directly via numerical integration, providing first-principles calculations of the consequences of QCD. The techniques of lattice QCD have been applied to such varied phenomena as the spectroscopy of heavy-quark hadrons, the tower of excited baryon states, flavor physics involving the CKM matrix, hadron decay constants and baryon axial couplings.
Using lattice QCD to study hadronic form factors will serve as valuable theoretical input for understanding hadronic structure within the QCD theory of the Standard Model.

The nucleon form factors have been calculated on the lattice by many groups, and calculations are still ongoing\cite{Liu:1994dr,Gockeler:2003ay,Alexandrou:2006ru,Hagler:2007xi,Alexandrou:2007xj,Gockeler:2007hj,Yamazaki:2007mk,Sasaki:2007gw,Lin:2008uz,:2010jn,Syritsyn:2009mx,Yamazaki:2009zq,Lin:2008mr,Hagler:2009ni}. Recently, lattice calculations have also been used to calculate transition form factors involving excited nucleons\cite{Lin:2008qv}.
However, the typical $Q^2$ range in lattice calculations of hadron form factors is less than $2.5\mbox{ GeV}^2$. When one attempts higher-$Q^2$ calculations, they suffer from poor signal-to-noise ratio; for example, see the case study for the pion in Ref.~\cite{Hsu:2007ai}.

In this work, we start to explore the possibility of reaching higher transfer momenta using lattice QCD by re-examining the conventional approach. To calculate the form factors, we need to calculate the three-point Green function, which requires an additional inversion of the fermion matrix over the whole lattice (with rank $12\times L^3\times T$). Since traditionally this type of operation is resource intensive, especially for light quark masses, the parameters are carefully tuned to maximize overlap with the ground-state nucleon on a specific gauge ensemble, and the same parameters are used throughout the whole calculation. This simplifies the three-point correlator analysis, since one only needs to consider one state in a given plateau region. However, when one increases the momentum of the nucleon state, the original fixed parameters are no longer optimal; thus, the signal dies out quickly after adding only a few units of discrete momentum on the lattice. We propose to keep multiple operators in the calculation with parameters tuned for both ground and excited states, but we also extend our analysis to account for these additional states. As a result, we can extract the best signals at each transfer momentum. Since our analysis explicitly treats multiple excited states, the ground state is safe from contamination. Further steps and details are addressed in Sec.~\ref{sec:setup}.
We have previously demonstrated the idea in an exploratory quenched lattice calculation\cite{Lin:2008gv} and extended it to a dynamical ensemble\cite{Lin:2009zzo}. In this work, we use multiple quark masses and improved statistics, giving a complete analysis for three pion-mass ensembles of both quenched and dynamical ($N_f=2+1$) configurations generated by the Hadron Spectrum Collaboration (HSC)\cite{Edwards:2008ja,Lin:2008pr}. Note that even in this study, we may suffer large systematic error due to the coarseness of the lattices ($a_s\approx 0.100$ and 0.123~fm, respectively). As finer lattices are generated by various lattice groups, one can apply the same approach to reach even higher $Q^2$ and significantly reduce systematic discretization errors.

Since the elastic form factors contain information about the spatial structure of the nucleon, we can convert our data into a description of its charge and magnetization densities. Due to the relativistic effects of the transferred momentum on the wavefunction of the nucleon, we cannot use a simple three-dimensional Fourier transformation without some recourse to models. Instead, we use the model-independent formulation of Ref.~\cite{Miller:2007uy} in terms of densities in a two-dimensional plane transverse to an infinite-momentum boost.

In this work, we concentrate on the electromagnetic properties of nucleons. The structure of this paper is as follows: In Sec.~\ref{sec:setup}, we provide details concerning the configurations for both quenched and dynamical lattices. The operators and two- and three-point analysis procedure are given, as well as various checks of the method. We detail how we extract the electromagnetic form factors from lattice calculations.
In Sec.~\ref{sec:num}, we discuss the momentum dependence of the form factors.
We also extract the electric-charge radii, magnetic radii and magnetic moments for the nucleon and compare with other $N_f=2+1$ calculations. We extrapolate the form factors to the physical pion mass and compare the quenched results to dynamical. We use our results to describe the spatial dependence of these quantities in a model-independent way as transverse charge and magnetization densities.
Our conclusions and ideas for future improvements to the calculation are presented in Sec.~\ref{sec:conclusion}.

\section{Methodology and Setup}\label{sec:setup}

In this work, we report on a calculation of nucleon form factors using anisotropic lattices, including both dynamical $N_f=2+1$ and quenched $N_f=0$.

%%%%%%% dynamical lattices %%%%%%%
The 2+1-flavor anisotropic lattices used in this calculation were generated by the Hadron Spectrum Collaboration (HSC)\cite{Edwards:2008ja,Lin:2008pr}. These lattices use Symanzik-improved gauge action with tree-level tadpole-improved coefficients, yielding a leading discretization error at $O(a_s^4,a_t^2,g^2 a_s^2)$.
In the fermion sector, they have anisotropic clover action\cite{Chen:2000ej}; the gauge links in the fermion action are 3-dimensionally stout-link smeared with smearing weight $\rho=0.14$ and $n_\rho=2$ iterations. The renormalized gauge and fermion anisotropies are around $\xi=3.5$ (that is, $a_s=3.5 a_t$), and the inverse of the spatial lattice spacing is about 1.6~GeV. For more details concerning the lattices and their action parameters, please see Ref.~\cite{Edwards:2008ja}.
From the HSC ensembles, we use the $16^3 \times 128$ lattices with pion masses of 875, 580 and 450~MeV. Quark propagators on the lattices are evaluated for source and sink operators with five Gaussian smearing parameters: $\sigma \in \{0.5, 1.5, 2.5, 3.5, 4.5\}$. Six, four and two time sources (respectively, from light to heavy pion mass) are used, and a total of around 200 configurations are used from each ensemble. The quark propagators are calculated under antiperiodic boundary conditions in the time direction, while the spatial ones remain periodic. We construct hadronic two-point correlators from all possible source-sink smearing combinations; however, for three-point correlators, we reduce the computational burden by keeping only the diagonal source-sink smearing-operator combinations.

%%%%%%% quenched lattices %%%%%%%
We also use quenched $16^3 \times 64$ lattices with anisotropy $\xi=3$, using Wilson gauge action with $\beta=6.1$ and stout-link smeared\cite{Morningstar:2003gk} Sheikholeslami-Wohlert (SW) fermions\cite{Sheikholeslami:1985ij} with smearing parameters $\{\rho,n_\rho\}=\{0.22,2\}$. The parameter $\nu$ is nonperturbatively tuned using the meson dispersion relation, and the clover coefficients are set to their tadpole-improved values. The inverse spatial lattice spacing is about 2~GeV, as determined by the static-quark potential, and the simulated pion masses are about 480, 720 and 1100~MeV. In total, we use 400, 200 and 200 configurations respectively at each pion mass.
On the quenched lattices, we use only three Gaussian smearing parameters: $\sigma \in \{0.5, 2.5, 4.5\}$. For both two-point and three-point hadronic correlators, we calculate all 9 possible source-sink smearing combinations.

%%%%%%% operators %%%%%%%

We construct correlators with the quantum numbers of the nucleon using baryonic interpolating operators of the form
\begin{eqnarray}
 \chi^N (x) &=& \epsilon^{abc} [q_1^{a\mathrm{T}}(x)C\gamma_5q_2^b(x)]q_1^c(x),
 \label{eq:lat_B-op}
\end{eqnarray}
where $C$ is the charge conjugation matrix, and $q_1$ and $q_2$ are one of the quarks $\{u,d\}$. For example, in the case of the proton, we want $q_1=u$ and $q_2=d$.
Two-point correlators are derived from these interpolating fields as
\begin{equation}
C_{AB}(t,t_0,\vec{p})=\sum_\mathbf{x}e^{i\vec{p}\cdot\vec{x}} \langle \Gamma \chi^N_A(\mathbf{x},t)\chi^N_B(\mathbf{0},t_0)^\dagger\rangle  , \nonumber
\label{eq:two-pt-correlators1}
\end{equation}
where $\vec{p}$ is the baryon momentum, the spin projection $\Gamma=\frac{1+\gamma_4}{2}$ and $A$ and $B$ index over the different smearing parameters. Eq.~\ref{eq:two-pt-correlators1} can be decomposed in terms of energy eigenstates:
\begin{eqnarray}
\Gamma^{(2)}_{AB}(t;\vec{p})= \sum_n \frac{E_n+m_n}{2E_n} Z_{n,A}
Z_{n,B} e^{-E_n(\vec{p})t},
 \label{eq:two-pt-correlators2}
\end{eqnarray}
where $n$ indexes over the basis of nucleon energy eigenstates.
(These states are defined to be normalized as $\langle 0 |(\chi^N)^\dagger|p,s\rangle = Z u_N(\vec p,s)$ with nucleon spin-1/2 interpolating field $\chi^N$); the spinors in Euclidean space satisfy
\begin{eqnarray}
\sum_{s} u_N(\vec p,s)\bar u_N(\vec{p},s)&=& \frac{E(\vec{p})\gamma^t
-i\vec\gamma\cdot \vec{p} + m}{2 E(\vec{p})}. \label{eq:spinor}
\end{eqnarray}

\begin{figure}
\begin{center}
\includegraphics[width=0.45\textwidth]{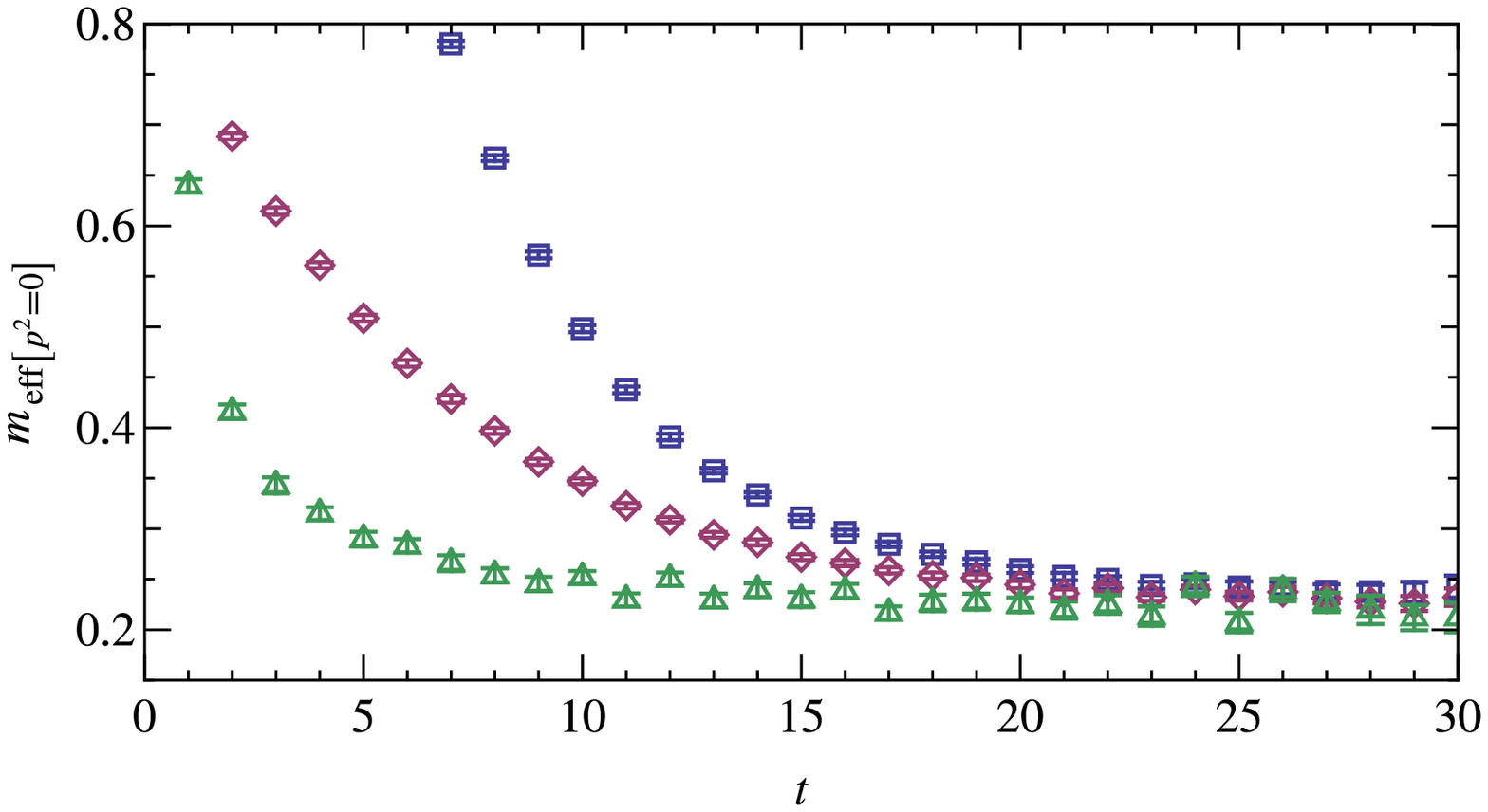}
\includegraphics[width=0.45\textwidth]{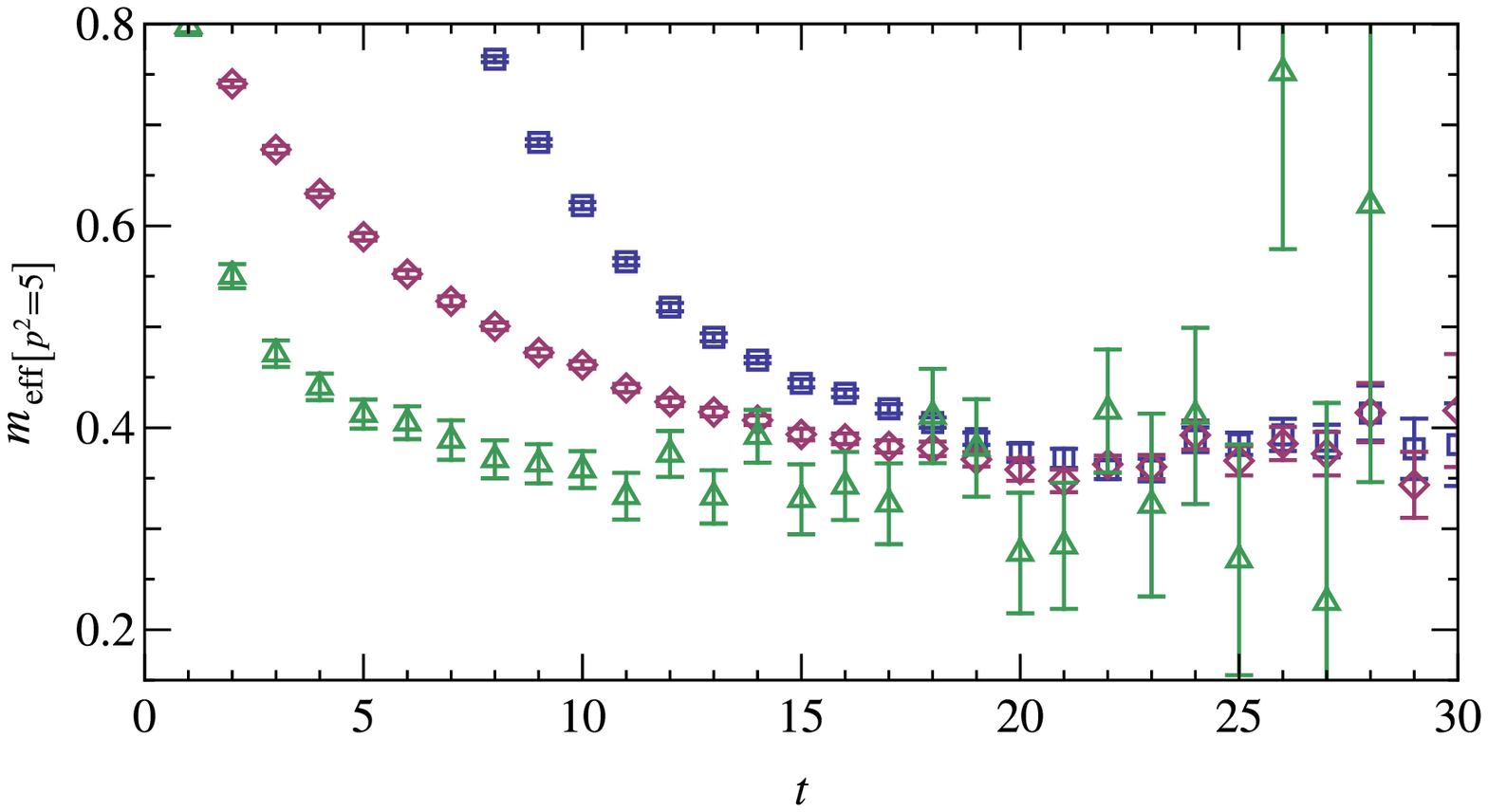}
\end{center}
\vspace{-0.7cm}
\caption{Nucleon effective-mass plots with ($|\vec{p}|^2=0$ and 5 in units of $\frac{2\pi}{L}a^{-1}$) from the $m_\pi=450$~MeV ensemble. The square, diamond and triangle points correspond to smeared-smeared correlators with $\sigma=0.5$, 1.5 and 4.5 respectively. }\label{fig:mNeff}
\end{figure}

Conventionally, one would choose a smearing parameter that optimizes the signal of the zero-momentum ground-state two-point correlators and carry out the form-factor calculation at various transfer momenta. We present an example from the dynamical 450-MeV ensembles. The left-hand side of Fig.~\ref{fig:mNeff} shows the effective mass plot of the zero-momentum two-point nucleon correlator with all diagonal (with $A=B$) Gaussian smearing parameters. Examining the behavior of these correlators, one usually chooses the smearing parameter that contains the least excited-state signal, since this makes the analysis of the ground state simpler. In this example, the correlator with Gaussian smearing parameter of $\sigma=4.5$ is a good candidate: the excited-state signals die out around $t=10$, giving enough data points to extract ground-state form factors from $t \in [10,30]$, if we put the sink around $t=40$ ($\approx 1$~fm source-sink separation). If we increase the magnitude of the momentum, say to $|\vec{p}|^2=5$ in units of $\frac{2\pi}{L}a^{-1}$, we can see the signal from the $\sigma=4.5$ correlator decays significantly. This is not surprising, since this smearing parameter was chosen to ``filter out'' higher-energy contributions. So when increasing the momenta, the signal for the ground state starts to disappear from broadly smeared sources, while it remains clear for  smaller values of $\sigma$. To improve the quality of the signal at higher momenta, we should use multiple $\sigma$ and explicitly subtract any excited-state contributions (ideally more than one) to make sure that the ground state will be free from them.

%%%%%%%%%%%%%%%  2-pt analysis  %%%%%%%%%%%%%%%
To extract form factors from three-point correlators, we need $E_n$ and $Z_n$ as inputs from analyzing two-point correlators. Here, we apply the variational method\cite{Luscher:1990ck} to extract the principal correlators corresponding to pure energy eigenstates from our matrix of correlators.
The $N \times N$ Gaussian smeared-smeared correlation matrix (with $N=5$ for dynamical and 3 for quenched) can be approximated as
\begin{equation}
	C_{ij}=\sum_{n=1}^N v_i^{n*} v_j^n e^{-t E_n}
\end{equation}
with eigenvalues
\begin{equation}
	\lambda_n(t,t_r)=e^{-(t-t_r)E_n}
\end{equation}
by solving the generalized eigensystem problem
\begin{equation}
	C(t)V=\lambda(t,t_r)C(t_r)V.
\end{equation}
where $V$ is the matrix of eigenvectors and $t_r$ is a reference time slice. The resulting 5 eigenvalues (principal correlators) $\lambda_n(t,t_r)$ are then further analyzed to extract the energy levels $E_n$.
In practice, we are only interested in the lowest two; the extra higher states provide buffers against contamination by higher excited states due to the lack of the orthogonality in our operators.
Since they have been projected onto pure eigenstates of the Hamiltonian, the principal correlator should be fit well by a single exponential and double checked for the consistency of the obtained energies. The leading contamination due to higher-lying states is another exponential having higher energy; we use a two-state fit to help remove this contamination.
The overlap factors ($Z_n$) between the interpolating operators and the $n^{\rm th}$ state are derived from the eigenvectors obtained in the variational method. Since we diagonalize the correlator matrix independently on every time slice, $Z_n$ is a function of time; even though the time-dependence is mild, we choose the best value of $Z_n$ by minimizing the difference between the correlator matrix reconstructed from $Z$ and $E$ and the original two-point correlator data.

\begin{figure}
\begin{center}
\includegraphics[width=0.45\textwidth]{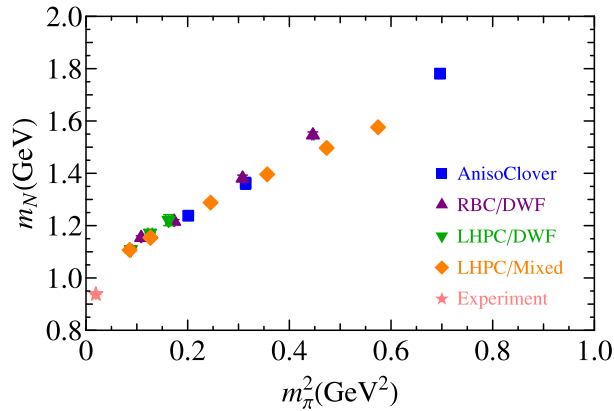}
\end{center}
\vspace{-0.7cm}
\caption{Summary of nucleon masses used by all currently existing $N_f=2+1$ nucleon electromagnetic form-factor calculations~\cite{:2010jn,Yamazaki:2009zq,Syritsyn:2009mx,Lin:2008mr}}\label{fig:all21mN}
\end{figure}

The nucleon masses from the dynamical ensembles used in this work are summarized in Fig.~\ref{fig:all21mN}; also shown are other $N_f=2+1$ nucleon masses used in published nucleon form-factor calculations.
Figure~\ref{fig:all21mN} summarizes the pion and nucleon masses used by various groups who have calculated nucleon isovector Dirac and Pauli radii using $N_f=2+1$.
``AnisoClover'' uses anisotropic lattice with clover actions; the pion masses ranges from 450--850 MeV with spatial lattice spacing and size around 0.123 and 2~fm. %\cite{Lin:2010xx}.
``RBC/DWF'' carried out calculations using domain-wall fermions (DWF) on ensembles with pion masses of 330--670~MeV, spatial lattice spacing 0.114~fm and box size 2.7~fm\cite{Yamazaki:2009zq}.
``LHPC/DWF'' also used DWF ensembles with the same spatial volume but at a smaller lattice spacing (0.084~fm) than ``RBC/DWF'', %2.7~fm
focusing on the pion-mass region 300--400~MeV\cite{Syritsyn:2009mx}.
``LHPC/Mixed'' used a staggered-fermion sea, DWF valence with pion masses 290--760~MeV and $a=0.124$~fm, $L=2.5$~fm; they include one additional point with lattice size 3.5~fm for a 350-MeV pion\cite{:2010jn}.
In this work, we emphasize calculation of larger-$Q^2$ quantities, necessitating the use of rather heavy quark mass inputs. In future calculations, we plan to have lighter quark masses and larger volumes. We can see from the plots that our nucleon masses fall nicely onto the trend outlined by other $N_f=2+1$ ensembles. Due to the higher quark masses used in this calculation, we do not find noticeable finite-volume effects despite our small $\approx 2$~fm box size.

%%%%%%%%%%%%%%%  3-pt operators  %%%%%%%%%%%%%%%

To calculate the nucleon electromagnetic form factors, we first calculate the matrix element $\langle \chi^N (\vec{p}_f) | V^\mu | \chi^N (\vec{p}_i) \rangle$, where $V^\mu=\overline{q}\gamma_\mu q$ is the vector current with $q$ being either an up or down quark, and $\vec{p}_{\{i,f\}}$ are the initial and final nucleon momenta. We integrate out the spatial dependence and project the baryonic spin, leaving a time-dependent three-point correlator of the form
\begin{eqnarray}\label{eq:general-3pt}
\Gamma^{(3),T}_{\mu,AB}(t_i,t,t_f,\vec{p}_i,\vec{p}_f) &=&
Z_V \sum_n \sum_{n^\prime} f_{n,n^\prime}(p_f,p_i,E_n^\prime,E_n,t,t_i,t_f)\nonumber \\
&\times& \sum_{s,s^\prime}
T_{\alpha\beta} u_{n^\prime}(\vec{p}_f,s^\prime)_\beta
\langle N_{n^\prime}(\vec{p}_f,s^\prime)\left|V_\mu\right|N_n(\vec{p}_i,s)\rangle\overline{u}_n(\vec{p}_i,s)_\alpha,
\end{eqnarray}
where $f_{n,n^\prime}(p_f,p_i,E_n^\prime,E_n,t,t_i,t_f)$ contains kinematic factors involving the energy $E_n$ and overlap factors ($Z_n$) obtained in the two-point variational method, $n$ and $n^\prime$ are the indices of different energy states and $Z_V$ is the vector-current renormalization constant (which is set to its nonperturbative value).
The projection $T$ used on the quenched lattices is $T_{\rm mix}=\frac{1}{4}(1+\gamma_4)(1+i\gamma_5\gamma_3)$, and the dynamical lattices use $T_{4}=\frac{1}{4}(1+\gamma_4)$ and $T_{53}=\frac{1}{4}(1+\gamma_4)(i\gamma_5\gamma_3)$.
The source-sink separation ($t_f-t_i$) is 34 and 39 time slices in lattice units on the quenched and dynamical lattices respectively. (These give source-sink separations about 1.13 and 1.36~fm.)
We use four final momenta ($\vec{p_f}=\frac{2\pi}{L}\{0,0,0\}a^{-1}$, $\frac{2\pi}{L}\{-1,0,0\}a^{-1}$, $\frac{2\pi}{L}\{-1,-1,0\}a^{-1}$, $\frac{2\pi}{L}\{-2,0,0\}a^{-1}$) and vary the initial momentum over all $\vec{p_i}=\frac{2\pi}{L}\{n_x,n_y,n_z\}a^{-1}$ with integer $n_{x,y,z}$ and $n_x^2+n_y^2+n_z^2 \leq 10$.
By fitting the time dependence of the three-point correlators to the form of Eq.~\ref{eq:general-3pt} with $n$ and $n^\prime$ restricted to 0 and 1, we extract the ground state and matrix elements involving the first excited state. In this work, we only concentrate on the ground-state matrix element with $n=n^\prime=0$.

With smeared fermion actions, it has been seen on three-flavor anisotropic lattices with tree-level tadpole-improved fermion-action coefficients that the nonperturbative coefficient conditions are automatically satisfied\cite{Edwards:2008ja,Lin:2007yf}. Similar behavior has also been observed in another quenched study\cite{Hoffmann:2007nm}, where the nonperturbative coefficients or renormalization constants in a smeared fermion action differed from tree-level values by a few percent. So the local vector current used here is $O(a)$ on-shell improved with the improved coefficient set to its tree-level value.

The form factors are then extracted from the vector-current matrix elements for any nucleon state $N$ through
\begin{eqnarray}
\label{eq:Vector-roper}
\langle N\left|V^{\rm }_\mu\right|N\rangle_{\mu}(\vec{q}) &=&
{\overline u}_N(\vec{p_f})\left[ F_1(q^2) \gamma_{\mu}
+\sigma_{\mu \nu}q_{\nu}
\frac{F_2(q^2)}{2 m_N}\right]u_N(\vec{p_i}), %e^{-iq\cdot x},
\end{eqnarray}
with $q=p_f-p_i$. 
By imposing the same projection matrix $T$, vector-current matrix elements, $\langle N\left|V_\mu\right|N\rangle$ (with $n=n^\prime=0$) and momenta used in the lattice calculation in Eq.~\ref{eq:general-3pt}, we obtain a series of equations corresponding to the various lattice correlators. The overdetermined system of linear equations allows solution for the Dirac and Pauli form factors $F_{1,2}$.

\begin{figure}
\begin{center}
\includegraphics[width=0.5\textwidth]{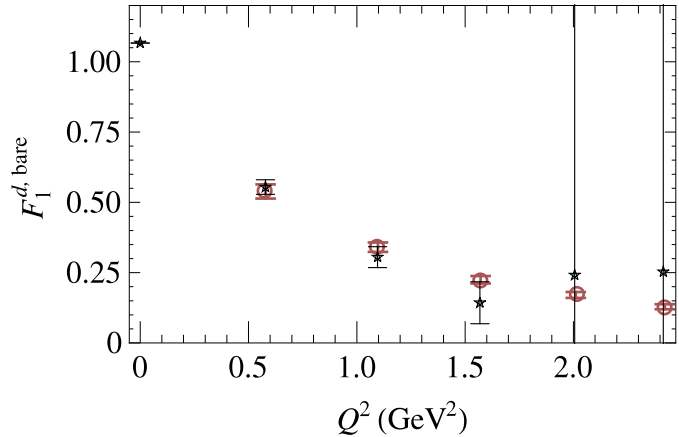}
\end{center}
\vspace{-0.7cm}
\caption{Unrenormalized $d$-quark contribution to the nucleon Dirac form factor $F_1^d$ obtained from the ratio approach (stars) and the method adopted by this work (circles) on a quenched ensemble with $m_\pi=720$~MeV}\label{fig:fNN-comp}
\end{figure}

To make sure our analysis is correct, we check our procedures against the traditional one, the ``ratio'' method, where one takes ratios of three- and two-point correlators:
\begin{eqnarray}\label{eq:Ratio_GP}
R_{V_\mu} &=& \frac{Z_V
      \Gamma^{(3),T}_{\mu,GG}(t_i,t,t_f;\vec p_i,\vec p_f)}{\Gamma^{(2),T}_{GG}(t_i,t_f;\vec p_f)}
%      \nonumber \\     &\times&
 \sqrt{\frac{\Gamma^{(2),T}_{PG}(t,t_f;\vec p_i)}{\Gamma^{(2),T}_{PG}(t,t_f;\vec
      p_f)}}\nonumber \\
 &\times&     \sqrt{\frac{\Gamma^{(2),T}_{GG}(t_i,t;\vec p_f)}{\Gamma^{(2),T}_{GG}(t_i,t;\vec p_i)}}
%        \nonumber \\ &\times&
   \sqrt{\frac{\Gamma^{(2),T}_{PG}(t_i,t_f; \vec p_f)}{\Gamma^{(2),T}_{PG}(t_i,t_f;\vec p_i)}},
\end{eqnarray}
with $P$ and $G$ corresponding to a point source and a Gaussian-smeared source. (We can also use another Gaussian smearing to replace the point sources.) In this ratio, the exponential time dependence is canceled and one only needs to fit a constant. Note that this method only assumes that the region where one extracts the matrix elements contains only ground-state signal. Otherwise, the matrix element will contain the contribution of unwanted excited-nucleon states. Here we demonstrate the method using the data from the quenched calculation with 720-MeV pion mass, $\vec{p_f}=\{0,0,0\}$. We use the largest Gaussian smearing ($\sigma=4.5$) three- and two-point data, which has good overlap with ground state and not very much excited-state signal in the two-point effective-mass.
Fig.~\ref{fig:fNN-comp} shows that both methods have consistent $d$-quark contributions to the Dirac form factor at pion mass 720~MeV. The star points are obtained using the conventional ratio method, and the circular points are from our analysis combining $3\times3$ three-point correlators. Note that we not only get consistent numbers for $F_1$, but at large momenta, our fitting approach dramatically improves the signal. This is because when one tries to improve the ground-state signal (normally done by examining the hadron effective-mass time dependence at zero momentum) with a choice of smearing operator, one wipes out not only the excited states but also the higher-momentum states. Therefore, when one tries to project onto these higher momenta, there is no doubt that the signal-to-noise ratio will worsen. Since we are explicitly considering excited states in our analysis, our sources which have good overlap with higher-momenta will be only lightly contaminated by excited-state signals.

\section{Numerical Results}\label{sec:num}

In this section, we compare the quenched and dynamical nucleon form factors at $Q^2$ as large as $6\mbox{ GeV}^2$ and examine their dependence on the pion mass. 
We also consider the charge radii and magnetic moments; these quantities are commonly derived from the electromagnetic form factors near or at $Q^2=0$. 
Using calculations on ensembles at different pion masses, we extrapolate the form factors and their ratios to the physical pion mass. 
Later in the section, we study the large-momentum region and look at the charge distribution.

In Sec.~\ref{sec:setup}, we described how the Dirac ($F_1$) and Pauli ($F_2$) form factors are obtained from lattice calculations. Another common set of form factor definitions, widely used in experiments, are the Sachs form factors; these can be related to the Dirac and Pauli form factors through
\begin{eqnarray}\label{eq:sachs}
G_E(Q^2) &=& F_1(Q^2) - \frac{Q^2}{4m_B^2}F_2(Q^2) \\
G_M(Q^2) &=& F_1(Q^2) + F_2(Q^2).
\end{eqnarray}

In this work, we only calculate the ``connected'' diagram, which means the inserted quark current is contracted with the valence quarks in the baryon interpolating fields. The ``disconnected'' contributions are notoriously noisy and difficult to calculate. Previous works have attempted to estimate their contribution through an indirect approach by applying ``charge symmetry'' to the connected ones (including the the octet family)\cite{Leinweber:1995ie,Leinweber:1999nf}. For $G_M^s$ (the strange contribution to the magnetic form factor) at small transfer momenta, Adelaide-JLab Collaboration got $-0.046(19)$ using quenched lattice data with chiral perturbatively corrected vacuum\cite{Leinweber:2004tc} and another dynamical calculation obtained $G_M^s=-0.06(3)$ using mixed action\cite{Lin:2009qs} with pion mass as low as 360~MeV. The same approach can be applied to the electric form factor, yielding $G_E^s(Q^2=0.1\mbox{ GeV}^2)=-0.0049(56)$ in the same dynamical calculation\cite{Lin:2009qs}.
Recently, another direct calculation from $\chi$QCD Collaboration\cite{Doi:2009sq} using the $\mathbb{Z}_4$ stochastic method (with unbiased subtractions) on $N_f=2+1$ clover fermion lattices, found  $G^s_M = -0.015(23)$ and $G^s_E(Q^2=0.1\mbox{ GeV}^2) = 0.0022(19)$.

Another independent work done by the Boston group using multigrid techniques on 2-flavor Wilson anisotropic lattices with 400~MeV pions also observed $G^s_E$ consistent with zero over $Q^2$ ranging 0.1--1 GeV$^2$\cite{Babich:2009rq}.
Both direct and indirect approaches indicate the disconnected contributions to the electromagnetic form factors are small, $O(10^{-2})$, at low momentum transfer. One would expect the disconnected contributions to become yet smaller at large momentum transfer, where the strong coupling between the loop and the baryon becomes weaker.
The light-quark disconnected contribution turns out to be only moderately larger than the strange one.  Using the isospin symmetry approach, the ratio of the strange-quark disconnected contribution to the light-quark was estimated to be 0.16(4) and 0.14(4) for electric and magnetic form factors, respectively\cite{Leinweber:2004tc,Leinweber:2006ug}. %hep-lat/0601025
We are not aware of any direct calculations of the disconnected quark ratios in the electromagnetic form factors; however, they exist in many other quantities. For example, the strange-to-light ratio of the first moment of the quark momentum fraction is 0.88(7)\cite{Deka:2008xr}, and the ratios for angular and orbital momentum contributions to the nucleon spin are also around 1\cite{Mathur:1999uf}. %hep-ph/9912289
It should be safe to estimate that the light disconnected contribution to the electromagnetic form factors is on the same order as the strange disconnected ones: $O(10^{-2})$ at low momentum transfer.

We expect the disconnected contributions to become yet smaller as the momentum transfer increases.
From the mesonic cloud point of view, the disconnected diagram is suppressed by an additional factor of $1/Q^2$ relative to the connected diagrams; thus, its relative contribution is suppressed at large momentum transfer. 
For a form factor whose typical magnitude in the low-momentum region is $O(1)$ (such as proton form factors and neutron magnetic form factor), the disconnected contributions are smaller than the statistical errors at most momentum points and the systematics from sources such as the nonzero lattice spacing.
However, the disconnected contributions could significantly affect small-magnitude form factors such as the neutron electric form factor.
We nevertheless include results from this channel for comparison to other lattice results but remind readers that the neutron electric form factor has larger uncertainty than the other form factors in this calculation.

\begin{figure}
\includegraphics[height=.33\textwidth]{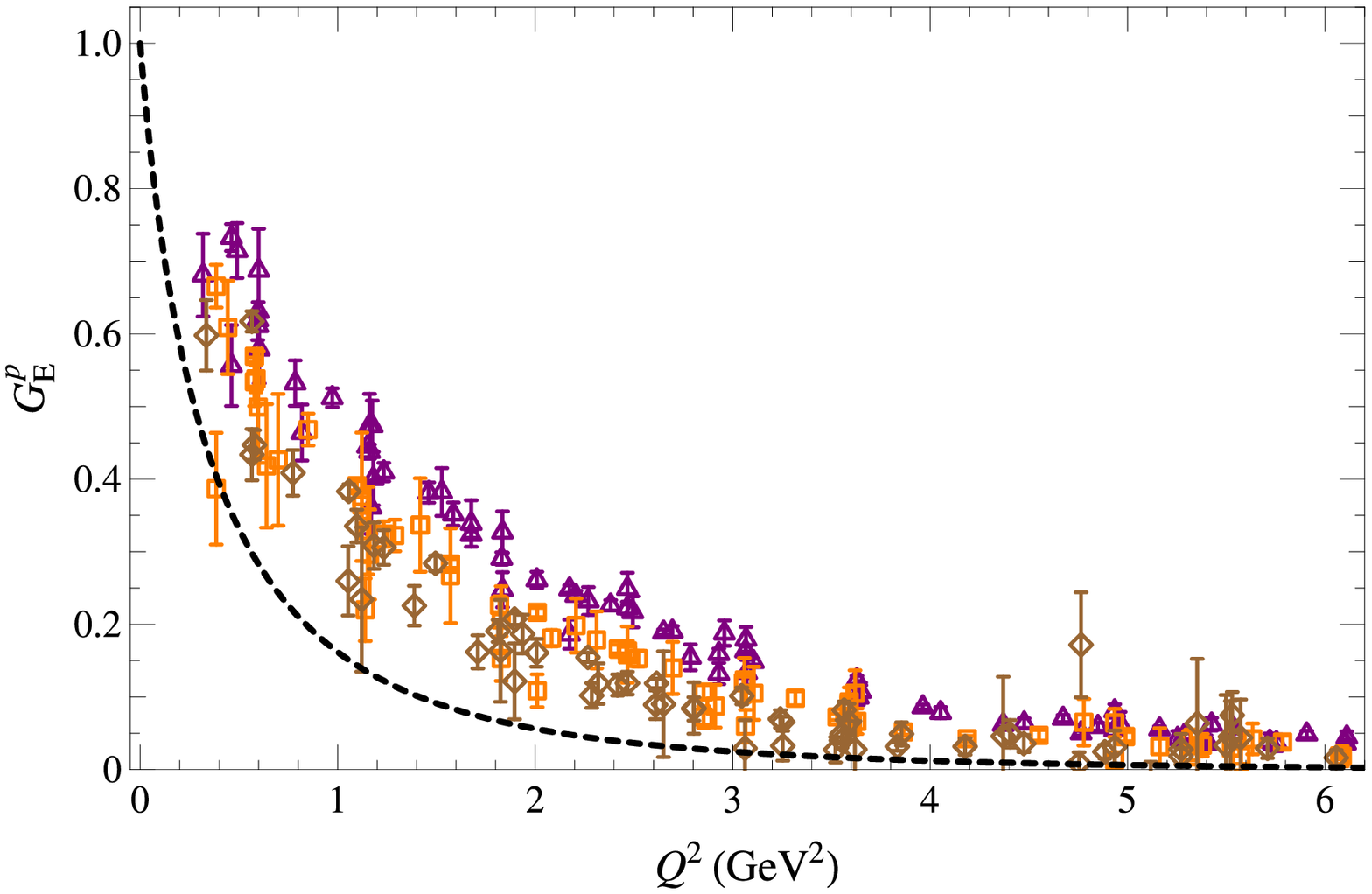}
\includegraphics[height=.33\textwidth]{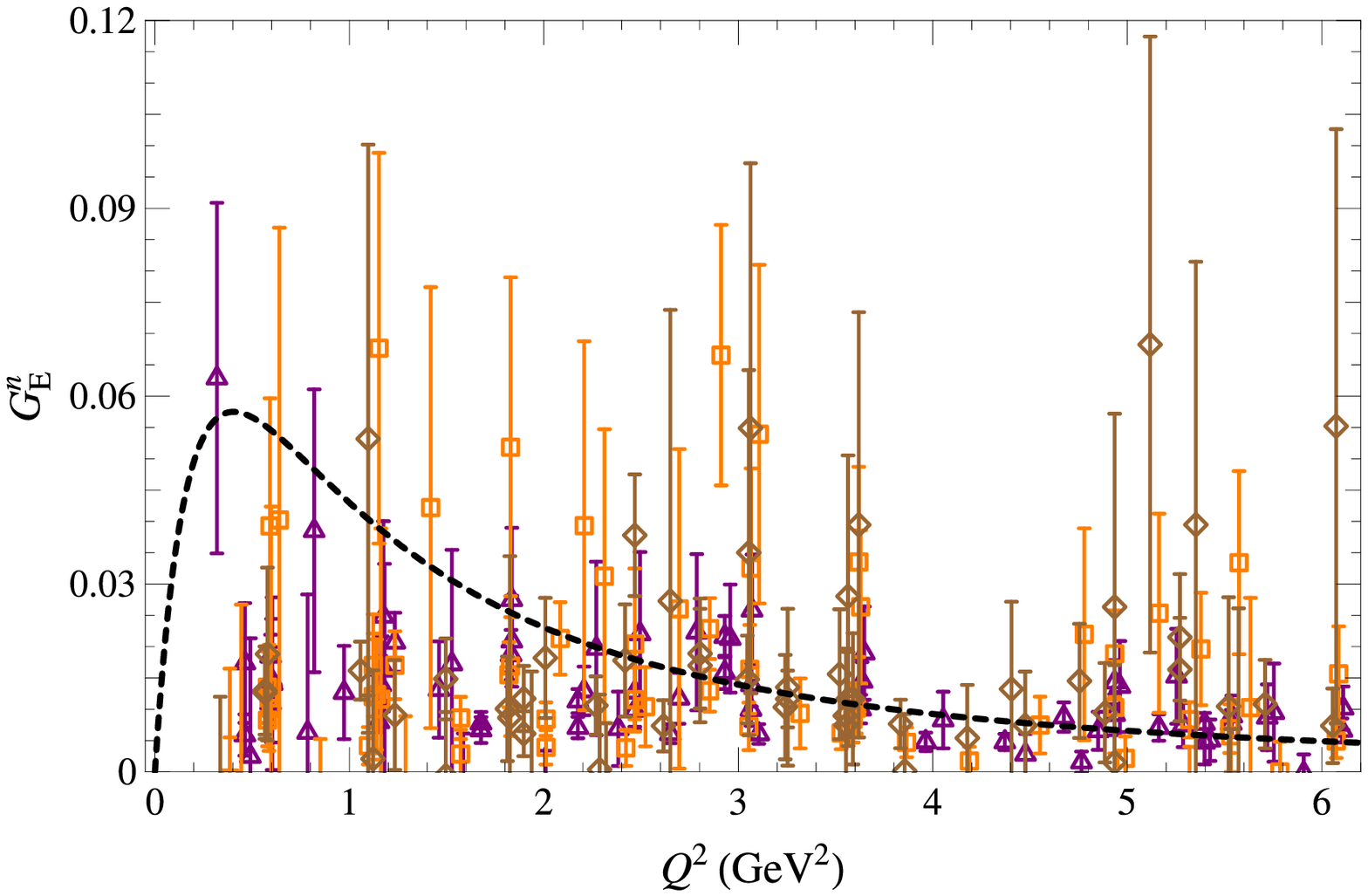}
\includegraphics[height=.33\textwidth]{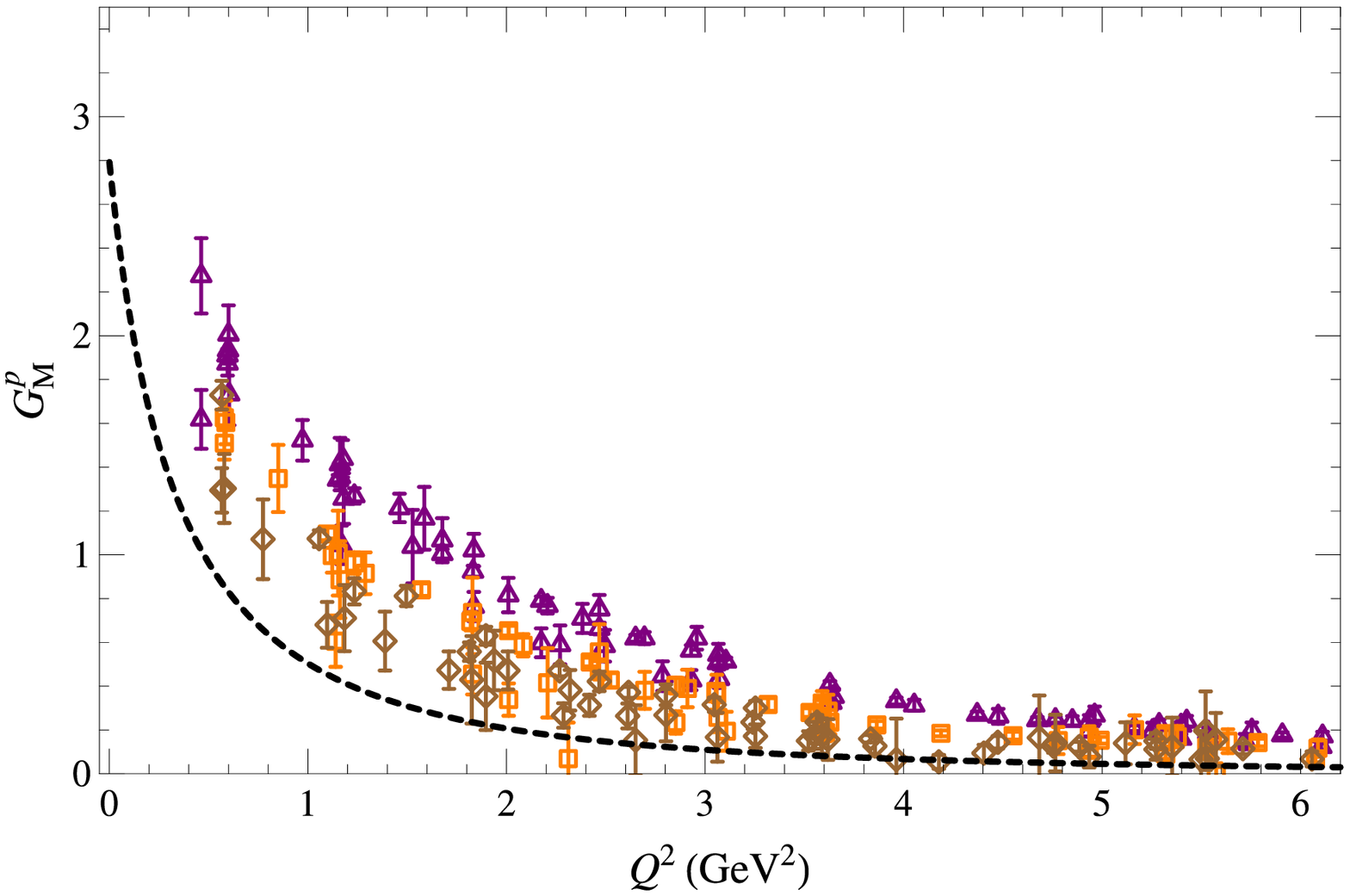}
\includegraphics[height=.33\textwidth]{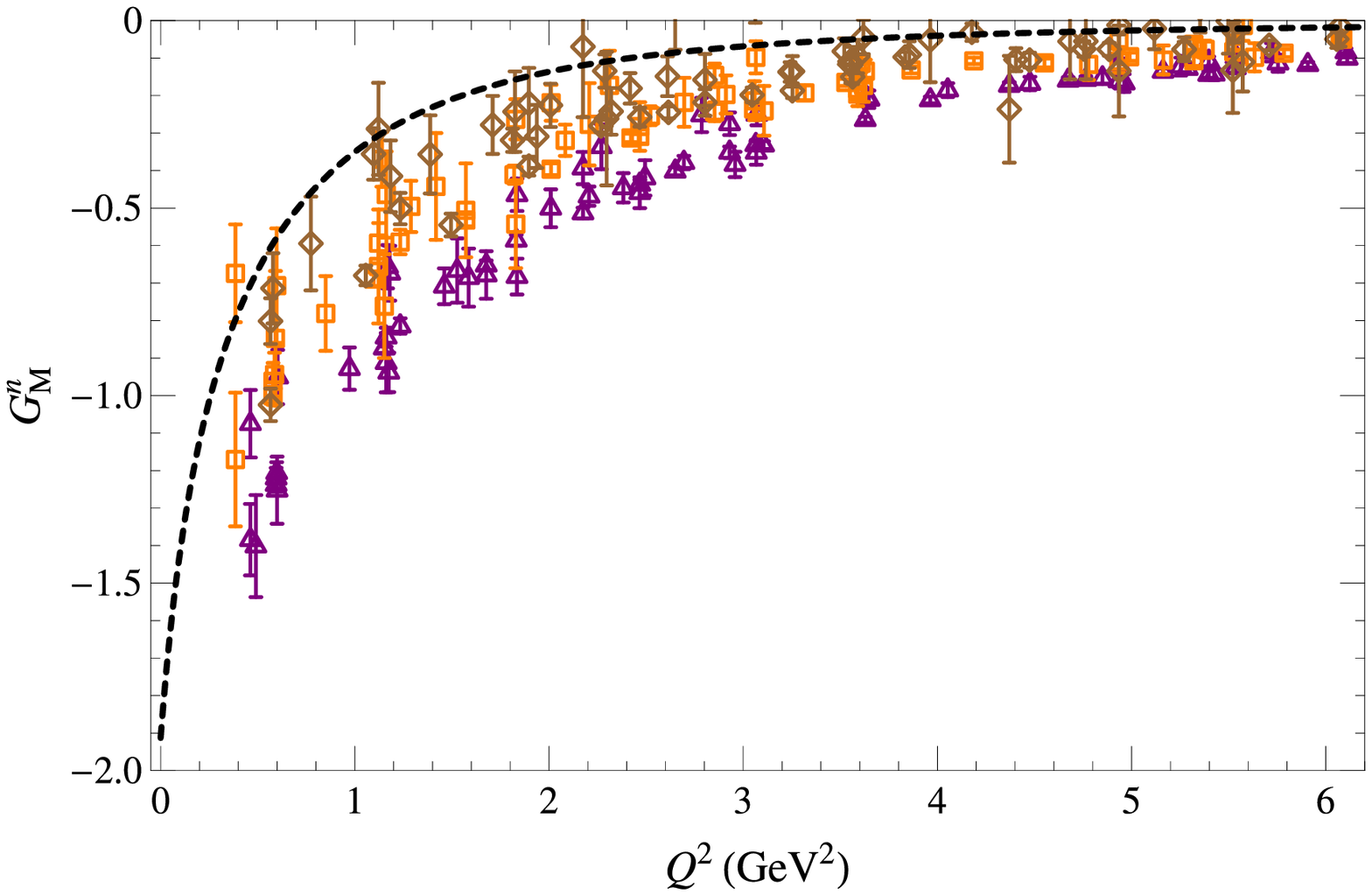}
\includegraphics[height=.33\textwidth]{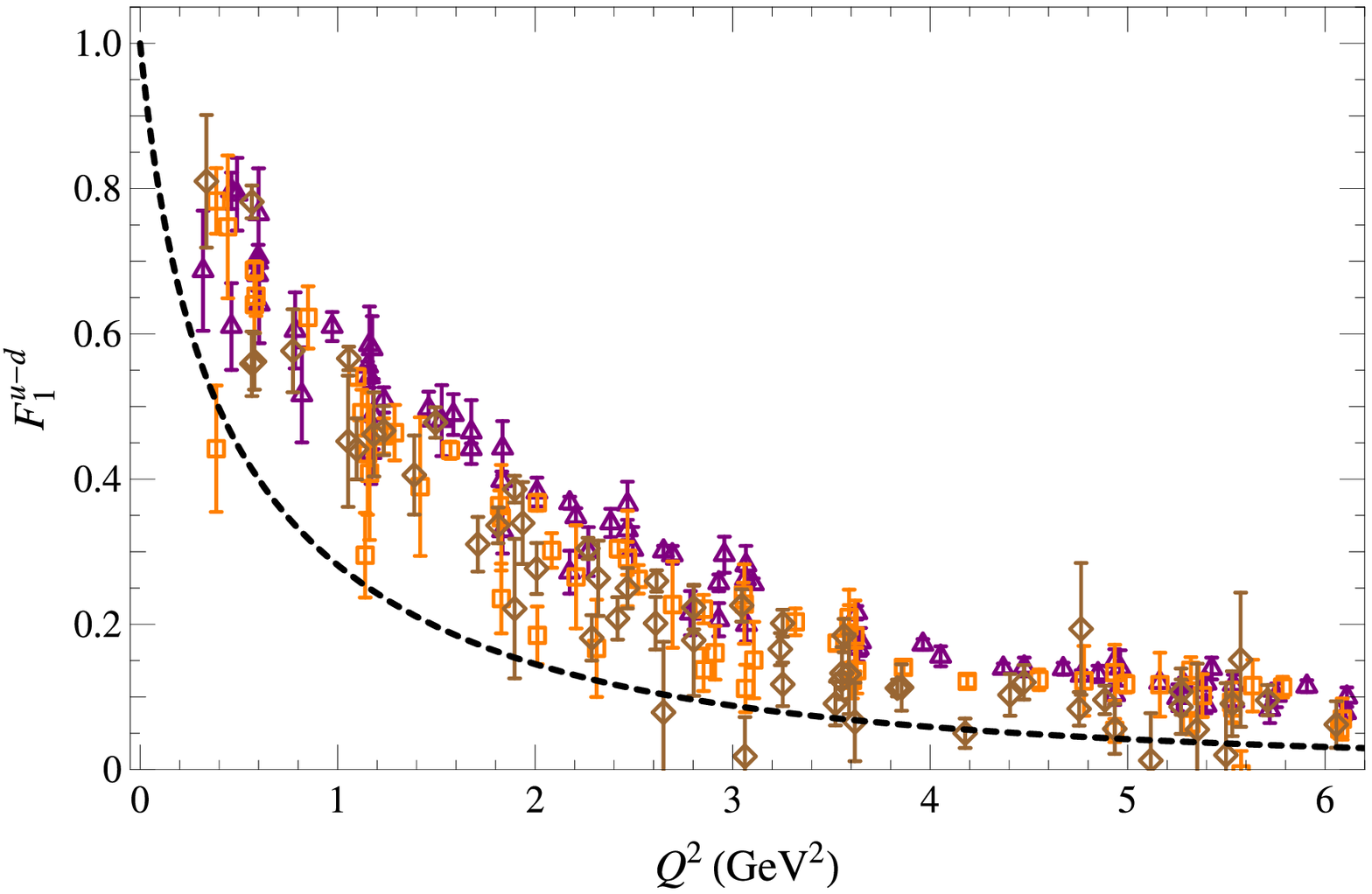}
\includegraphics[height=.33\textwidth]{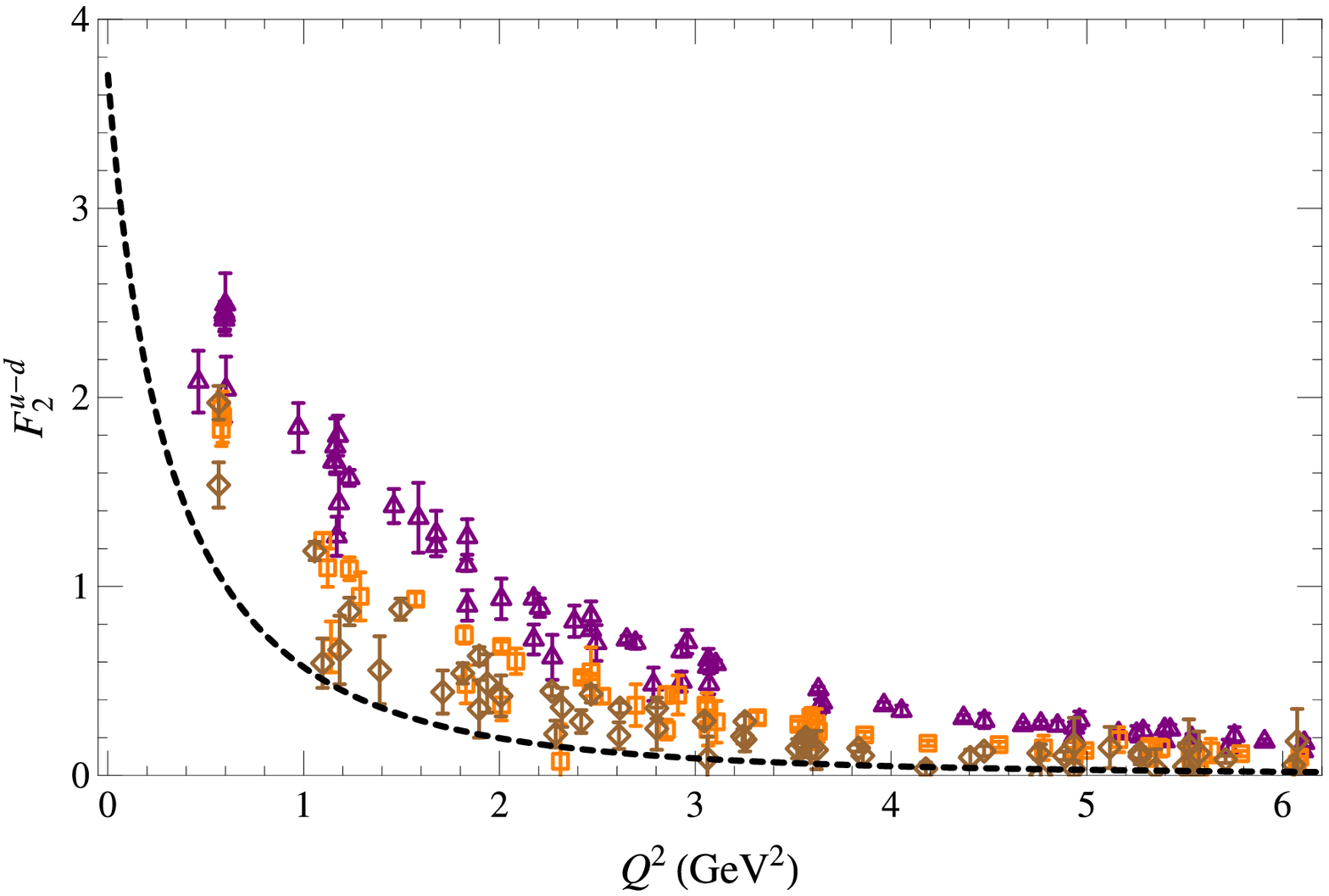}
\vspace{-0 cm} \caption{Nucleon form factors on quenched configurations with pion masses of 480 (brown), 720 (orange) and 1080 (purple) MeV. The dashed lines are a parametrization of experimental form-factor data\cite{Arrington:2007ux,Kelly:2004hm}
} \label{fig:GEGM}
\end{figure}

\begin{figure}
\includegraphics[height=.33\textwidth]{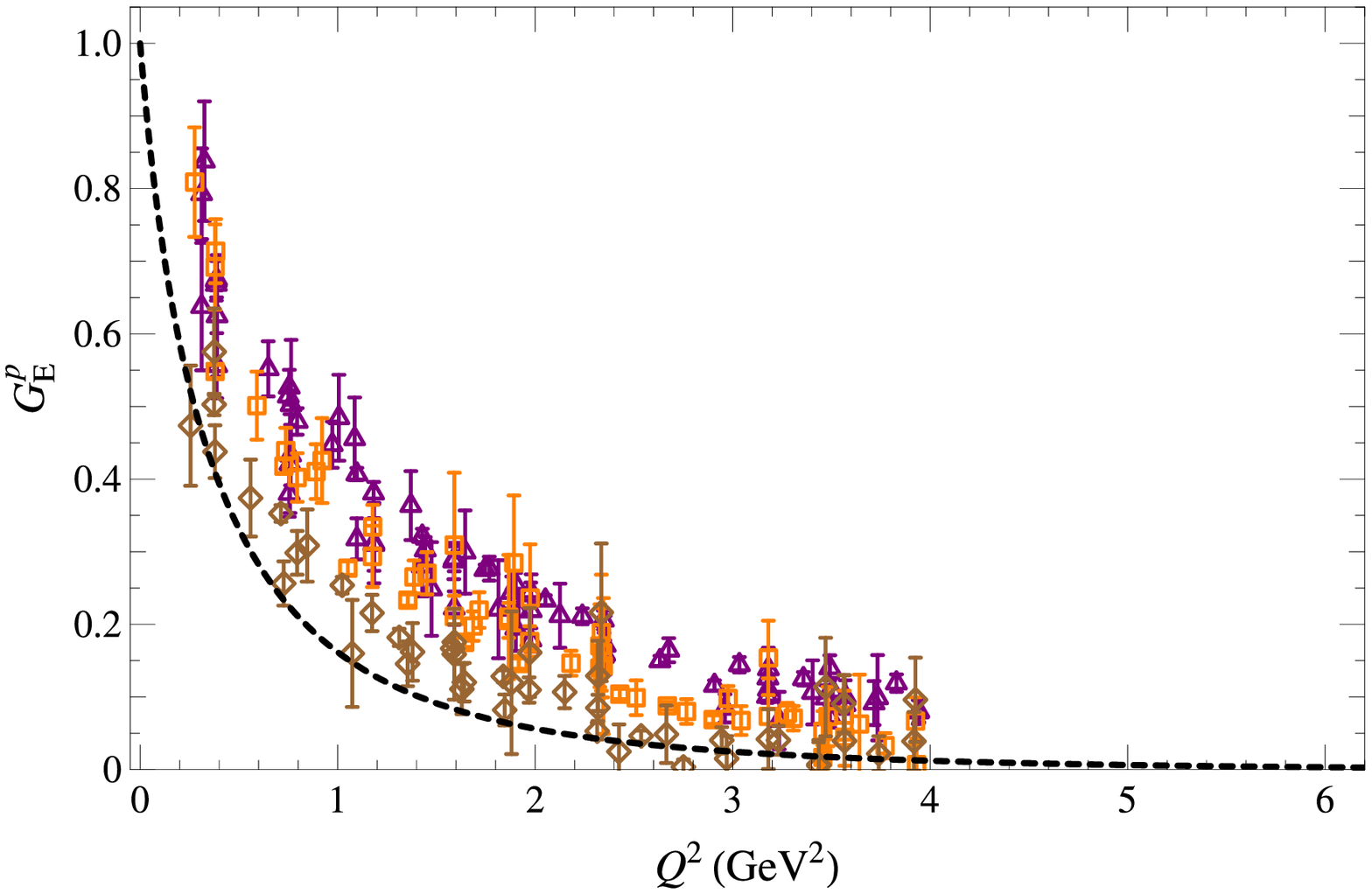}
\includegraphics[height=.33\textwidth]{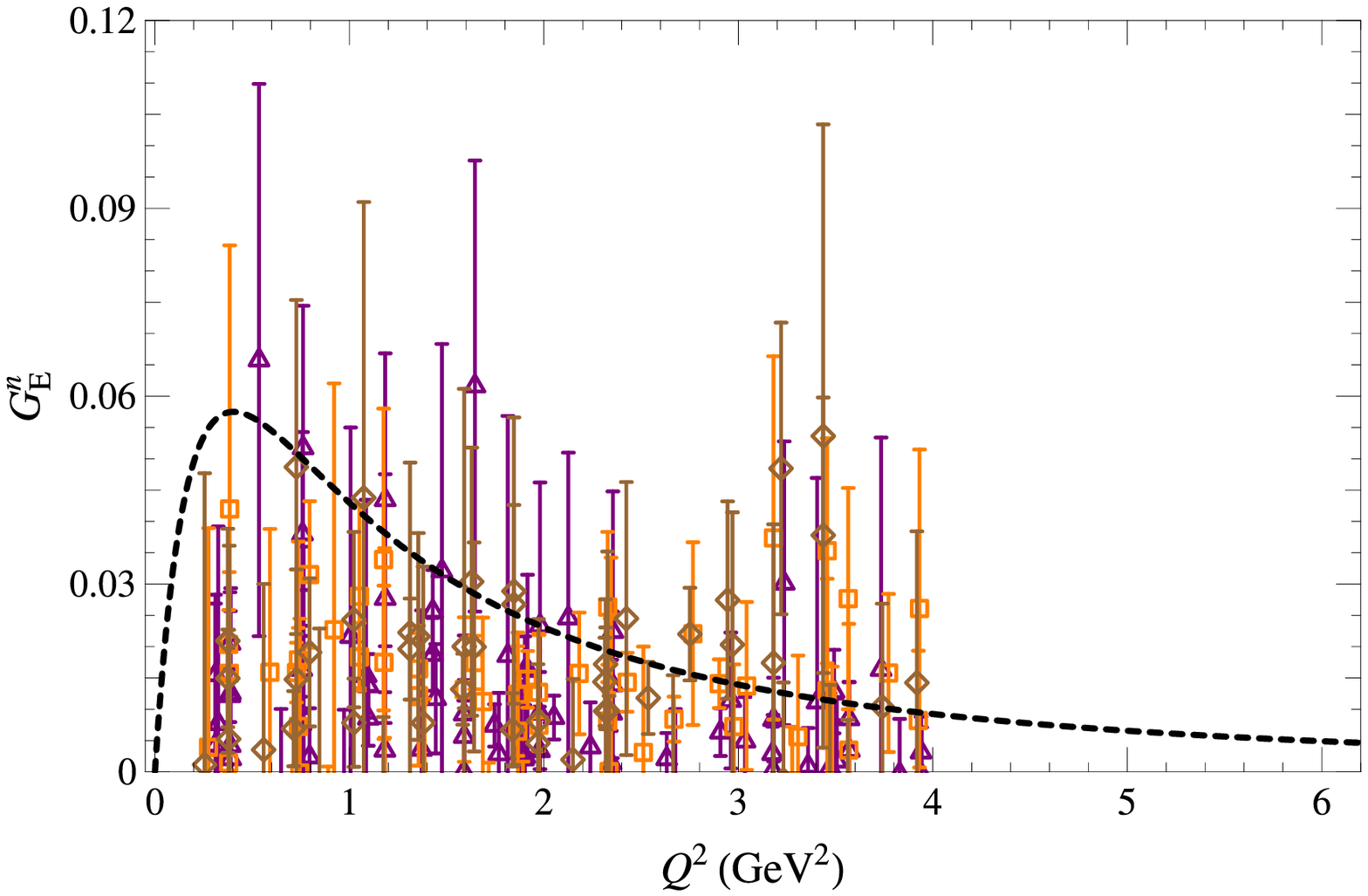}
\includegraphics[height=.33\textwidth]{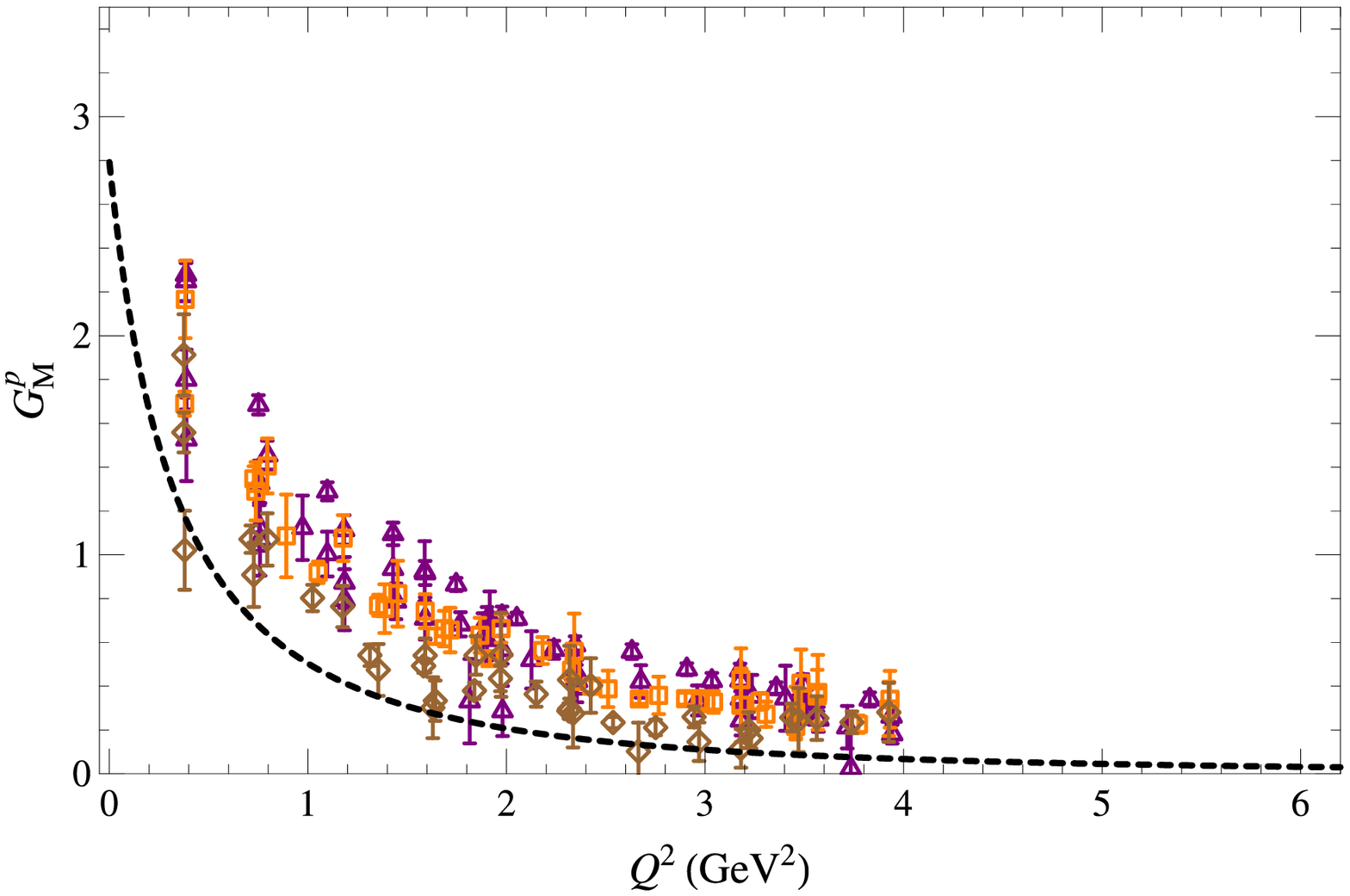}
\includegraphics[height=.33\textwidth]{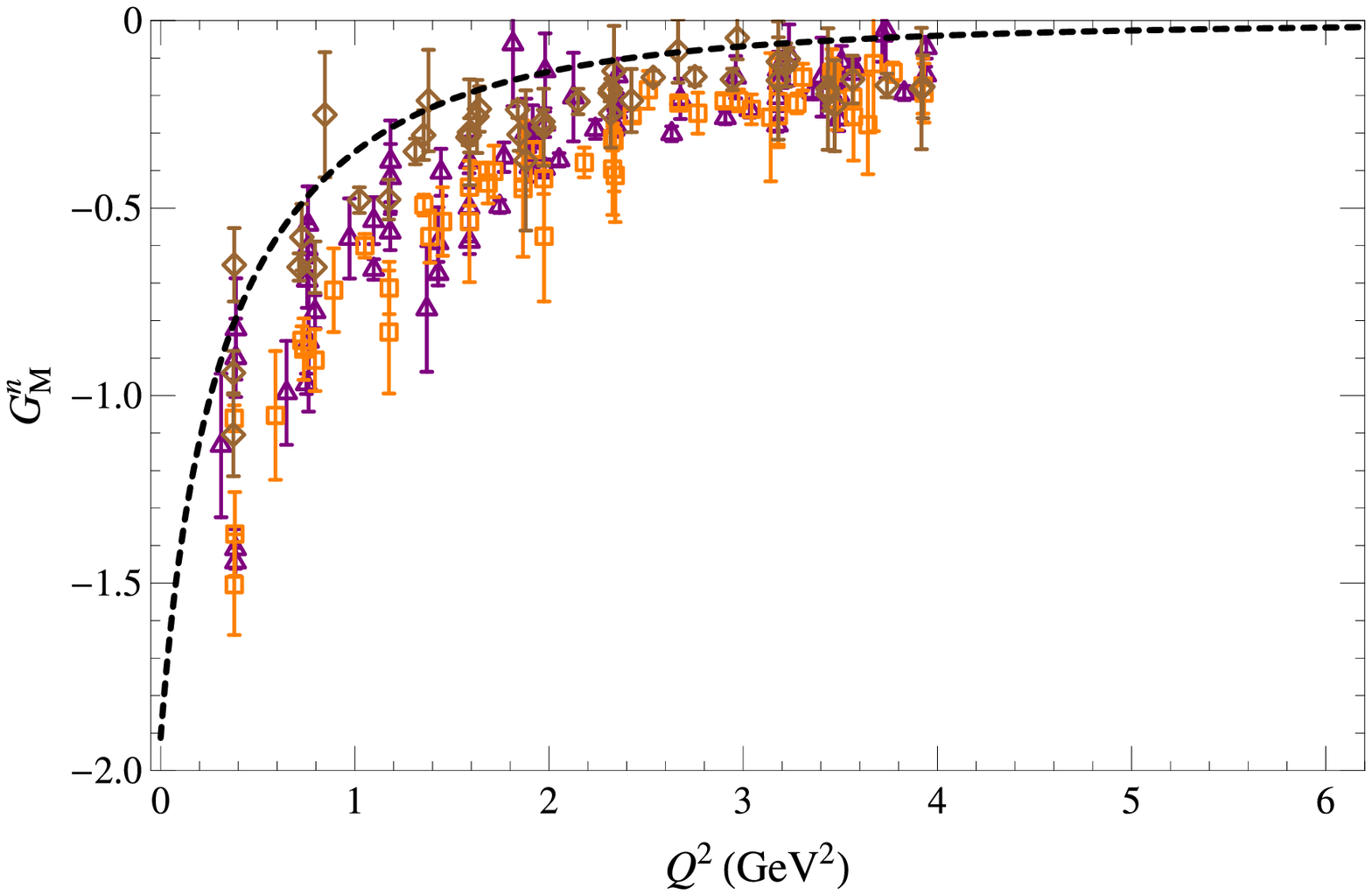}
\includegraphics[height=.33\textwidth]{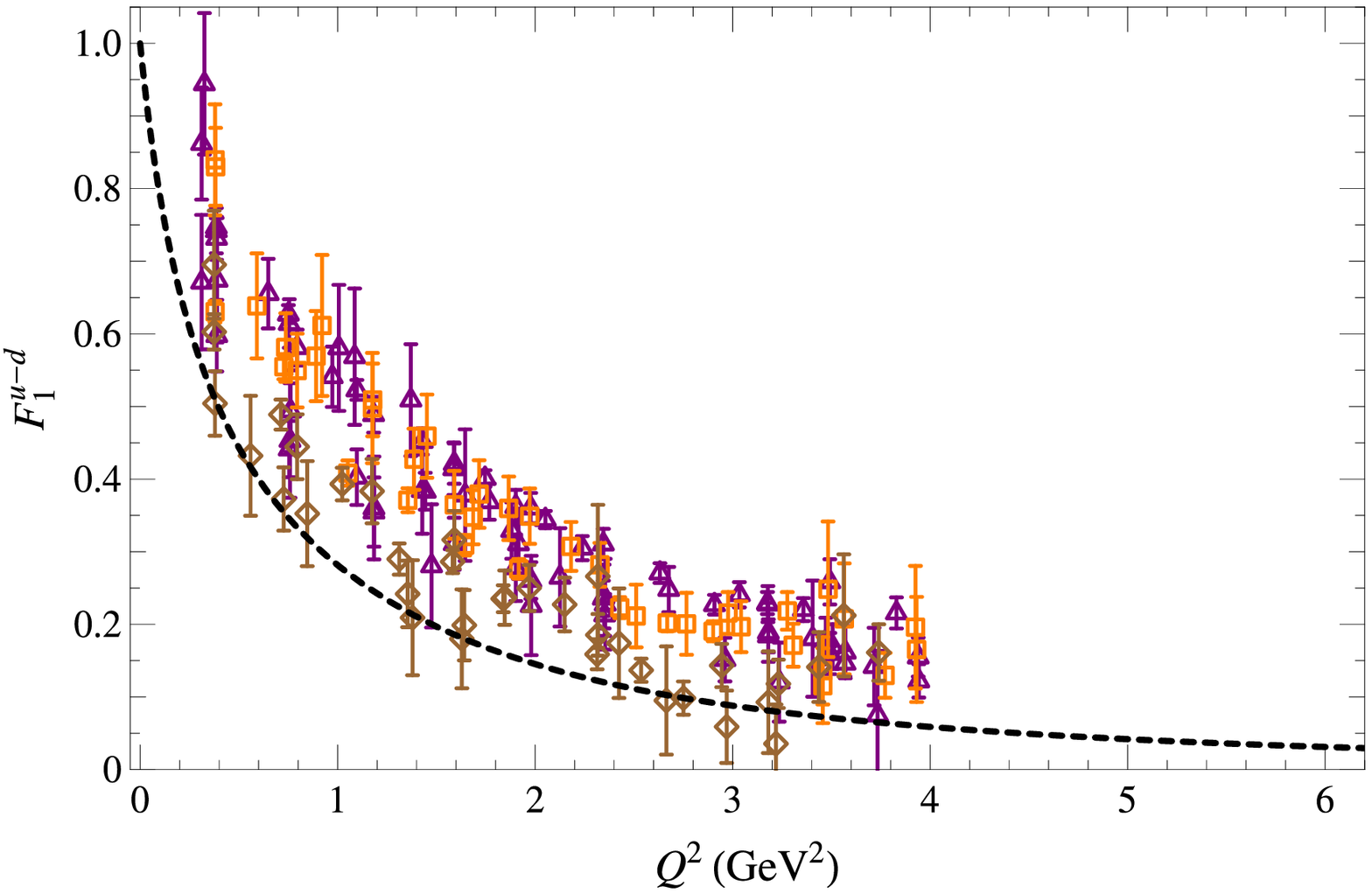}
\includegraphics[height=.33\textwidth]{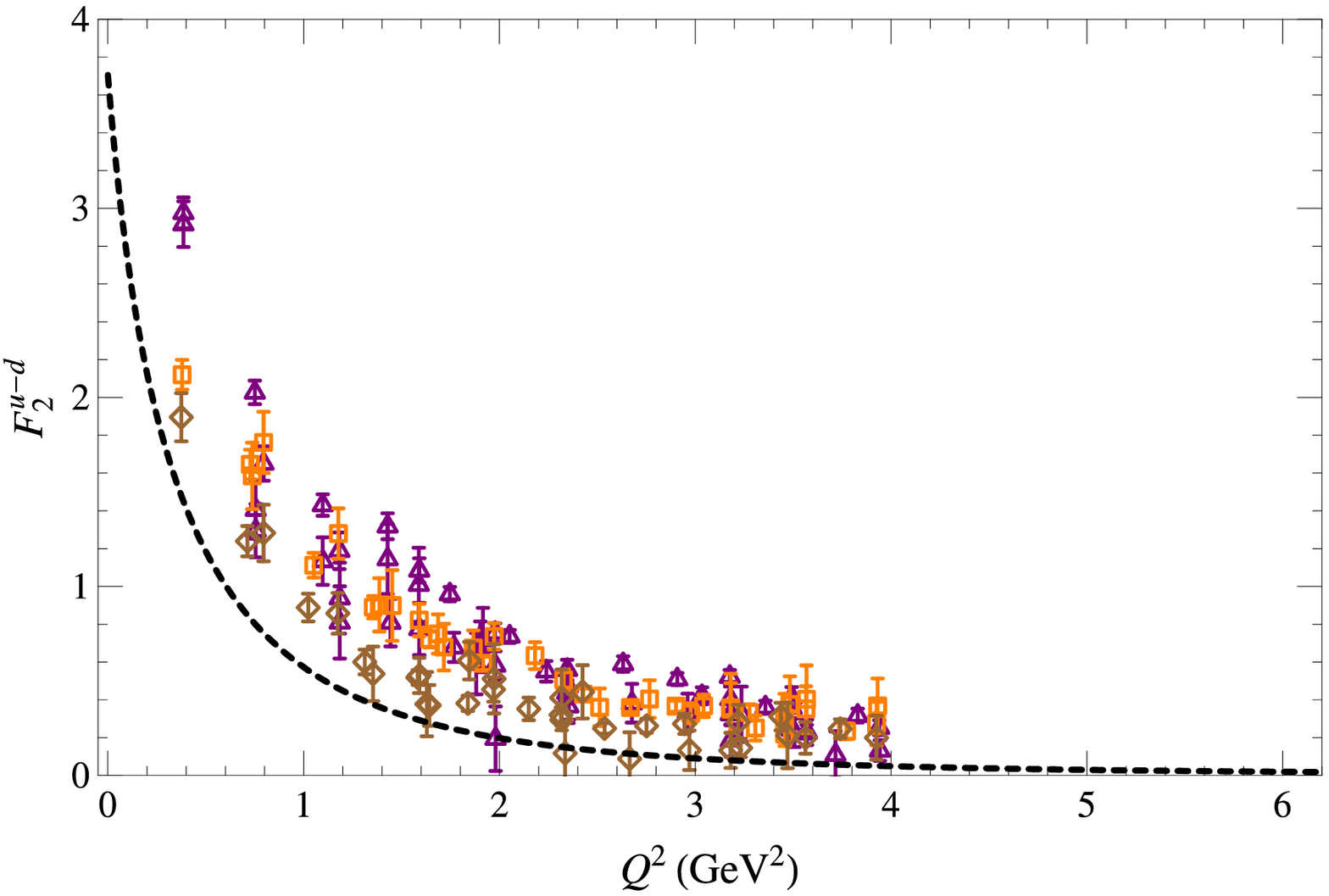}
\vspace{-0 cm} \caption{Nucleon form factors on 2+1 configurations with pion masses of 450 (brown), 580 (orange) and 875 (purple) MeV. The dashed lines are a parametrization of experimental form-factor data\cite{Arrington:2007ux,Kelly:2004hm}
} \label{fig:GEGM-dyn}
\end{figure}

Figures~\ref{fig:GEGM} and \ref{fig:GEGM-dyn} summarize the quenched and dynamical calculations of the electric $G_E^{p,n}$, magnetic $G_M^{p,n}$ and isovector Dirac and Pauli $F_{1,2}^v$ form factors as functions of momentum transfer $Q^2$; the different symbols (colors) indicate different quark masses (or equivalently, pion masses) used as input to the calculation, as indicated in the captions of the figures. The dashed lines on these plots are taken from a parametrization of the experimental data\cite{Arrington:2007ux,Kelly:2004hm}.
In the case of proton, the parametrization used should be valid up to around 6~GeV$^2$ for $G_E$ and 25~GeV$^2$ for $G_M$. %\cite{Arrington:2007ux}
The neutron form factors are only accurately known to 4.5~GeV$^2$ for $G_M^n$
and $1.5\mbox{ GeV}^2$ for $G_E^n$;
we extrapolate all the experimental parametrizations through 6~GeV$^2$ for comparison.
Later in this section, we will discuss the extrapolation of the lattice data to the physical pion mass and the form-factor ratios.

From the experimental parametrization of the Sachs form factors $G_{E,M}^{p,n}$, we can obtain Dirac and Pauli form factors $F_{1,2}$ by reversing the definitions in Eq.~\ref{eq:sachs}:
\begin{eqnarray}
F_1(Q^2) &=& \frac{G_E+\tau G_M}{1+\tau}\\
F_2(Q^2) &=& \frac{-G_E+ G_M}{1+\tau}
\end{eqnarray}
for both proton and neutron form factors, where $\tau=Q^2/(2M)^2$. The isovector $F^v$ form factors are just $F^p-F^n$. We can further estimate the individual quark contributions $F_{1,2}^{u,d}$ in terms of $F_{1,2}^{p,n}$ as
\begin{eqnarray}
F_{1,2}^u (Q^2) &=& F_{1,2}^n+2F_{1,2}^p\\
F_{1,2}^d (Q^2) &=&  2F_{1,2}^n+F_{1,2}^p.
 \label{eq:quark-pn}
\end{eqnarray}

All the form factors are nonperturbatively renormalized such that $G_E^p(0)=1$.
The largest available transfer momentum from the dynamical lattice is limited by the coarse lattice spacing. The inverse of the spatial lattice spacing is only around 1.6~GeV, while in the quenched case it is 2~GeV. Application of the same calculation procedure on finer lattices could easily extend the largest available transfer momentum.
The pion masses used are 1080, 720 and 480~MeV for the quenched ensembles and 875, 580 and 450~MeV for the dynamical ensembles.
Both quenched and dynamical form factors display a trend toward the experimental parametrization as the pion mass decreases except for $G_E^n$.
The quenched form factors have milder pion-mass dependence with around 300-MeV separation, compared with the dynamical calculation.
The lightest pion masses in the quenched and dynamical calculations are roughly the same, allowing us to see how the sea-quark contribution influences the form factors. It is immediately obvious in $G_E^p$ that the addition of the sea-quark degrees of freedom lowers the form factors toward the experimental line.
$G_E^n$ has the largest noise-to-signal; the magnitude is at the order of the disconnected contributions mentioned earlier this subsection which could be influencing the connected lattice points significantly.

The lattice data show some broad trends: As the lattice momentum used in any given point increases, the noise also increases. In this calculation, this means that the highest transfer-momentum points will have larger uncertainty, but also that there will be some points with smaller transfer momentum that have large uncertainty simply because they are constructed using large initial and final momenta. For purposes of clarity, we omit some of these points with very large errors; all points are included in the fits, although points with large error have little influence.


%%%%%%%%%%%%%%%%%%%%%%%%%%%%%%%%%%%%%%%%%%%%%%%%%%%%%%%%%%%%%%%%%%%%%%%%%%%%%%%
\subsection{Isovector Radii}\label{subsec:radii}

The size of the nucleon characterized by the effective charge and magnetic radii can be determined from the electromagnetic form factors.
We first examine commonly calculated quantities using Dirac and Pauli isovector currents, where the disconnected contribution is highly suppressed due to isospin symmetry.

The isovector Dirac and Pauli mean-squared radii can be extracted from the isovector electric form factors $F_{1,2}^v$ via
\begin{eqnarray}\label{eq:GEradii}
\langle r_{1,2}^2\rangle &=& (-6)\frac{d}{dQ^2}\left(\frac{F_{1,2}^v(Q^2)}{F_{1,2}^v(0)}\right)\Big|_{Q^2=0}.
\end{eqnarray}
Most groups have studied radii with the $Q^2$ dependence over ranges 0.5--2.0~GeV$^2$ among their own data and found the extracted radii to be independent (within the statistical error bars) of $Q^2$ choice\cite{:2010jn,Syritsyn:2009mx,Lin:2008mr}.

\begin{figure}
\begin{center}
\includegraphics[width=0.45\textwidth]{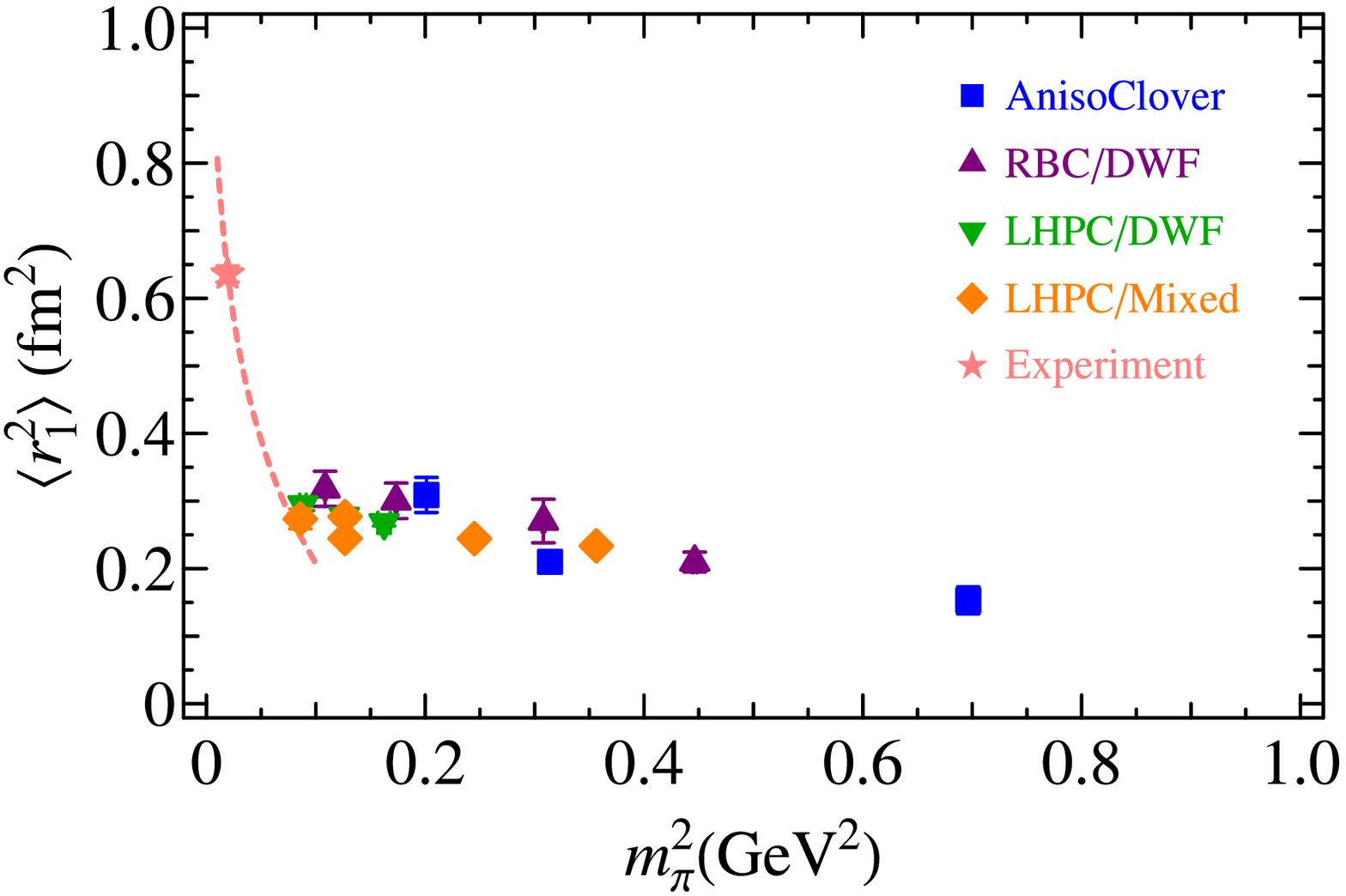}
\includegraphics[width=0.45\textwidth]{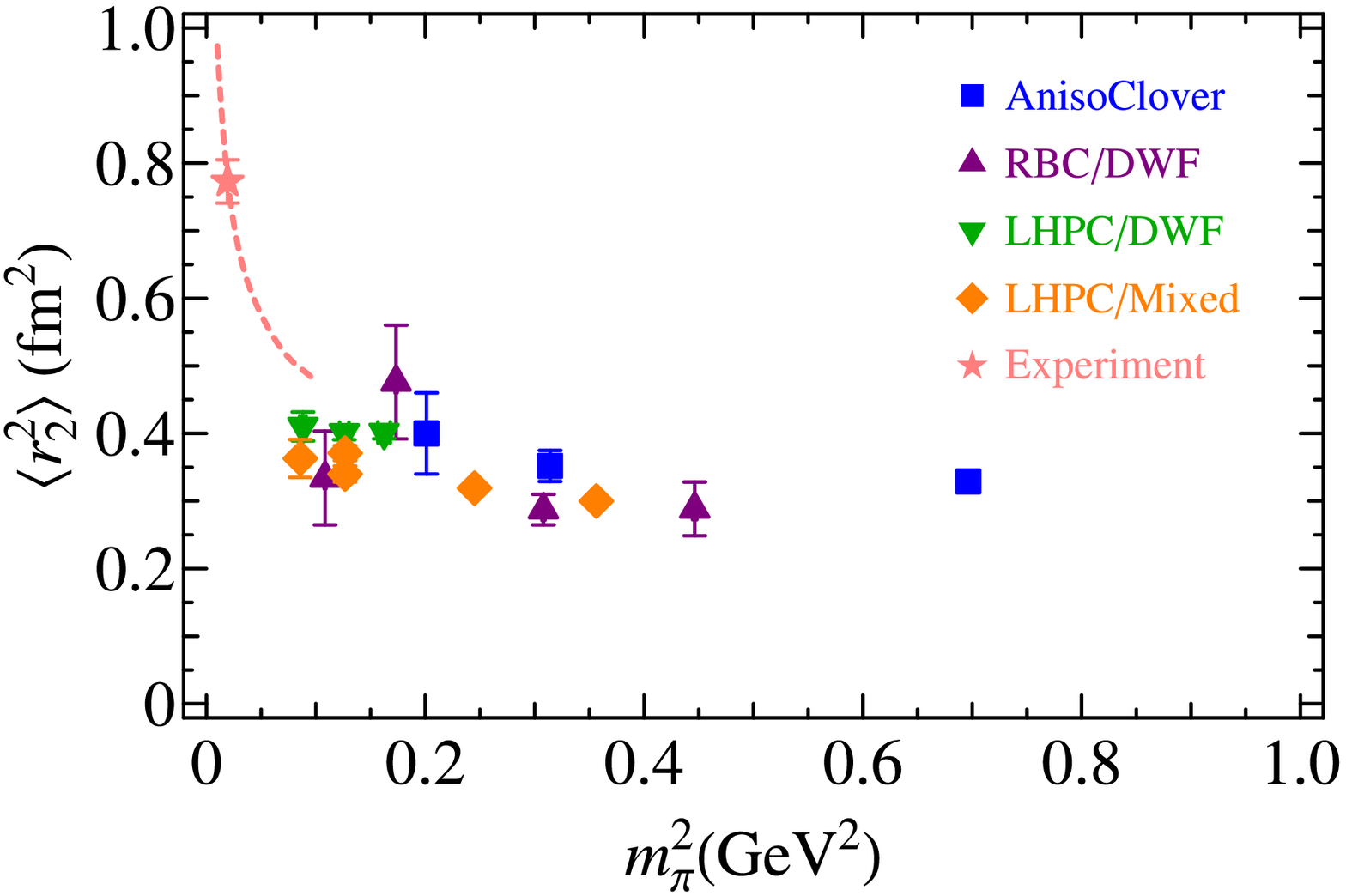}
\end{center}
\vspace{-0.7cm}
\caption{Summary of the isovector Dirac and Pauli mean-squared radii from all currently existing $N_f=2+1$ nucleon electromagnetic form-factor calculations\cite{:2010jn,Yamazaki:2009zq,Syritsyn:2009mx,Lin:2008mr}. The dashed line indicates the leading-order HBXPT prediction.}\label{fig:all21-rv2}
\end{figure}

It is commonly agreed that a dipole extrapolation should be used for $F_1^v$, but opinions differ as to whether a dipole ($a (Q^2+b)^{-2}$) or tripole ($a (Q^2+b)^{-3}$) is preferred for $F_2^v$. Refs.~\cite{Syritsyn:2009mx,Lin:2008mr} found insignificant differences between results for either choice, while Ref.~\cite{:2010jn} observed some discrepancy and adopted the numbers from the tripole for $\langle r_{2}^2\rangle$. 
A summary of all the $N_f=2+1$ lattice calculations of the isovector radii can be found in Fig.~\ref{fig:all21-rv2}. Note that only the statistical errors are shown in this figure.

The Dirac and Pauli mean-squared radii from the dynamical ensembles are summarized in Fig.~\ref{fig:all21-rv2} along with other $N_f=2+1$ lattice calculations and lowest-order heavy-baryon chiral perturbation theory (HBXPT) using experimental inputs\cite{%Beg:1973sc,
Bernard:1998gv}. Our results are nicely in agreement with isotropic $N_f=2+1$ calculations having various sea and valence fermion actions; this demonstrates the universality of the lattice QCD calculations.

%%%%%%%%%%%%%%%%%%%%
\subsection{Magnetic Moments}\label{subsec:magnetic-moments}

Calculations in a finite volume (without using twisted boundary conditions or an external magnetic field) cannot give us the $Q^2\approx0$ Pauli $F_2$ or magnetic $G_M$ form factor. However, we are interested in the (anomalous) magnetic moments:
\begin{eqnarray}
\kappa_x &=& F_2^x(Q^2=0)\\
\mu^x &=& G_M^x (Q^2=0),
\end{eqnarray}
where $x$ can be the proton, neutron, or up or down quark flavors. Two approaches are considered:
The first method is to perform a dipole fit to the $Q^2$-dependence of the form factors.
However, due to the limited minimum momentum available (a constraint related to the lattice box size), the fit form can be poorly constrained in the near-zero $Q^2$ region, resulting in fit-instability in obtaining the magnetic moments.
The second method can be applied in the small $Q^2$ region by polynomial fitting to the ratio of the magnetic to the electric form factor, $G_M/G_E$ (or $F_2/F_1$). The two form factors are expected to be functions of $\tau=\frac{Q^2}{4m_N^2}$. When $Q^2$ is small, we can Taylor expand both form factors (except for the neutron):
\begin{eqnarray}
G_M/G_E &=&  {\mu/G_E(0)} (1+a_1 \tau) + O(\tau^2)\\
F_2/F_1 &=&  {\kappa/F_1(0)} (1+a_1 \tau) + O(\tau^2).
\end{eqnarray}
Both methods (dipole fit and linear fit to form-factor ratios) have been performed in a previous study on most of the octet baryons\cite{Lin:2008mr} and the latter method was found to have good consistency and improved stability.
The magnetic moments are related via $\mu_x=\kappa_x+e_x$, where $e$ is the electric charge and $x$ can be proton or up or down quark flavors.

\begin{figure}
\begin{center}
\includegraphics[width=0.45\textwidth]{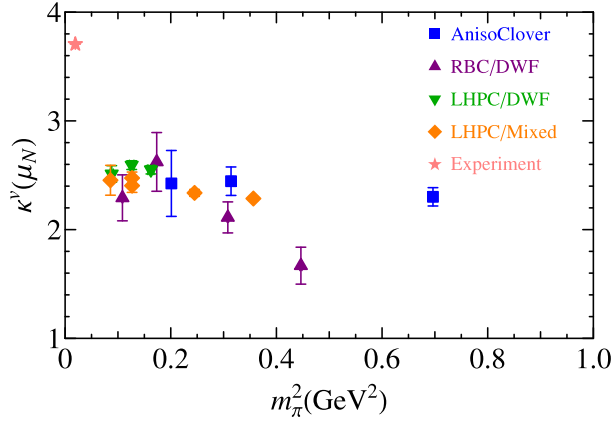}
\end{center}
\vspace{-0.7cm}
\caption{Summary of the normalized isovector anomalous magnetic moments from all currently existing $N_f=2+1$ nucleon electromagnetic form-factor calculations\cite{:2010jn,Yamazaki:2009zq,Syritsyn:2009mx,Lin:2008mr}, including this work}\label{fig:all21-normKappa2}
\end{figure}

Here we will only show the dipole form to extract the anomalous magnetic moments from isovector form factors; both approaches give results consistent within statistical errors.
To better compare our results with other studies, we convert $\kappa$ into the natural units of the nuclear magneton ($\mu_N=\frac{e}{2m_N}$) by multiplying by $\frac{m_N^{\rm phys}}{m_N^{\rm lat}}$. Our dynamical results are displayed in Fig.~\ref{fig:all21-normKappa2} along with other $N_f=2+1$ results. The $\kappa$ obtained from tripole fitting are consistent with dipole results; similar observations were made by other $N_f=2+1$ calculations\cite{:2010jn,Yamazaki:2009zq,Syritsyn:2009mx}. Once again, we observe the universal behavior among different groups and lattice-parameter choices. We also note that the lattice results display a rather mild dependence on the pion mass in the range from 300~MeV to 850~MeV, yet they are only 2/3 the experimental value.
We may again expect a rapid rise in the smaller-pion mass region if lattice QCD correctly reproduces the experimental values.
In Ref.~\cite{Lin:2008mr}, the magnetic moments of the octet baryons were studied with SU(3) next-to-leading-order (NLO) HBXPT formulae; they found large discrepancies from the lattice points (with the lightest pion at 350~MeV), suggesting that NNLO effects are significant.
We might have similar expectations for the isovector anomalous magnetic moments.

%%%%%%%%%%%%%%%%%%%%%%%%%%%%%%%%%%%%%%%%%%%%%%%%%%%%%%%%%%%%%%%%%%%%%%%%%%%%%%%%%
%\clearpage
\subsection{Chiral Extrapolation of Form Factors}\label{subsec:}

Phenomenologically, as done in Refs.~\cite{Arrington:2007ux,Kelly:2004hm}, it is common to describe the $Q^2$-dependence of the experimental $G_{E,M}$ form-factor data using a dimensionless parameter $\tau=\frac{Q^2}{4m_N^2}$ via an expansion form
\begin{eqnarray}
G_E^p \left(\mbox{or }\frac{G_M^{p,n}}{\mu_{p,n}}\right) &=&
	\frac{1+\sum_{i=1}^{k-2}a_i\tau^i}{1+\sum_{i=1}^{k}b_i\tau^i} \nonumber \\
G_E^n &=& \frac{a_1 \tau}{1+b_1\tau} G_D, \label{eq:GEM-fitform}
\end{eqnarray}
where $k$ is a selected integer and $G_D$ is $\frac{1}{(1+Q^2/(0.71\mbox{ GeV}^2)^2}$.
Note that in the continuum, there is no difference between extrapolating the data in terms of $Q^2$ or $\tau$, since the $m_N$ is fixed.
However, in a lattice calculation where the sea and valence quark masses vary, $m_N^{\rm lat}$ is a function of the pion mass. Using a dimensionless parameter should help to smooth the chiral extrapolation.
We could also extrapolate $F_{1,2}^{u,d}$ through a similar procedure as $G_{E,M}^{p,n}$, except that $F_2$ should have one fewer power of $\tau$; that is,
\begin{eqnarray}
F_{1} &=&
\frac{a_0+\sum_{i=1}^{k-2}a_i\tau^i}{1+\sum_{i=1}^{k}b_i\tau^i} \nonumber \\
\left(\frac{F_{2}}{\kappa}\right) &=& \frac{1+\sum_{i=1}^{k-3}a_i\tau^i}{1+\sum_{i=1}^{k}b_i\tau^i}. \label{eq:F12-fitform}
\end{eqnarray}

To extrapolate to the physical pion mass, we can use one of several different approaches: 
First, we could use a simultaneous fit to the $\tau$ and $m_\pi$ dependence where the fit parameters, $a_i$ and $b_i$ in Eqs.~\ref{eq:GEM-fitform} and \ref{eq:F12-fitform} are in terms of $m_\pi^2$. 
A second approach is to fit data for each lattice momentum as a function of $Q^2(m_\pi^2)$.
Each data point $f(Q^2,m_\pi^2)$ composes with the same kinematic momentum combination can then be extrapolated to the physical point, where we can further apply Eqs.~\ref{eq:GEM-fitform} and \ref{eq:F12-fitform} to fit the $Q^2$ dependence.

\begin{figure}[!h]
\begin{center}
\includegraphics[width=0.45\textwidth]{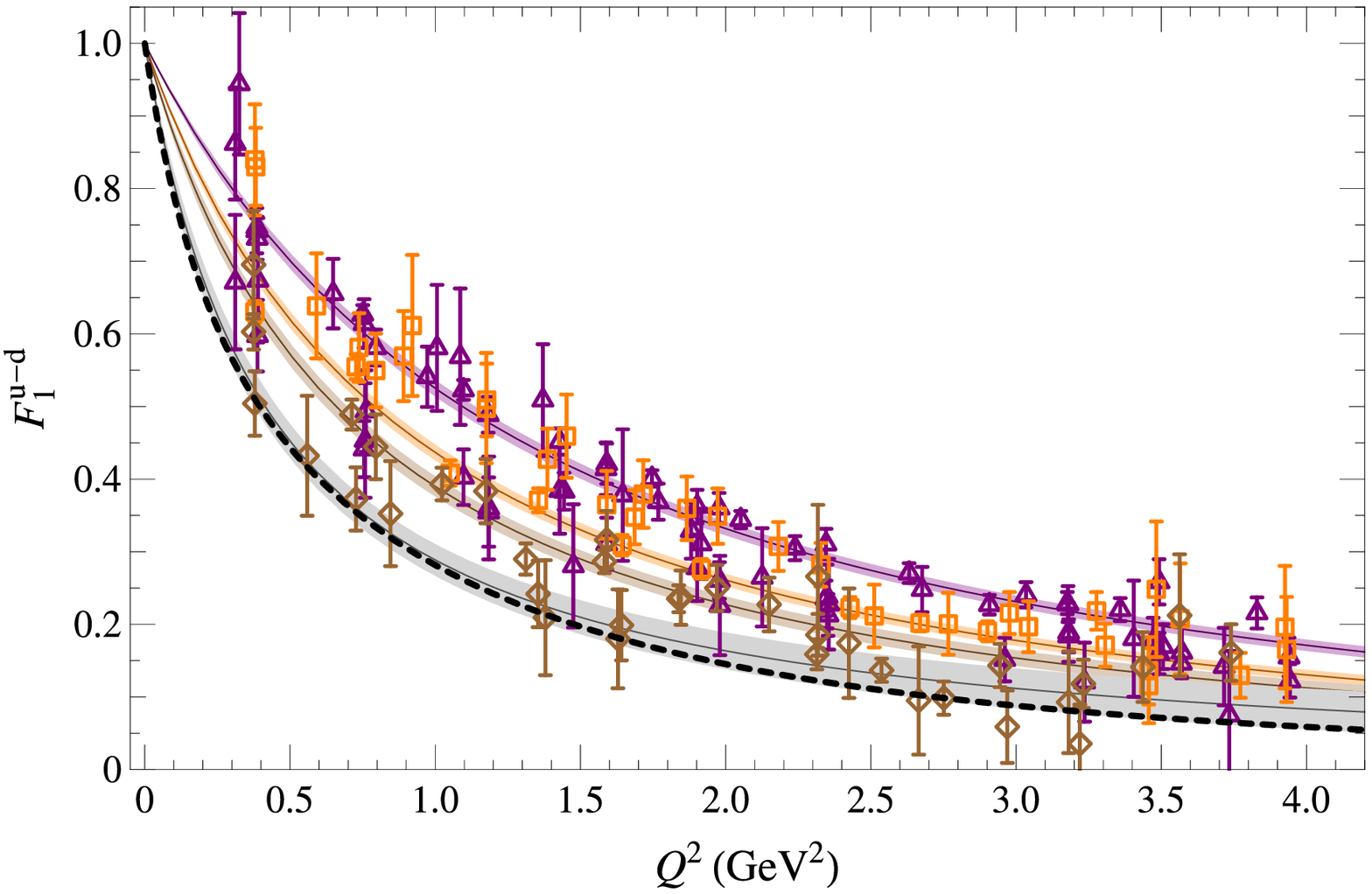}
\includegraphics[width=0.45\textwidth]{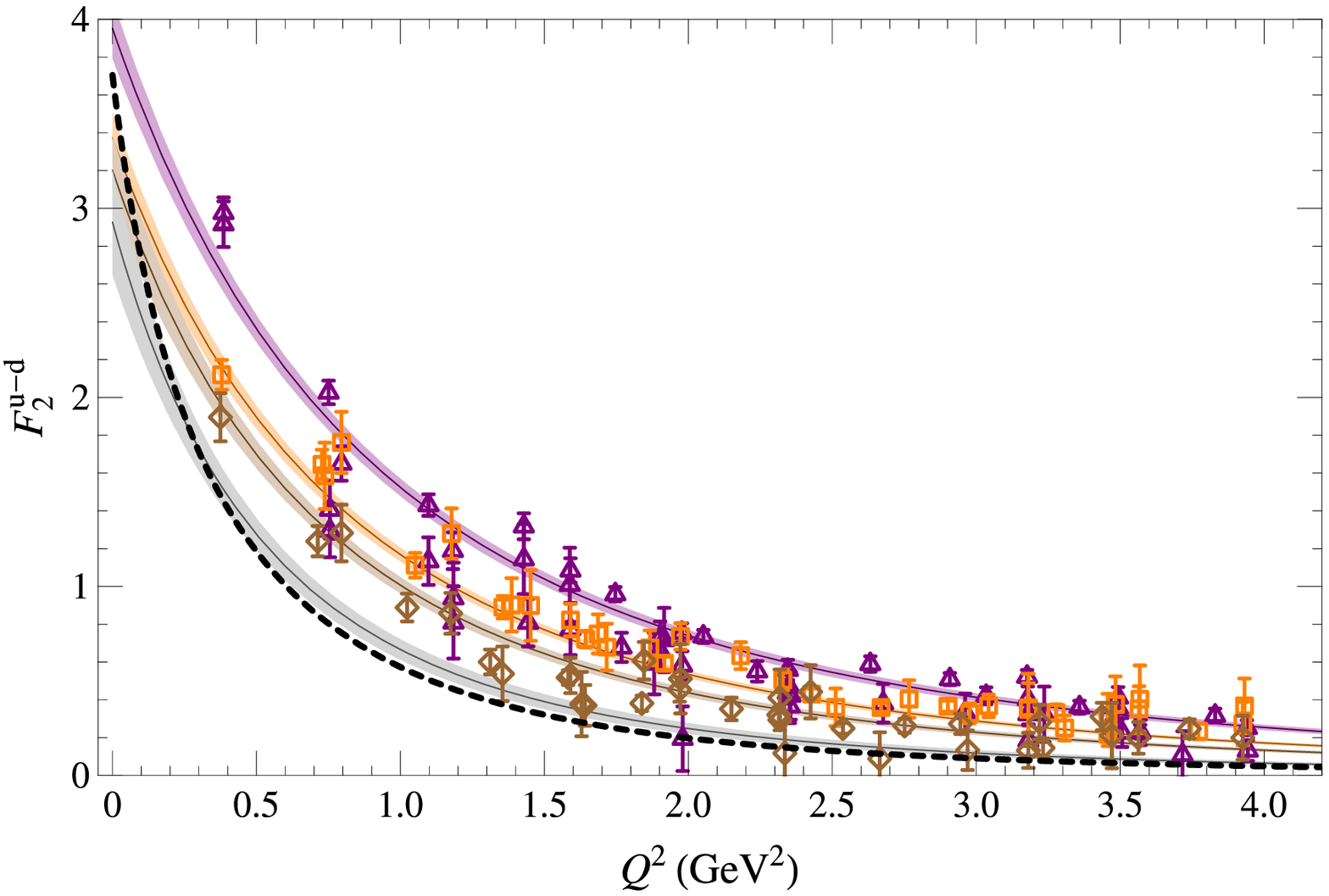}
\end{center}
\vspace{-0.7cm}
\caption{Isovector Dirac and Pauli form-factor extrapolation using $N_f=2+1$ calculations in this work. The shaded bands show the simultaneous fit at the pion mass corresponding to each data set, where the color of the band matches the color of the points. The lowest gray band is the extrapolation to the physical pion mass.}\label{fig:isoF12}
\end{figure}

In this work, we simultaneously fit the $m_\pi$ and $Q^2$ (or $\tau$) dependence of the lattice data, expanding each fit parameter in terms of the pion mass: $a_i=a_i^{(0)}+a_i^{(1)} m_\pi^2$.
The pion masses on these ensembles are heavy enough that the linear ansatz for squared-pion-mass dependence should hold, simplifying the complicated extrapolation.
This doubles the number of fitting parameters, but increases the number of data points by a factor of the number of pion masses used (3 in our case).
Figure~\ref{fig:isoF12} shows an example of the extrapolations on the isovector Dirac and Pauli form factor in the dynamical ensembles. The $\chi^2/{\rm dof}$ for $F_{1,2}^v$ are 1.4 and 2.0, respectively. The points and their corresponding symbols are the same as those in Fig.~\ref{fig:GEGM-dyn} and now we have the $Q^2$ extrapolation line with statistical error band going through these lattice points. Further, the lowest line/band represents the extrapolated form factors at the physical pion mass.
We repeat a similar process for both quenched ($N_f=0$) and dynamical ($N_f=2+1$) on other form factors.

\begin{figure}[!h]
\begin{center}
\includegraphics[width=0.45\textwidth]{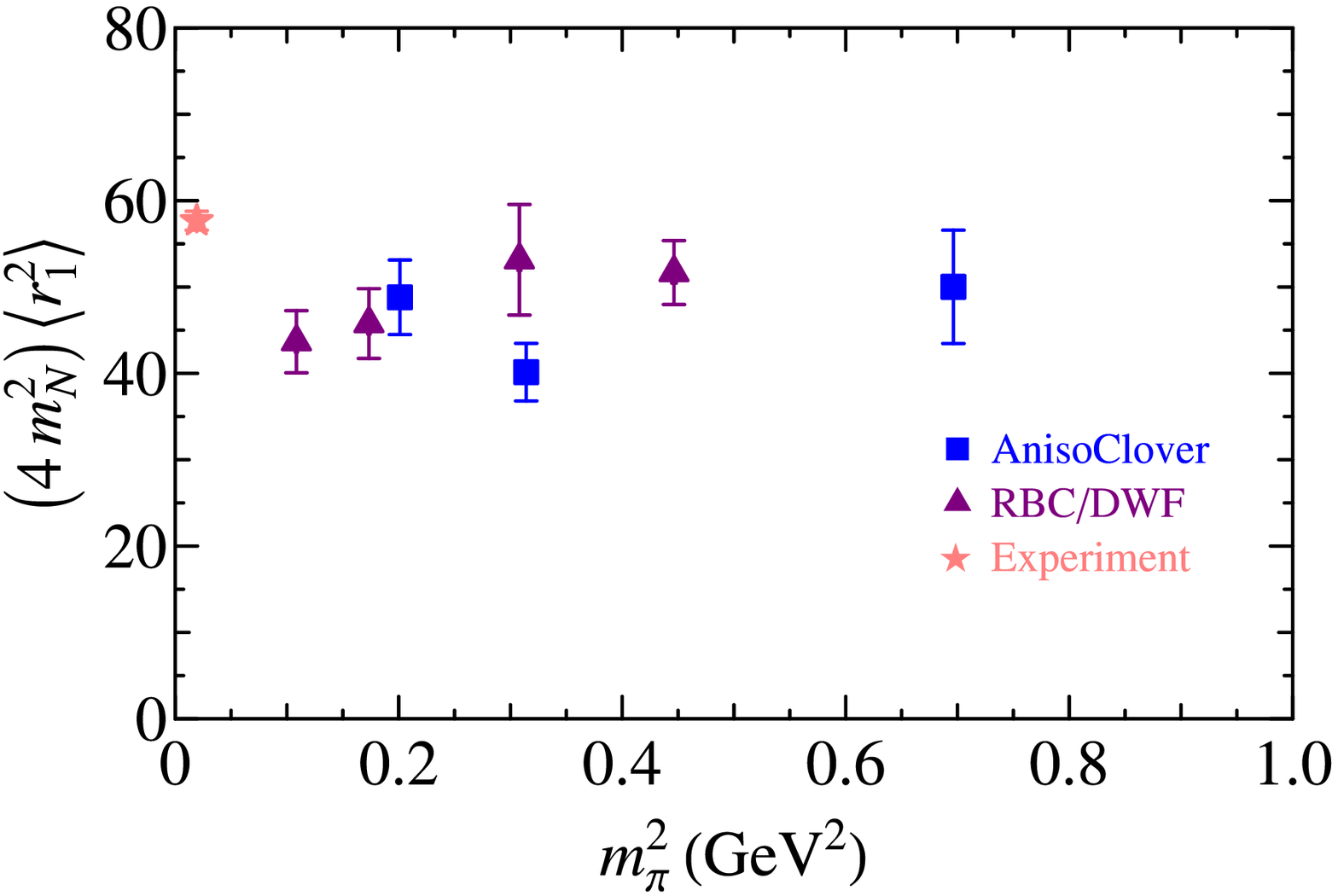}
\includegraphics[width=0.45\textwidth]{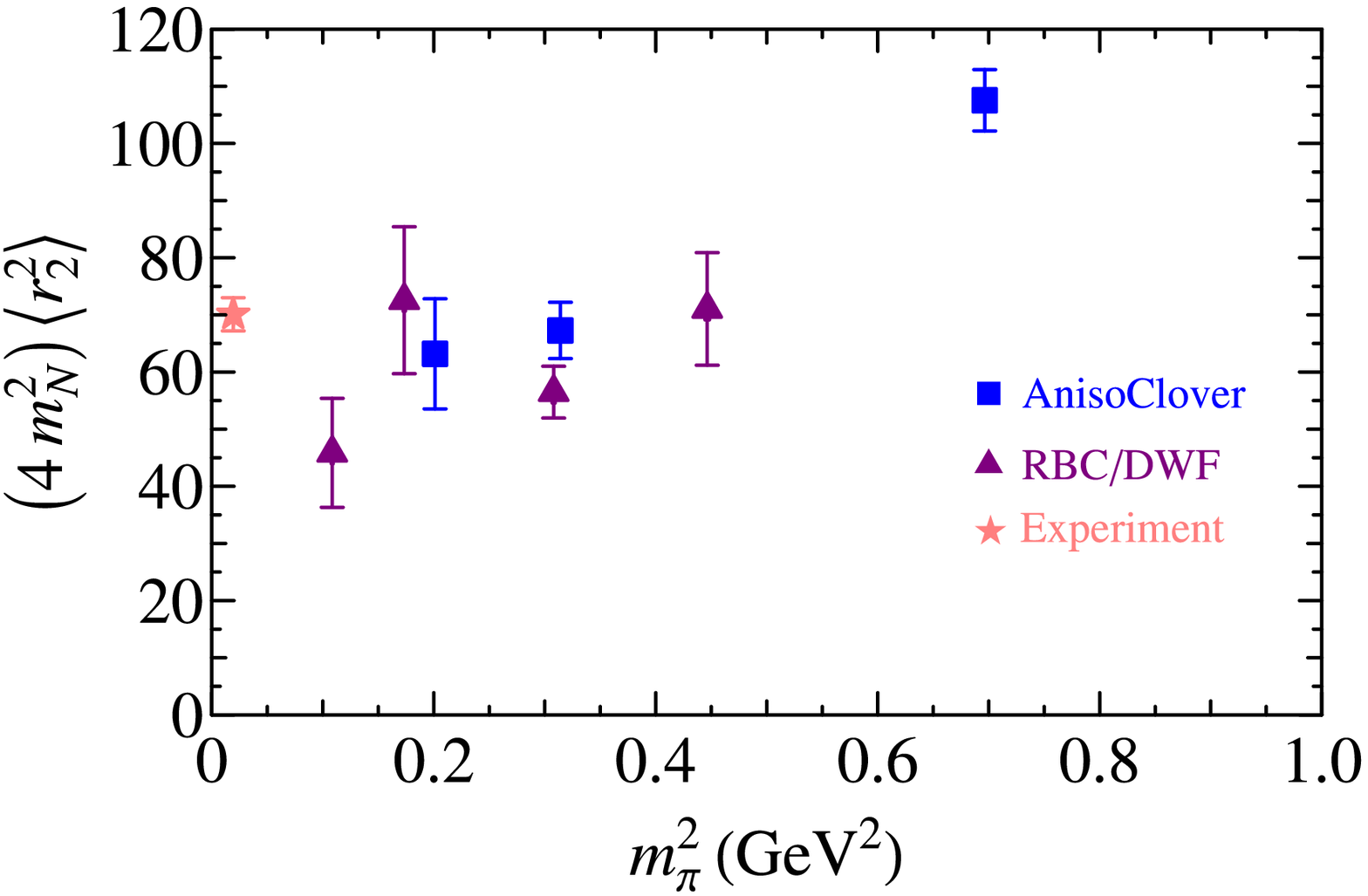}
\end{center}
\vspace{-0.7cm}
\caption{Summary of the ``scaled'' isovector Dirac and Pauli mean-squared radii from our and RBC/UKQCD  $N_f=2+1$ nucleon electromagnetic form-factor calculations\cite{Yamazaki:2009zq}}\label{fig:some-4mN2rv2}
\end{figure}

We find good agreement for the extrapolation to the physical pion mass in Fig.~\ref{fig:isoF12}; that is, the curvature (which corresponds to the charge radii) as a function of $Q^2$ fits nicely with interpolating forms derived from experimental data. This seems to contradict what we obtained in Subsec.~\ref{subsec:radii}.
Note that the difference in terms of $Q^2$ and $\tau$ dependence is a factor of $4(m_N^{\rm lat})^2$. This makes no difference in the continuum, since the nucleon mass is a fixed constant. However, in lattice-QCD calculations, where the quark masses are varied (as shown Fig.~\ref{fig:all21mN}), this could create a strong dependence.
The good agreement in the radii extrapolation might be due to cancellation between the chiral curvatures in the nucleon mass and the radii. We compare our results with RBC/UKQCD $N_f=2+1$ results and observe similar behavior.
Figure~\ref{fig:some-4mN2rv2} shows the same $N_f=2+1$ products of the Dirac and Pauli radii with $4(m_N^{\rm lat})^2$ from our and RBC/UKQCD $N_f=2+1$ results.
The results are encouraging; the notorious charge radii problem may be ameliorated by looking at dimensionless quantities. In the Dirac radii product case, the discrepancy is improved to within a few standard deviations and even smaller for the Pauli radii product. However, further investigation and improvement in the statistical errors of these calculations will be required to fully understand the nature of these dimensionless quantities.

\begin{figure}[!h]
\begin{center}
\includegraphics[width=0.45\textwidth]{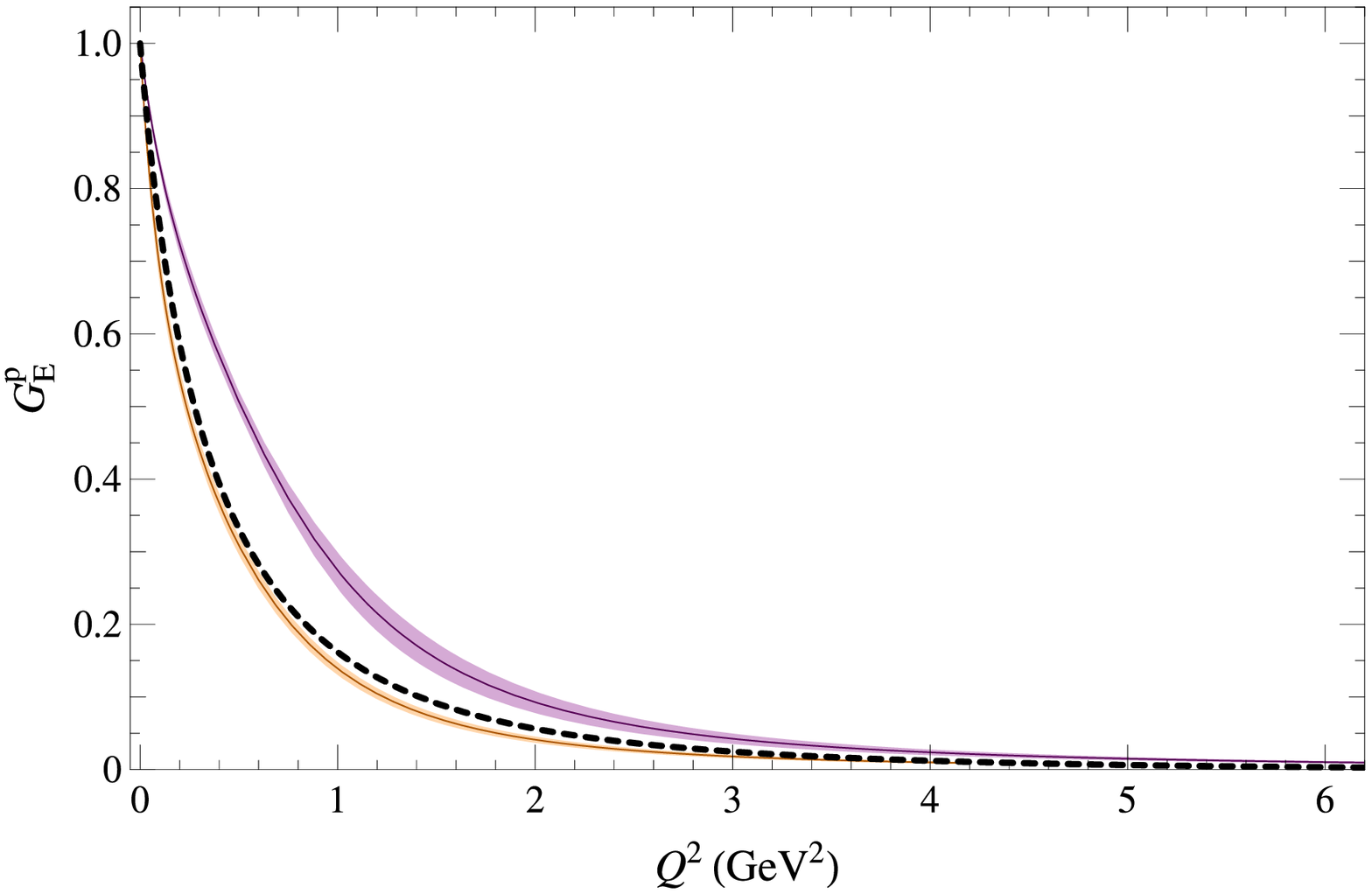}
\includegraphics[width=0.45\textwidth]{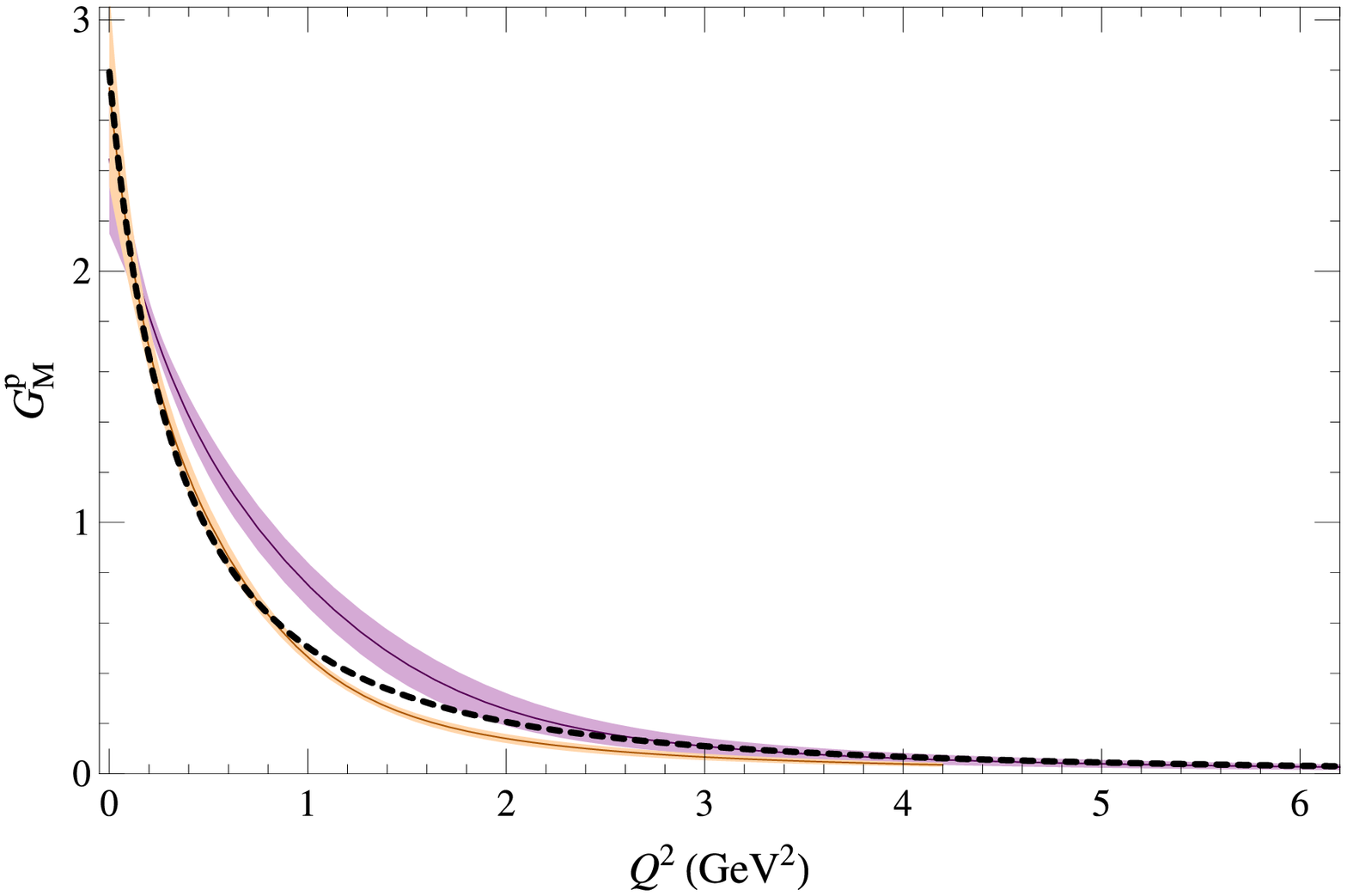}
\includegraphics[width=0.45\textwidth]{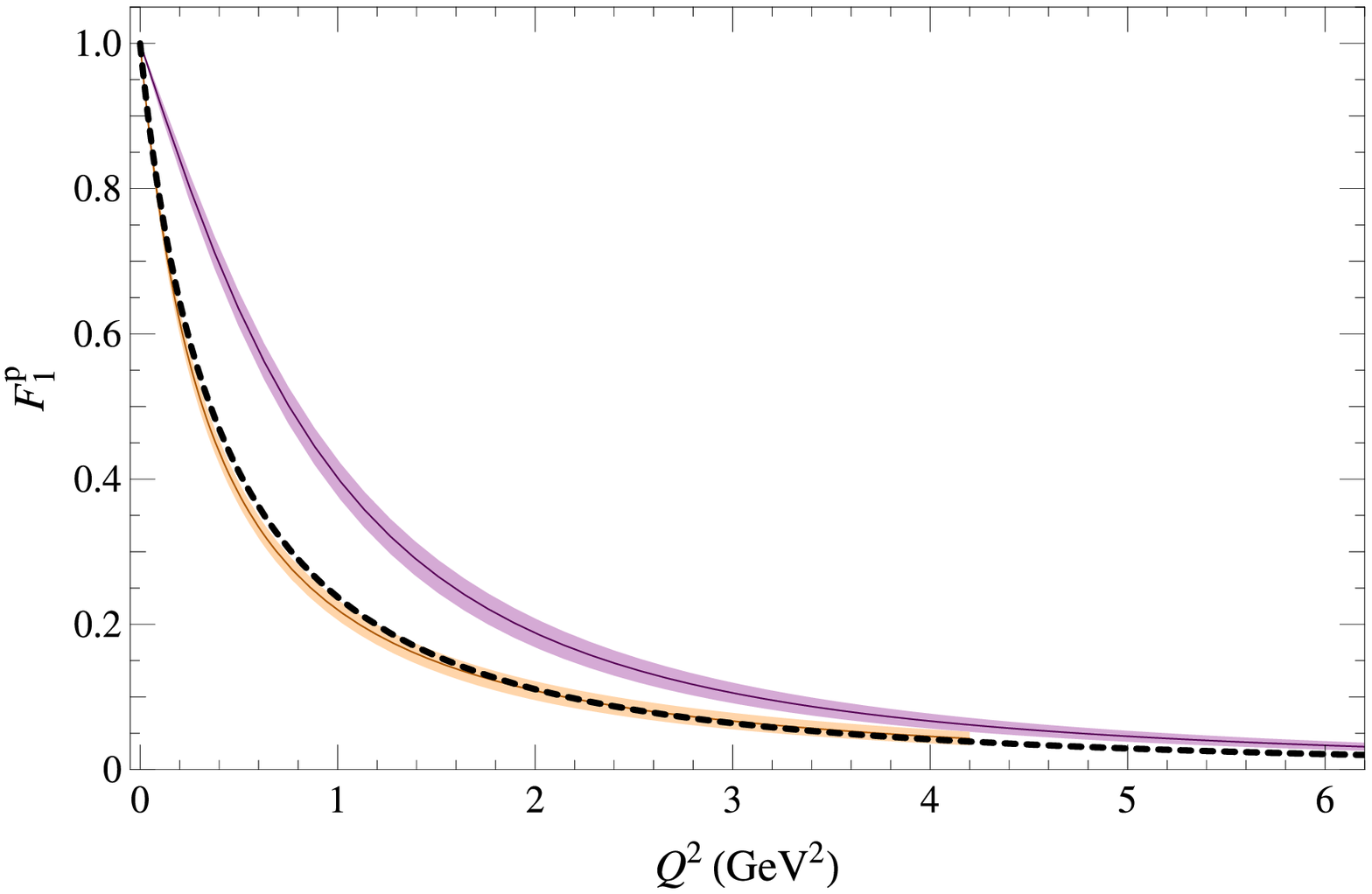}
\includegraphics[width=0.45\textwidth]{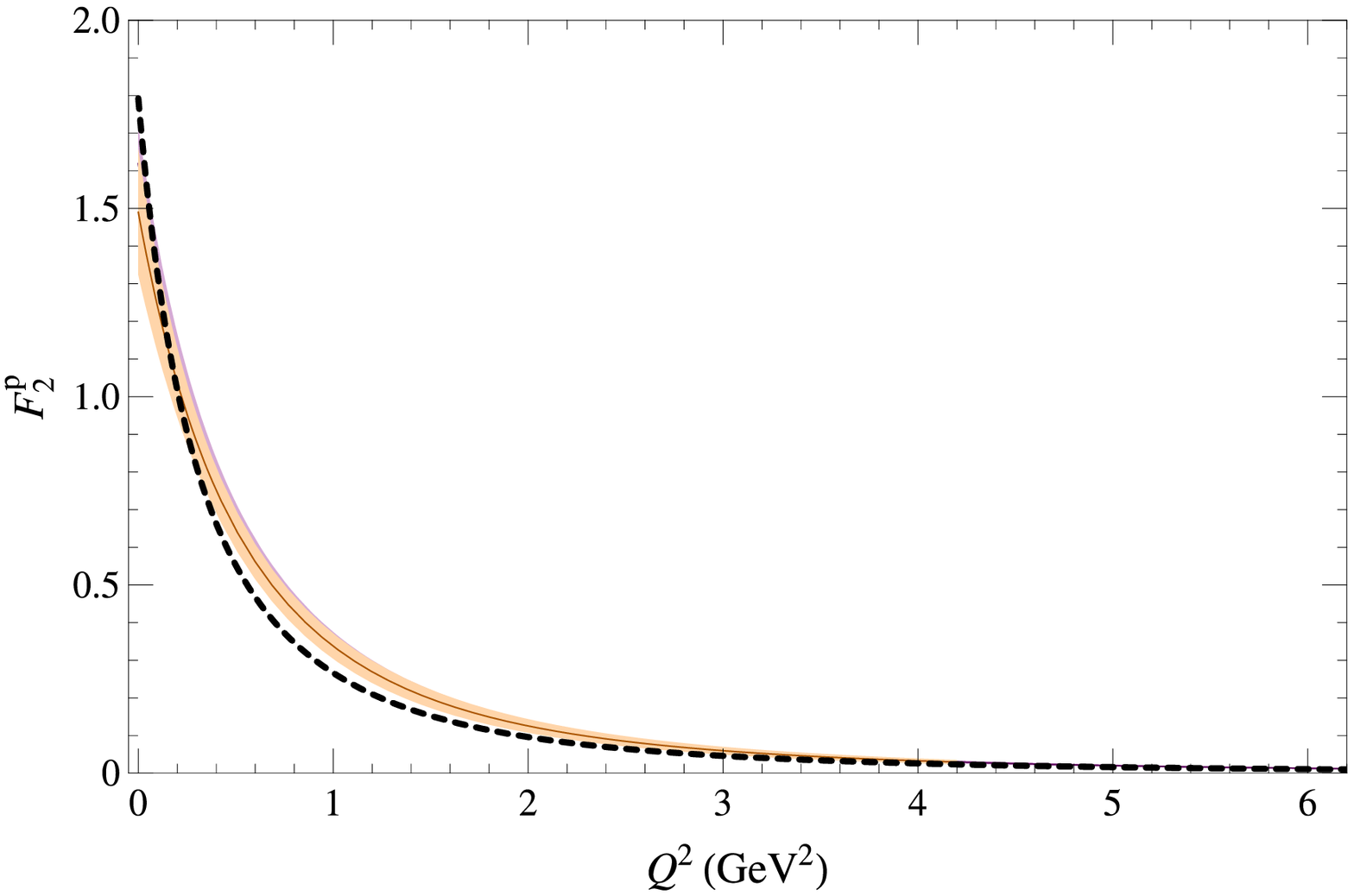}
\end{center}
\vspace{-0.7cm}
\caption{Proton form factors on both quenched (purple) and dynamical (orange) ensembles. The dashed line represents the experimental parametrization and the lines with bands are from the lattice calculations.}\label{fig:proton-FF}
\end{figure}

Figure~\ref{fig:proton-FF} shows the proton Sachs, Dirac and Pauli form factors at the physical pion mass from both quenched and dynamical ensembles. The dashed lines are the experimental parametrization as mentioned in earlier subsections. We find a significant difference in the results using quenched or dynamical ensembles, most strongly reflected in the electric and Dirac form factors. These form factors are better constrained due to the fixed data point at $Q^2=0$, and the dynamical extrapolations are only a couple of sigma away from the experimental line; the quenched results are consistently higher. The magnetic and Pauli form factor extrapolations, on the contrary, fail to reproduce the (anomalous) magnetic moments, similar to what we discuss in the magnetic moment subsection, and the extrapolations diverge around zero transfer momentum. To our surprise the dynamical and quenched results are consistent for the proton Pauli form factor. An examination of the original lattice data to make sure that this is not caused by the extrapolation verifies that the similarity comes from the data. This indicates that the Pauli form factor (for the proton) is not very sensitive to whether the QCD vacuum contains fermion degrees of freedom.

\begin{figure}[!h]
\begin{center}
\includegraphics[width=0.45\textwidth]{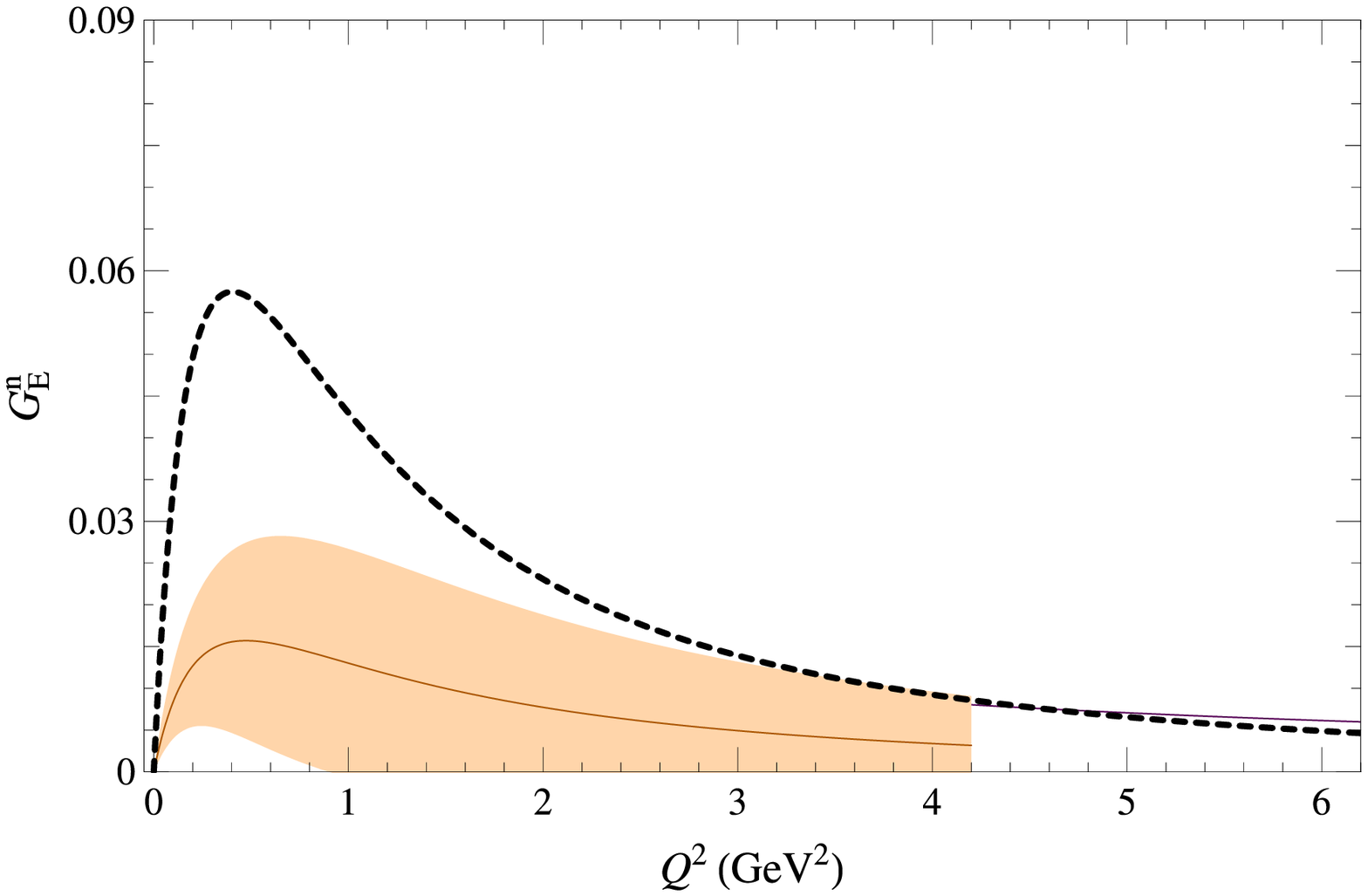}
\includegraphics[width=0.45\textwidth]{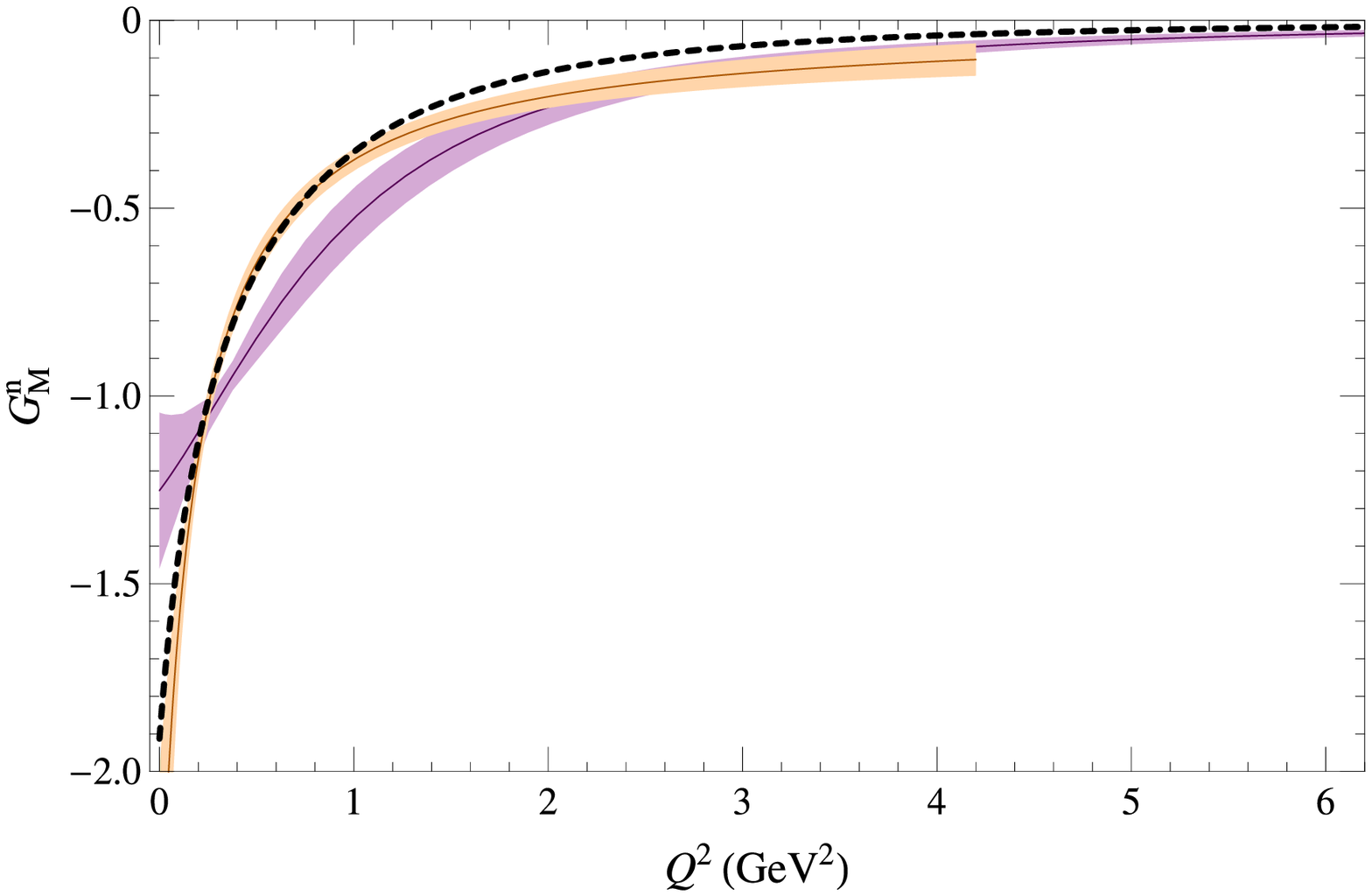}
\includegraphics[width=0.45\textwidth]{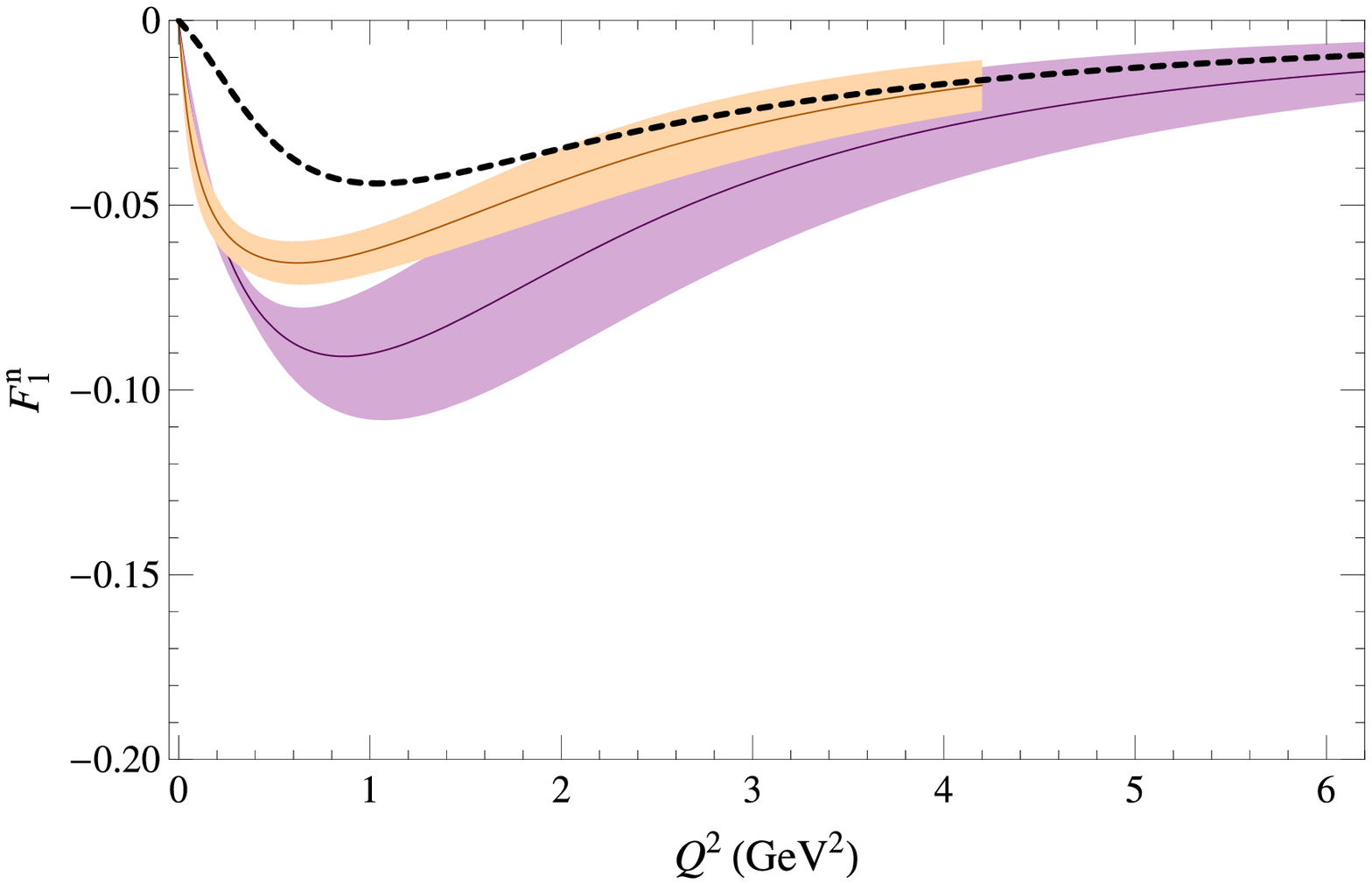}
\includegraphics[width=0.45\textwidth]{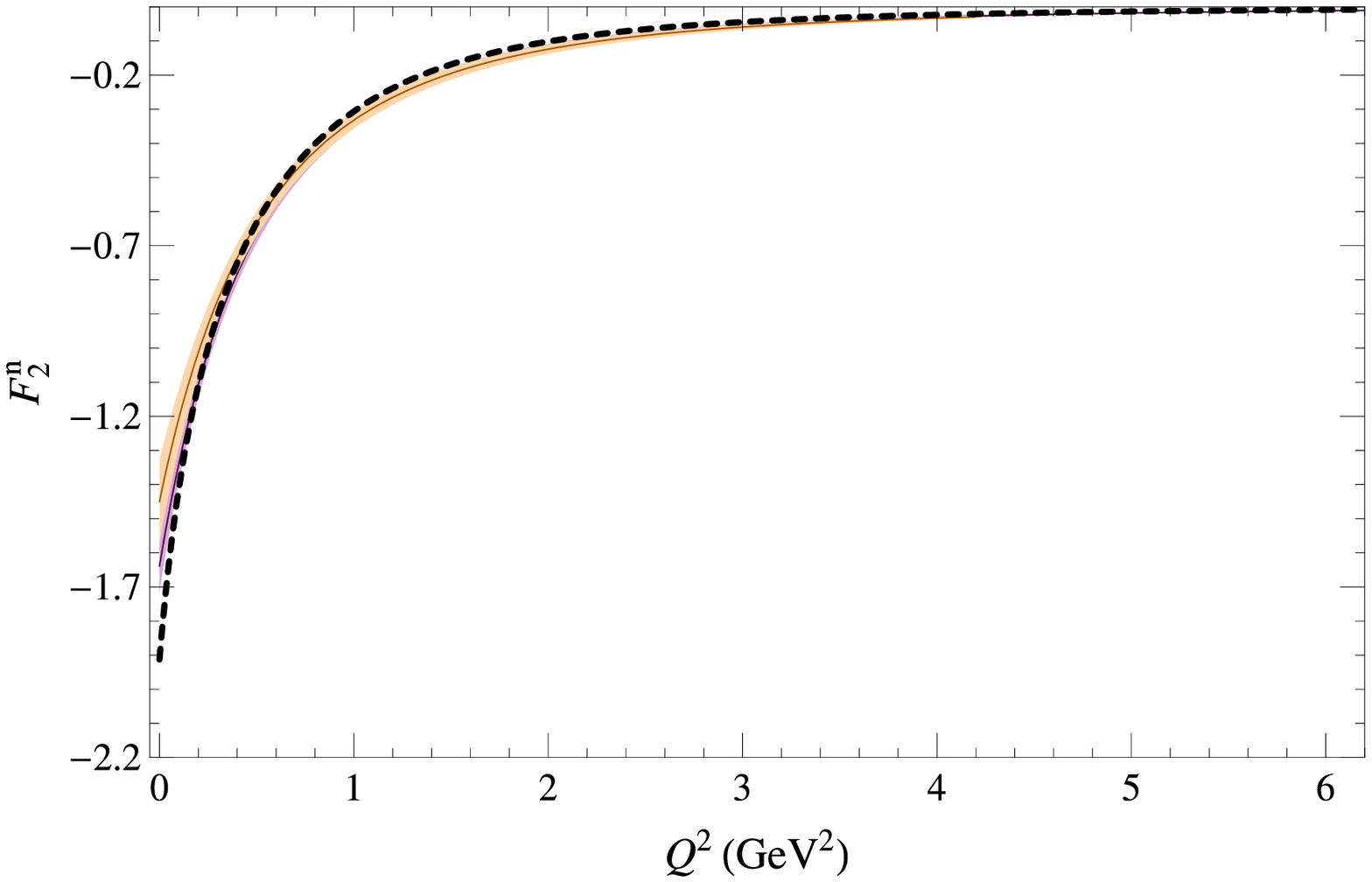}
\end{center}
\vspace{-0.7cm}
\caption{Neutron form factors on both quenched (purple) and dynamical (orange) ensembles. The dashed line represents the experimental parametrization and the lines with bands are from the lattice calculations.}\label{fig:neutron-FF}
\end{figure}

Figure~\ref{fig:neutron-FF} shows a similar plot for the neutron form factors. These include the neutron electric and Dirac form factors, although we again caution that there could be significant disconnected contributions that are ignored in this calculation. Therefore, it is not fair to compare just the connected contribution with the experimental data; we provide it as a guide line to remark on the differences.  The smallness of the quenched extrapolation error band is due to the large statistics and less-correlated data points, as seen in Fig.~\ref{fig:GEGM}. We note that as we anticipated, the connected-only contribution agrees fairly well at large transfer momentum; this suggests an interpretation assigning the remaining discrepancy at low transfer momentum to the omitted disconnected terms. The neutron magnetic and Pauli form factors suffer from similar problems to those from the proton, that the fits are less constrained and the quenched magnetic form factor has obviously divergent behavior. Also similar to the proton case, the neutron Pauli form factor shows indifference to the fermionic vacuum.

\begin{figure}[!h]
\begin{center}
\includegraphics[width=0.45\textwidth]{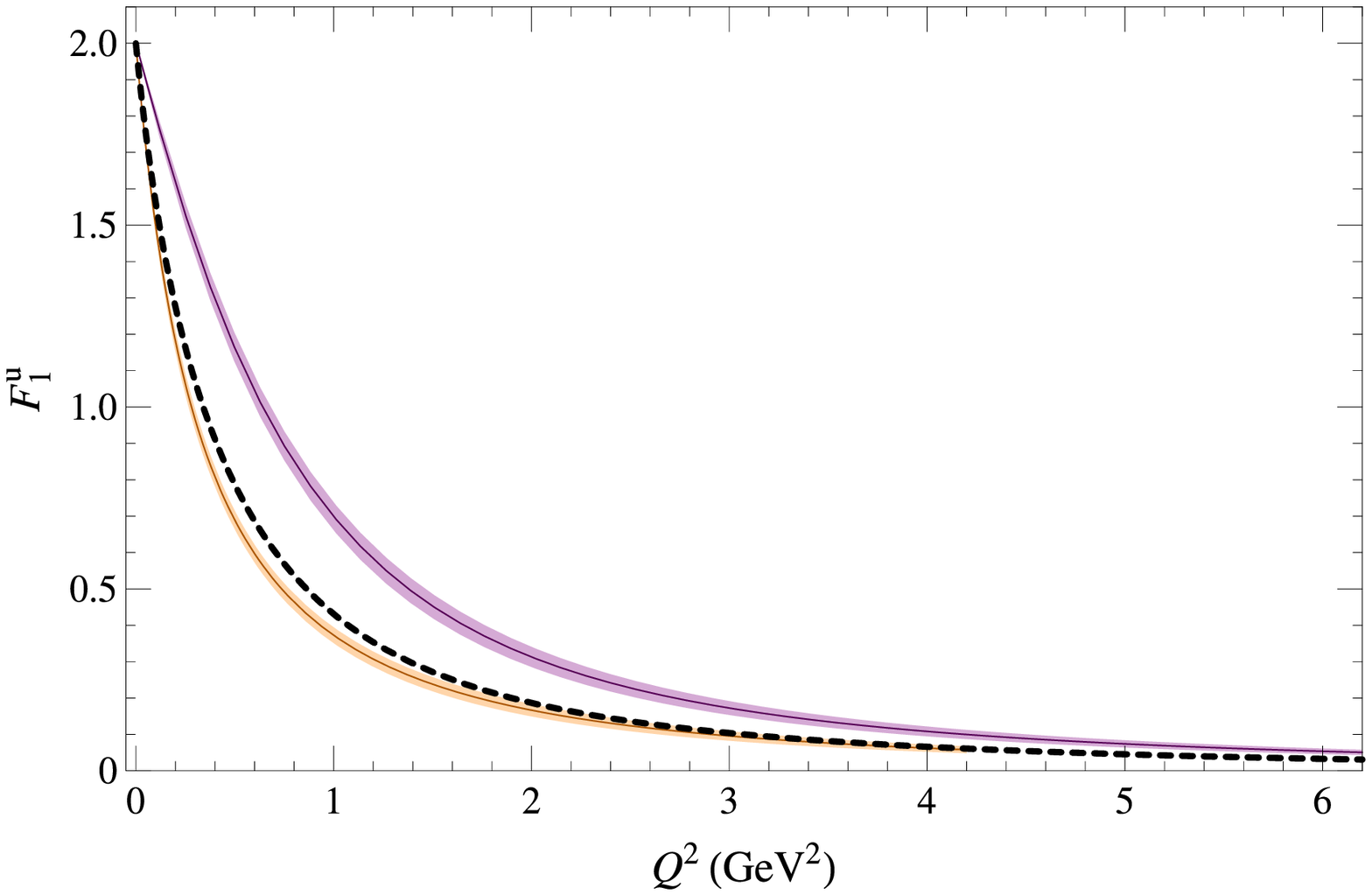}
\includegraphics[width=0.45\textwidth]{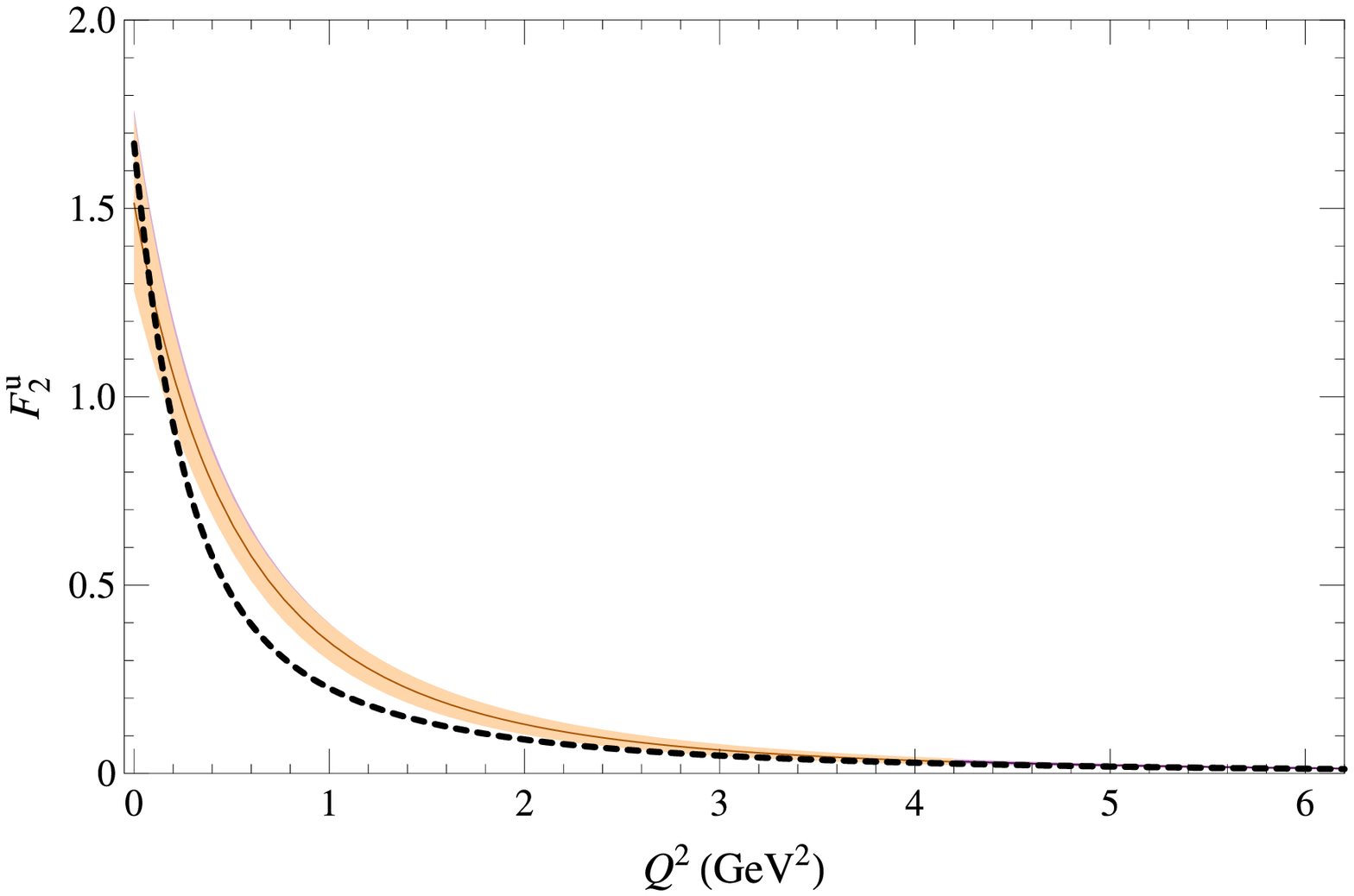}
\includegraphics[width=0.45\textwidth]{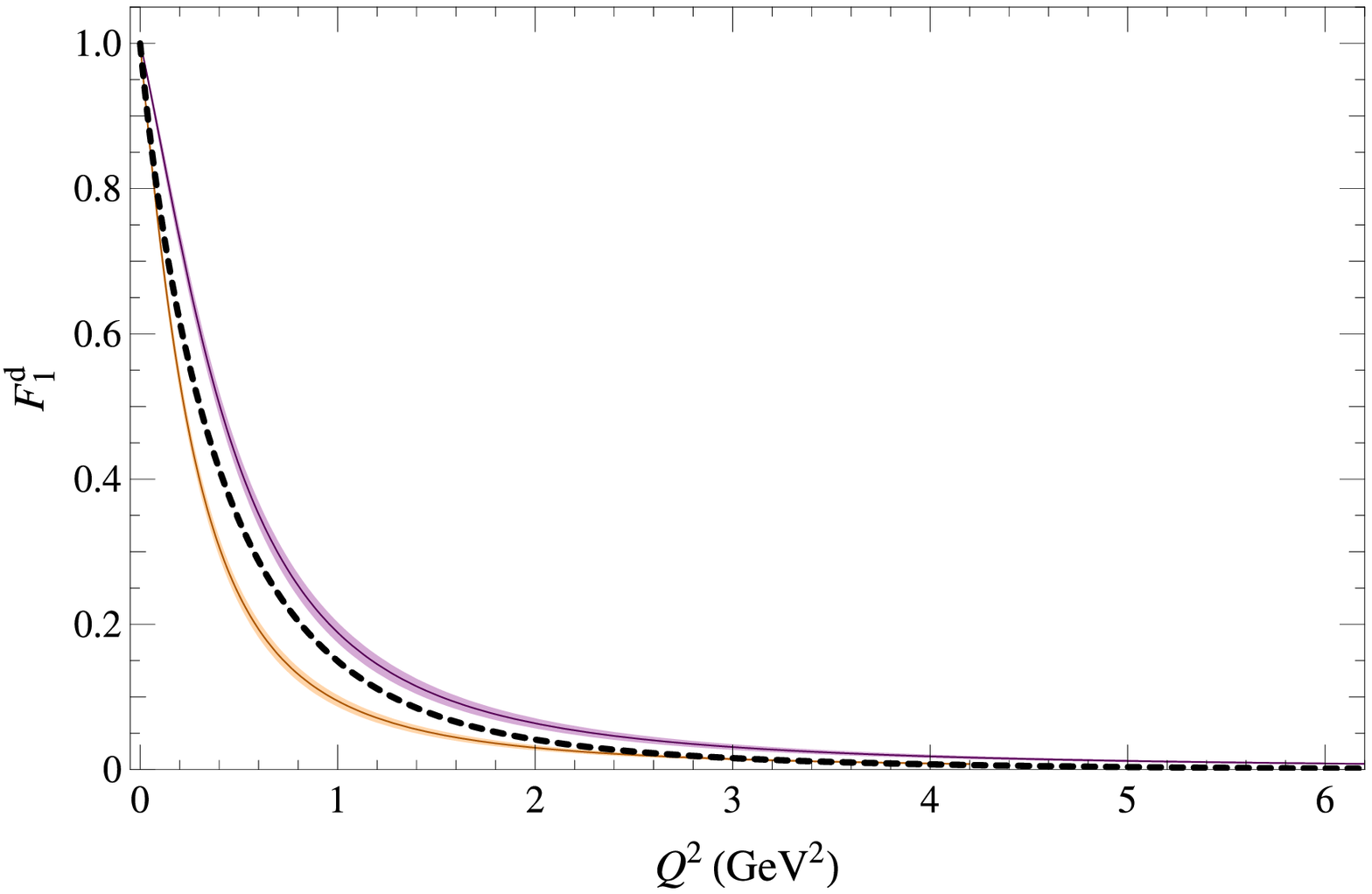}
\includegraphics[width=0.45\textwidth]{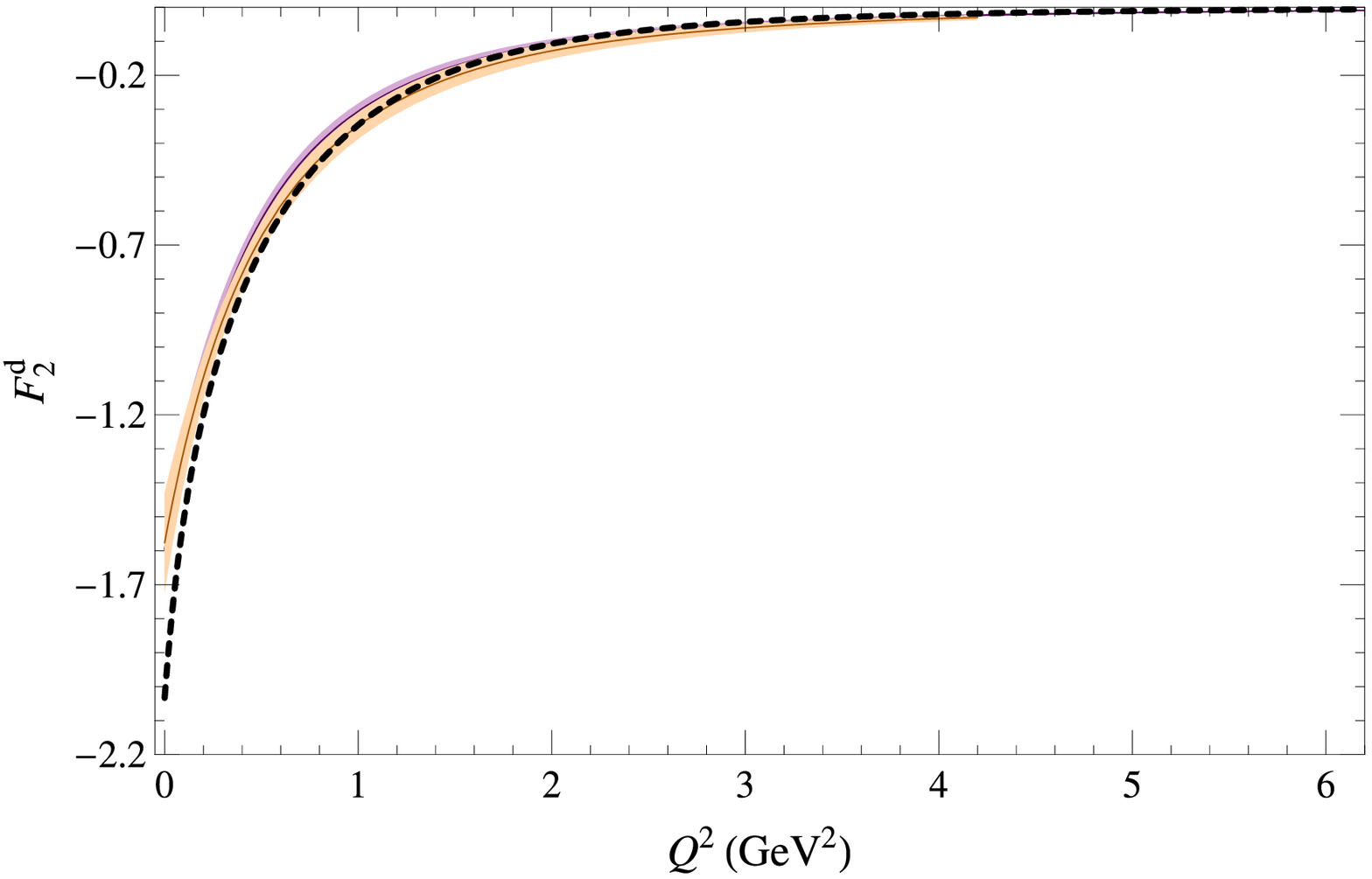}
\includegraphics[width=0.45\textwidth]{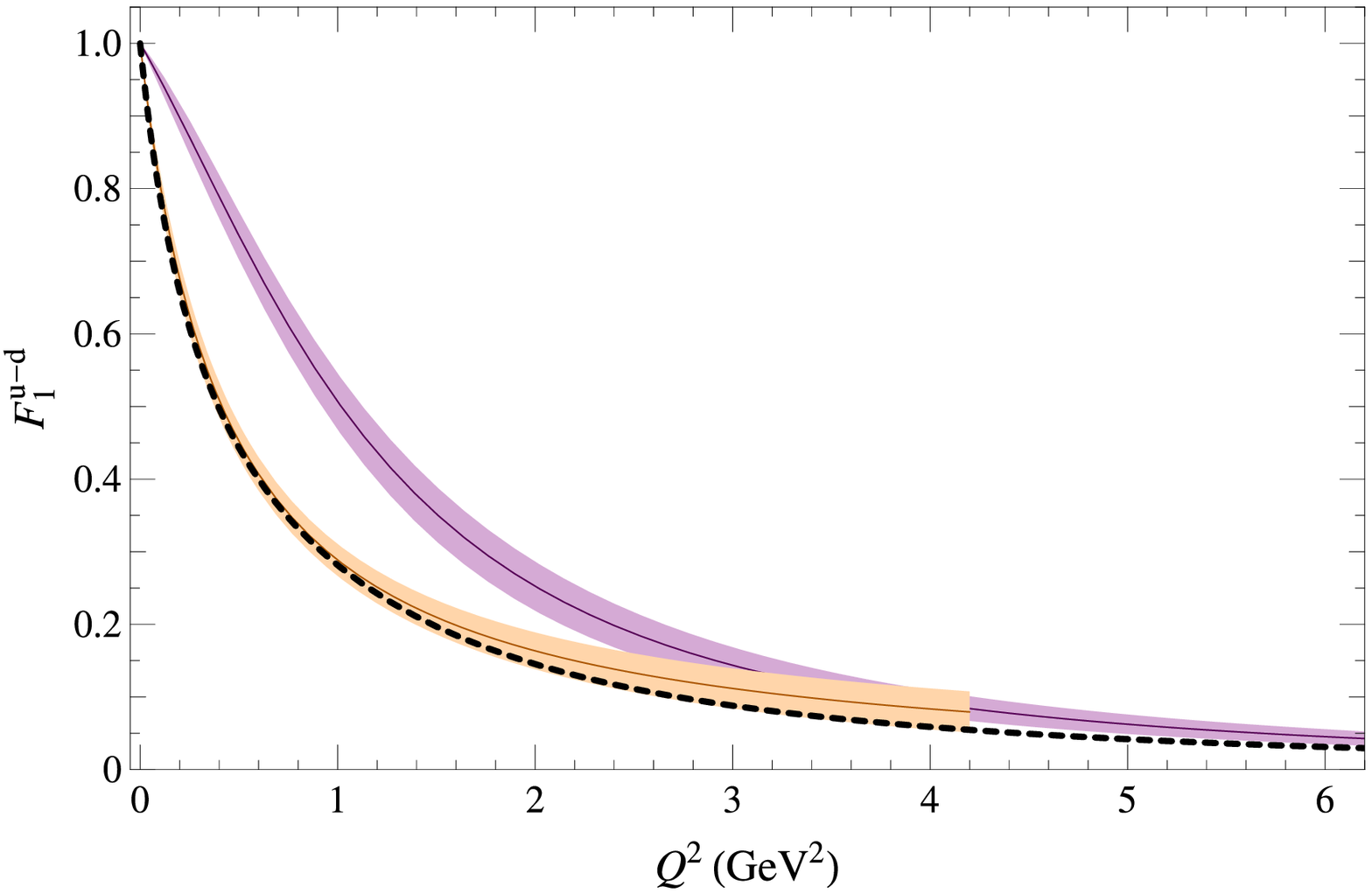}
\includegraphics[width=0.45\textwidth]{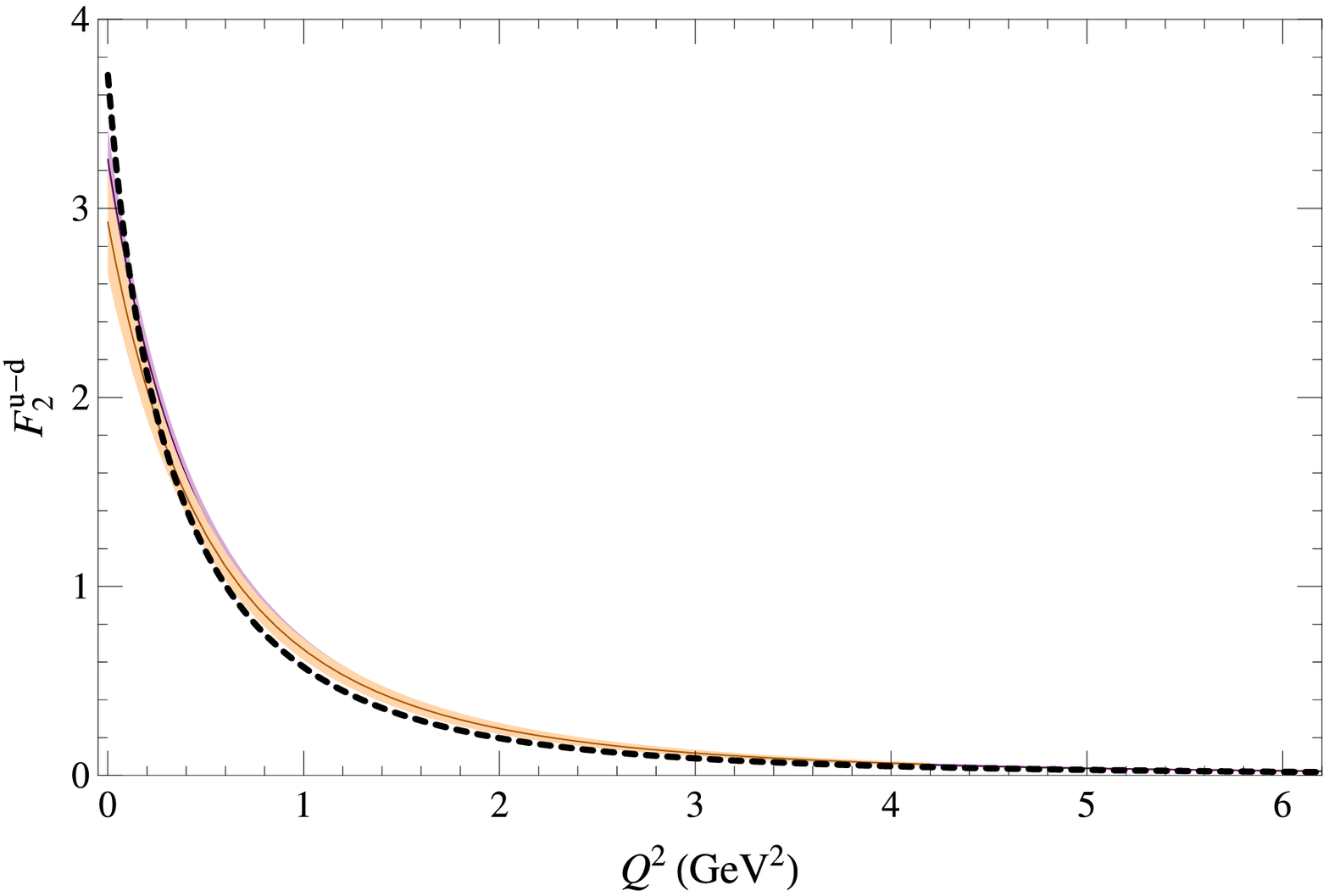}
\end{center}
\vspace{-0.7cm}
\caption{Quark contributions to the form factors on both quenched (purple) and dynamical (orange) ensembles. The dashed line represents the experimental parametrization and the lines with bands are from the lattice calculations.}\label{fig:quarks-FF}
\end{figure}

Figure~\ref{fig:quarks-FF} shows the individual quark and isovector contributions to the Dirac and Pauli form factors.
The dashed lines are constructed from the proton and neutron electric and magnetic experimental parameterizations form factors, as shown in Eq.~\ref{eq:quark-pn}.
On the lattice, we can explicitly pick out the individual quark contributions by varying the input quark currents.
Once again, we see the difference between quenched and dynamical is significant for the Dirac and negligible for the Pauli form factors.
The dynamical results for both up and down quarks are below the experimental reconstruction of the quark contributions, which may vary for future precision neutron form factors. However, the linear combinations of up and down for the proton and isovector form factors seem to cancel the difference and become agreeable with experiment.
The Pauli form factors are within two sigma of the experimental values except in the low-$Q^2$ region.

\subsection{Transverse Densities}\label{subsec:density}

Following the discussions of Ref.~\cite{Miller:2007uy} (and subsequent studies for other hadrons: pion density in Ref.~\cite{Miller:2009qu}, $N$-$P_{11}$ in Ref.~\cite{Tiator:2008kd} and deuteron in Ref.~\cite{Carlson:2008zc}; see Ref.~\cite{Miller:2010nz} for a detailed review), we now attempt to translate our extracted nucleon form factors into a description of the spatial structure of the nucleon. Although the form factors are typically thought of as being the Fourier transforms of the nucleon wavefunction, there is a complication due to the imparted momentum at the vertex. Since the incoming and outgoing states have different momentum, their wavefunctions are not identical. We can avoid this difficulty by only attempting to describe the spatial structure in the plane transverse to an infinite momentum boost. Thus the transverse charge density is defined as the Fourier transform of the form factor in such a plane:
\begin{equation}
\rho(\mathbf{b}) \equiv \int\!\frac{d^2\mathbf{q}}{(2\pi)^2}F_1(\mathbf{q}^2)e^{i\mathbf{q}\cdot\mathbf{b}},
\end{equation}
where bold vectors $\mathbf{b}$ and $\mathbf{q}$ lie in the transverse plane. Equivalently,
\begin{equation}
\rho(b) = \int_0^\infty\!\frac{Q\,dQ}{2\pi}J_0(bQ)F_1(Q^2),
\end{equation}
for scalar $b$, where $J_0$ is a Bessel function. We can perform this integral numerically, using the $F_1(Q^2)$ obtained by extrapolating our fit form to the physical pion mass.

\begin{figure}
\includegraphics[width=0.45\columnwidth]{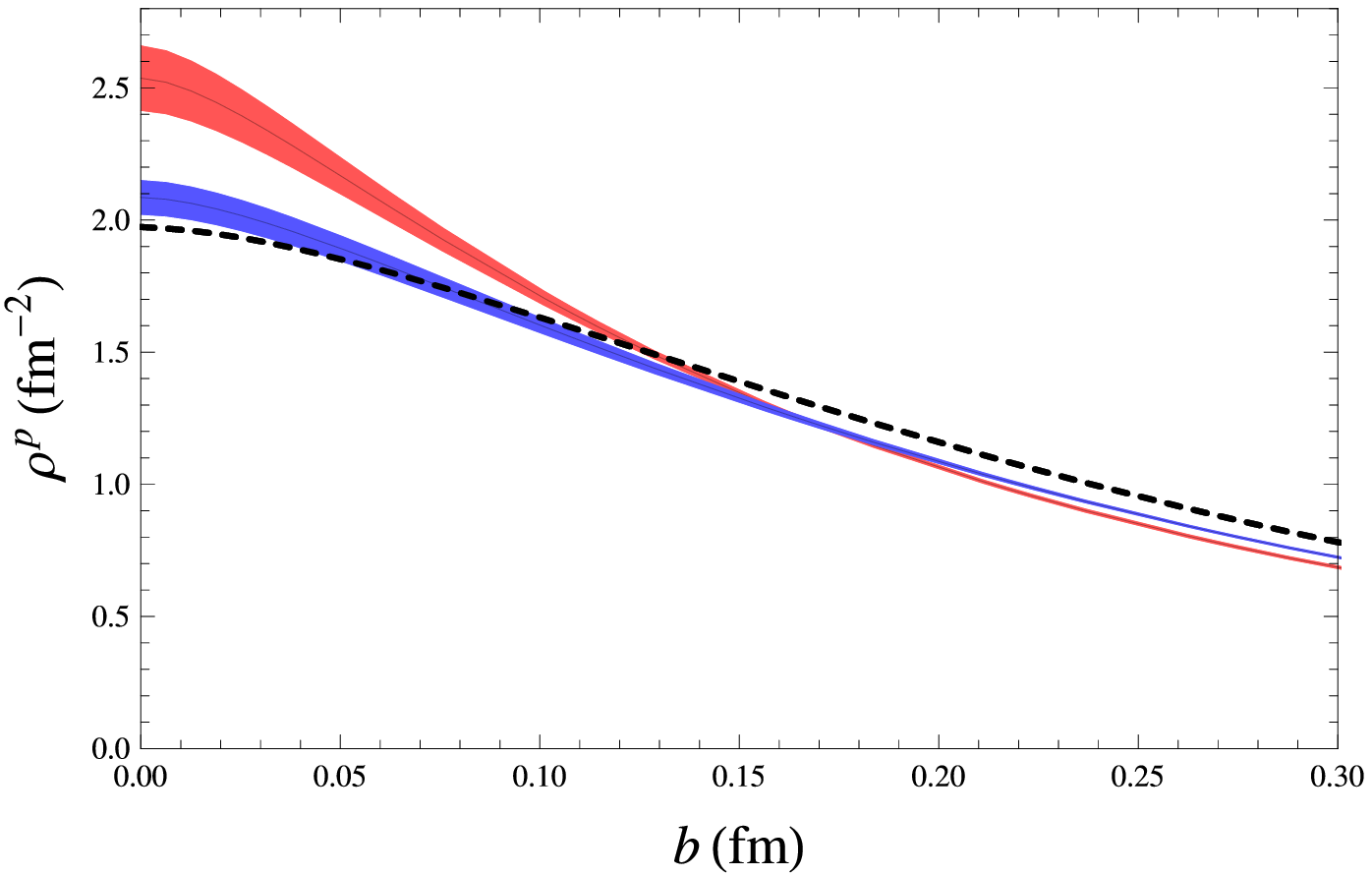}
\caption{The transverse charge density $\rho(b)$ of the proton in the central region with different cuts applied to the $F_1$ data prior to fitting. The lower blue band uses all available $Q^2$ from the 2+1-flavor ensembles; the upper red band uses only data with $Q^2<2\mbox{ GeV}^2$. The experimental result derived from the parametrization of Ref.~\cite{Arrington:2007ux,Kelly:2004hm} is shown as a dashed line.}
\label{fig:density-vs-cut}
\end{figure}

The nature of the Bessel integral, oscillatory and exponentially declining, means that the central core of the distribution is most strongly impacted by large-$Q^2$ form factors. If we wish to well characterize this part of the nucleon density functions, we need precise information about form factors in the upper range of transfer momentum. We demonstrate this effect by restricting the dynamical data set to the region $Q < 2$~GeV$^2$ (which is the upper limit of transfer momentum for many lattice QCD calculations). Then we compare this result to the density obtained from using all the available lattice $Q^2$. Figure~\ref{fig:density-vs-cut} shows in red the effect of omitting the highest $Q^2$ data (above 2~GeV$^2$), compared to the blue which uses all $Q^2$ from the dynamical ensembles. The impact is significant in the central core, as we anticipated; omitting information about large transfer momenta results in a deviation in the density around 25\%.

\begin{figure}
\includegraphics[width=0.45\columnwidth]{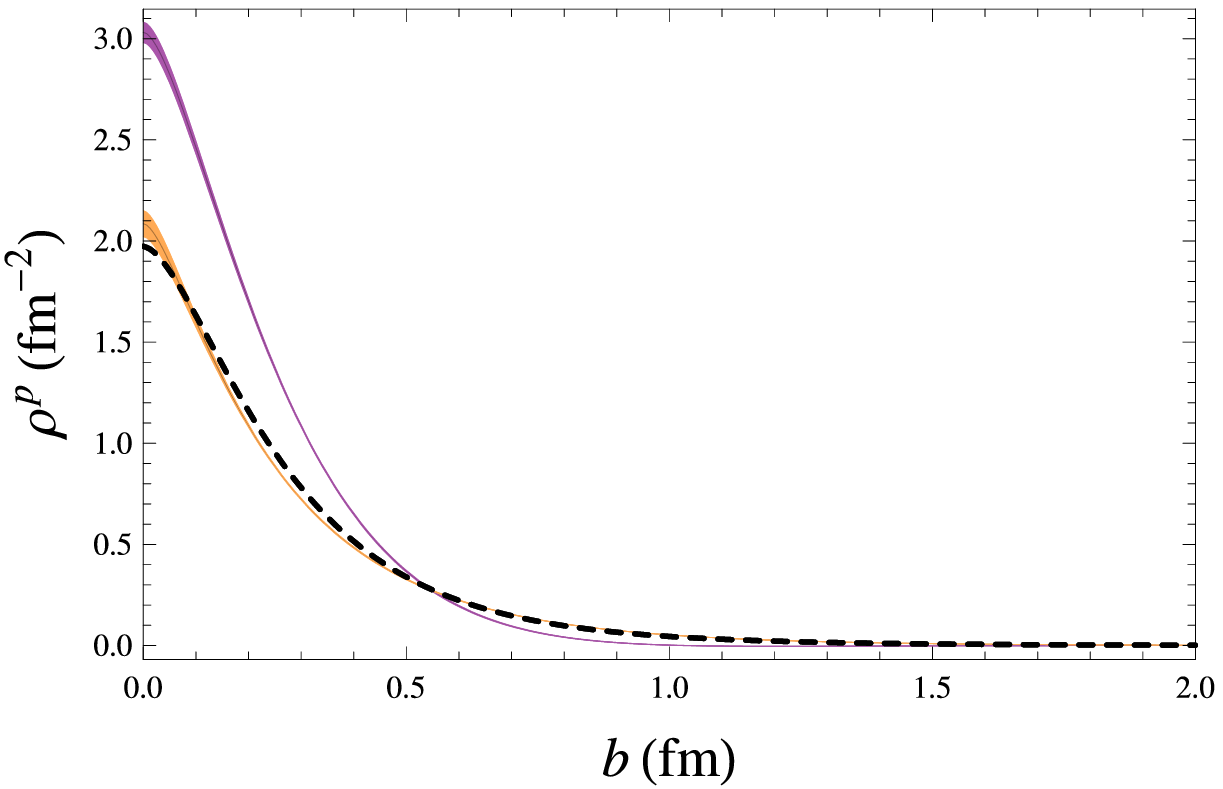}
\includegraphics[width=0.45\columnwidth]{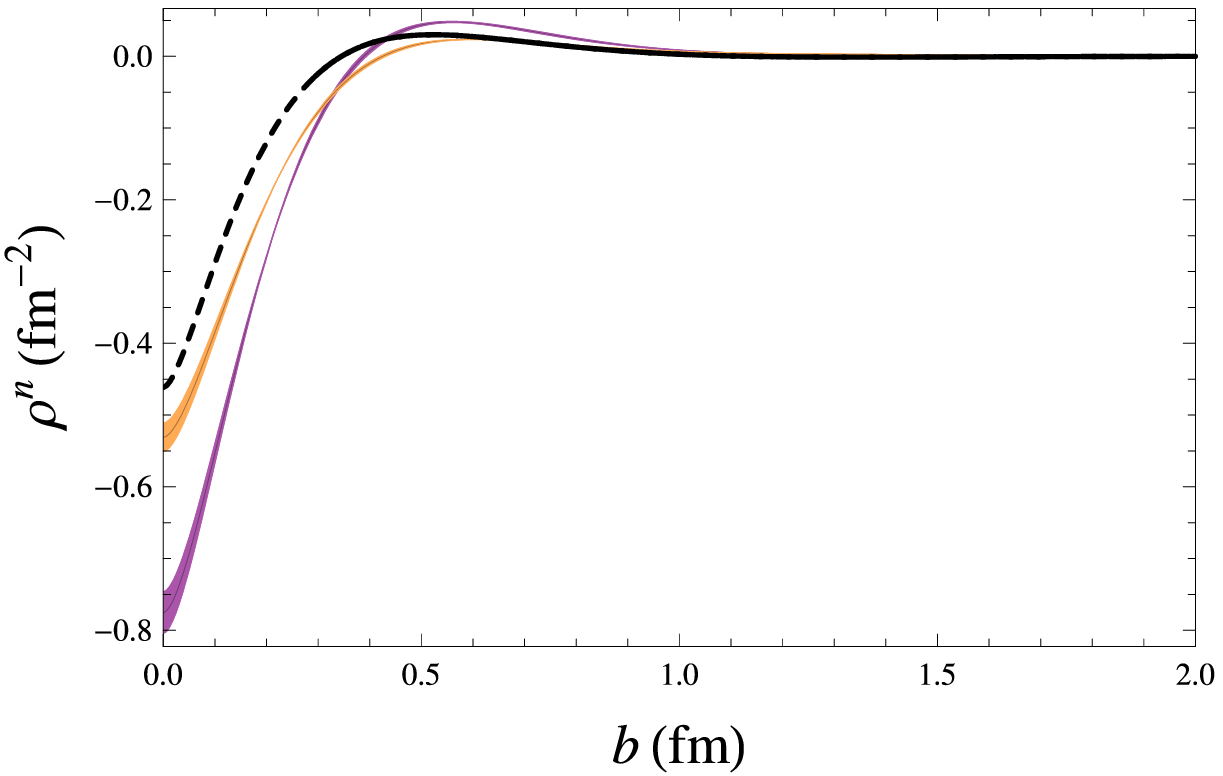}
\includegraphics[width=0.45\columnwidth]{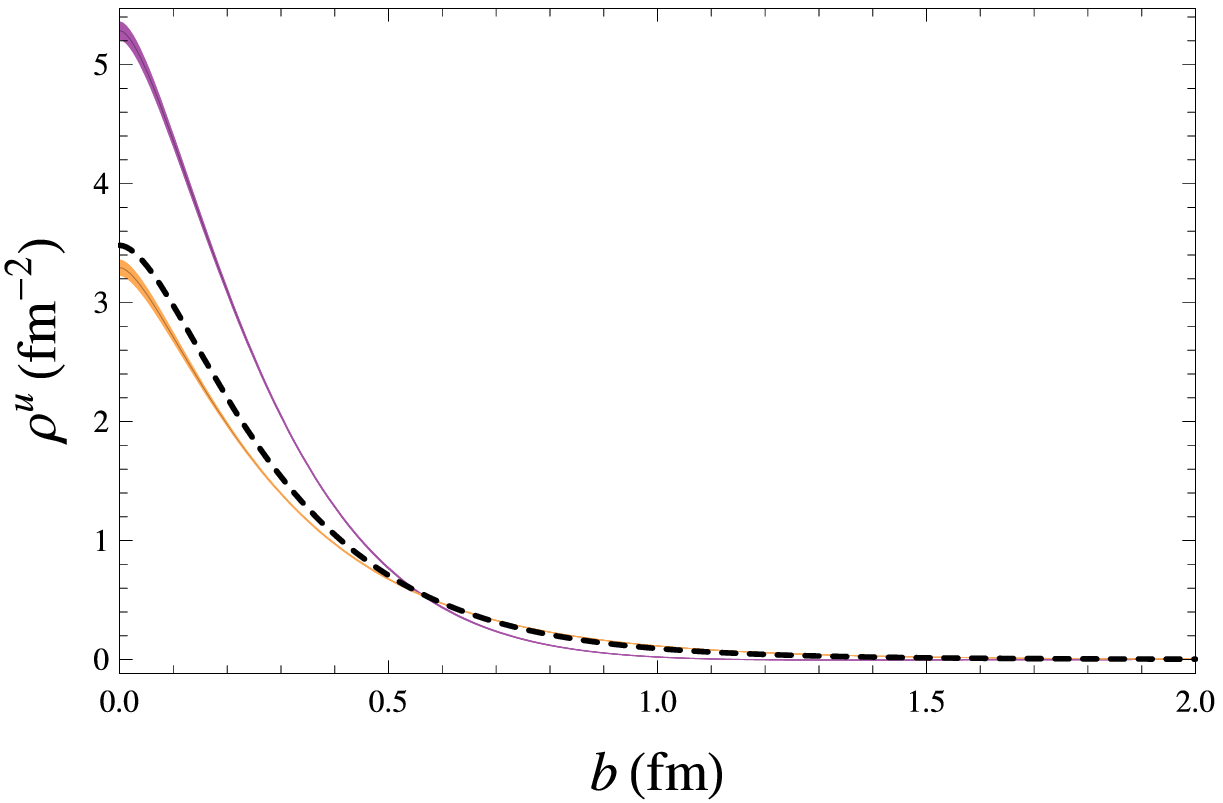}
\includegraphics[width=0.45\columnwidth]{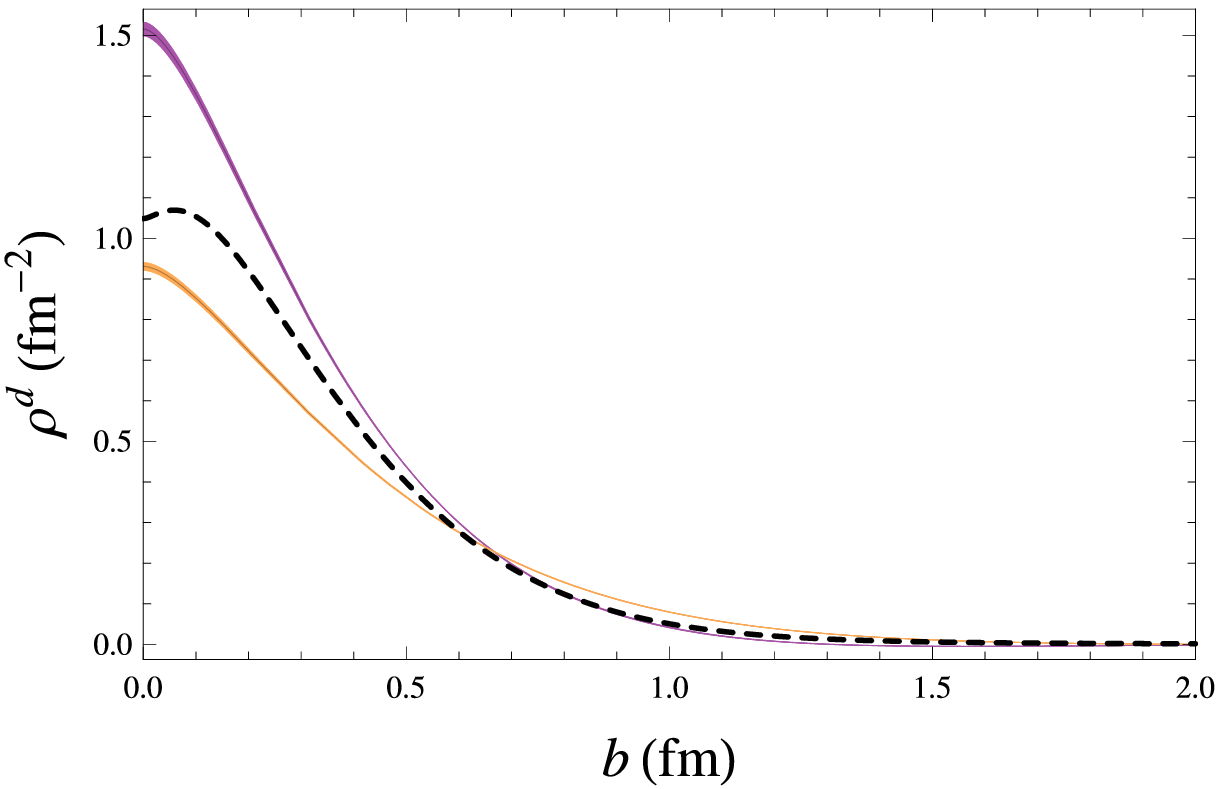}
\caption{The transverse charge density $\rho(b)$ of the proton (upper left), neutron (upper right), up (lower left) and down (lower right) quark
contributions, obtained by integrating the fitted $F_1$ form at the physical pion mass. Results from the dynamical 2+1-flavor ensembles are shown as orange bands. Results from quenched ensembles are shown as purple bands. The experimental result derived from the parametrization of Ref.~\cite{Arrington:2007ux,Kelly:2004hm} is shown as a dashed line.}
\label{fig:F1density}
\end{figure}

Our results for the transverse charge densities of the proton, neutron, up and down quark 
contributions, are shown in Fig.~\ref{fig:F1density},
along with the same quantity using a parametrization of experimental data from Ref.~\cite{Arrington:2007ux,Kelly:2004hm}.
Overall, we observe that the dynamical content of the vacuum appears to have a strong influence on the transverse charge densities in all cases; we see large differences between the quenched and dynamical results. In particular, the effect appears to increase the density of the core region for quenched ensembles. This difference could be plausibly assigned to the presence of a mesonic cloud in the dynamical case, which spreads out the charge distribution. Since the quenched ensembles do not include the effects of sea quarks, a nucleon in that environment does not couple to baryon-meson systems in the usual way.

In the proton channel, there is quite good agreement between our dynamical and experimental results. This might seem suprising, given the omission of the disconnected diagram from our calculation. However, it is a relatively small contribution to the proton Dirac form factor; we expect it to be at the level of a percent in this case. Since the quenched $F_1$ form factor lies above the experimental one throughout the entire $Q^2$ region we calculated, it accumulates a large contribution to the density near the core region. The density becomes smaller than the experimental value at larger distances due to the oscillatory nature of the Bessel function. Similar behavior also occurs in other channels.

The neutron density is expected to have the largest systematics due to the neglect of the disconnected contributions in this work. Even so, we observe only about a 25\% difference between the dynamical results and experiment, surprising considering that the disconnected value is on the same order as the neutron form factor itself. Once again, the difference becomes much more significant in the quenched case, and the even larger negative core charge density is consistent with suppression of mesonic contributions.

The experimental up and down densities are calculated using linear combinations of the experimental parametrizations of the proton and neutron Dirac form factors, as shown in Eqs.~\ref{eq:quark-pn}. The lattice ones are calculated directly using inserted vector currents with specific quark flavors. As in the proton case, the disconnected diagrams should be minimal. The agreement between the dynamical results and experiment for individual quark contributions is not as good as in the case of the proton; in particular, the down-quark density is smaller. It is also notable here that the up-quark distribution is more sharply peaked in the core than the down-quark distribution.

\begin{figure}
\includegraphics[width=0.45\columnwidth]{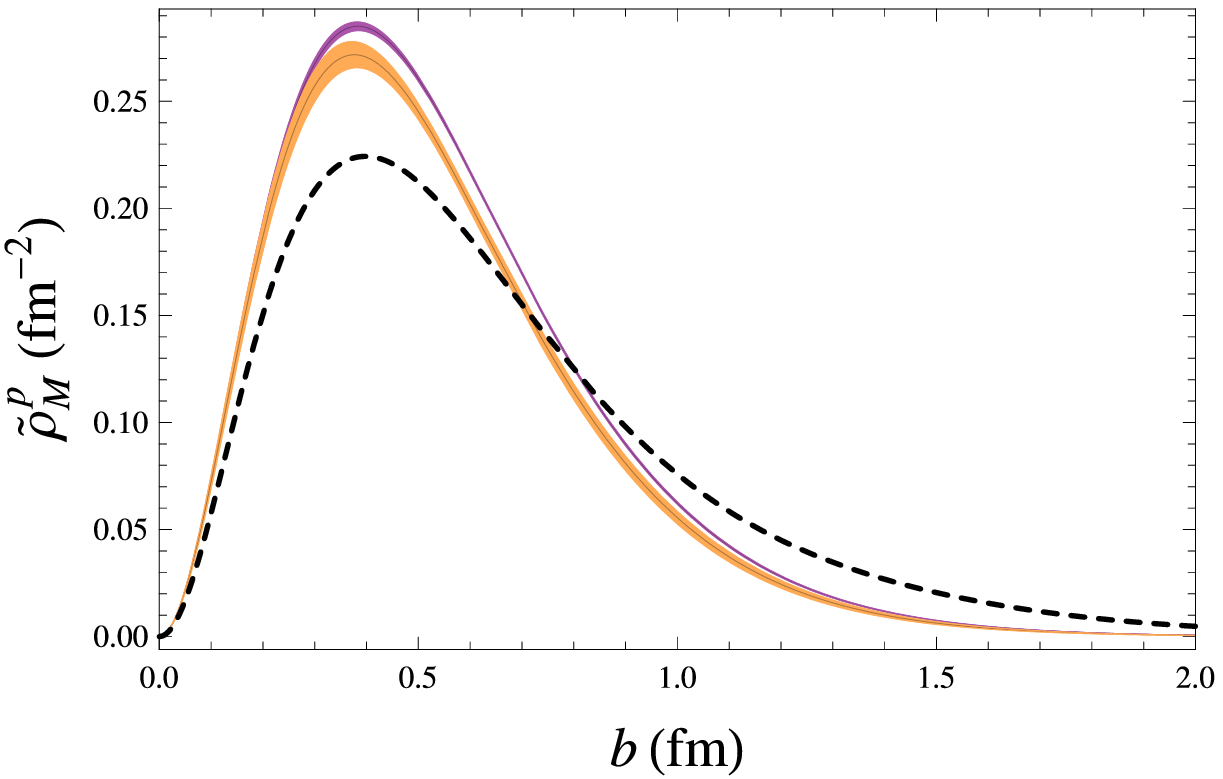}
\includegraphics[width=0.45\columnwidth]{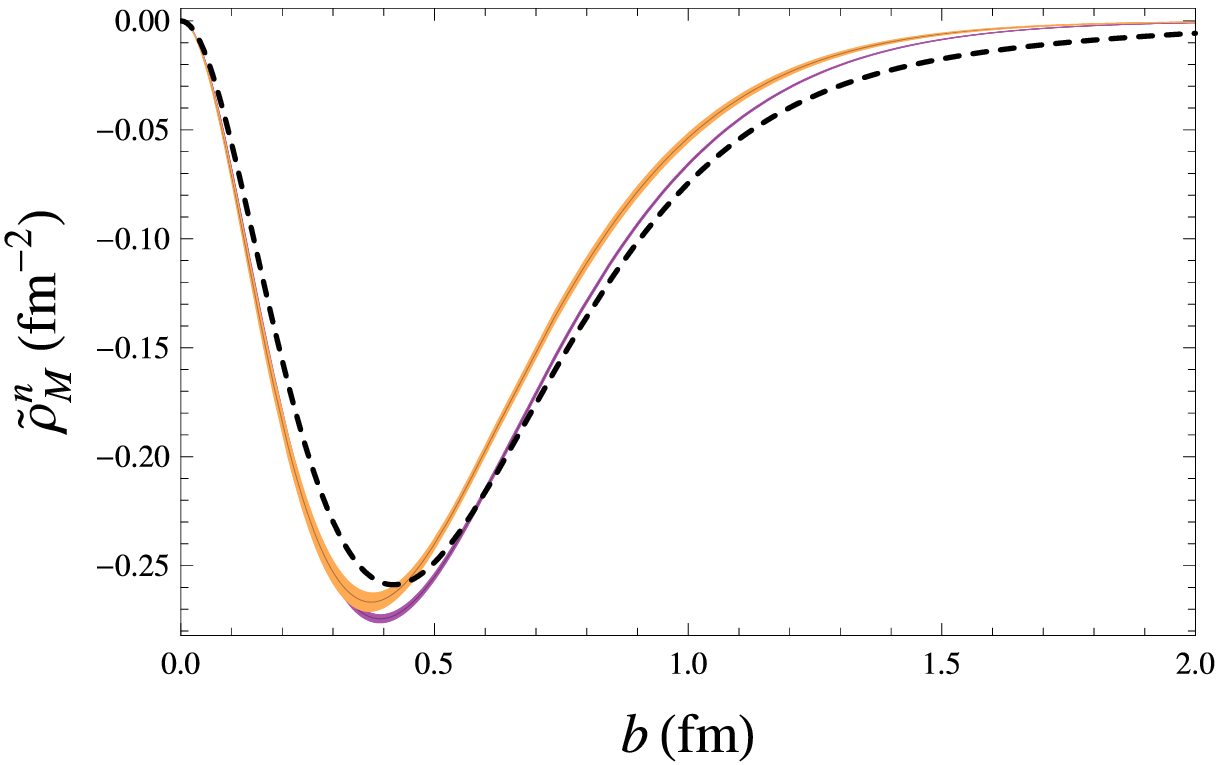}
\includegraphics[width=0.45\columnwidth]{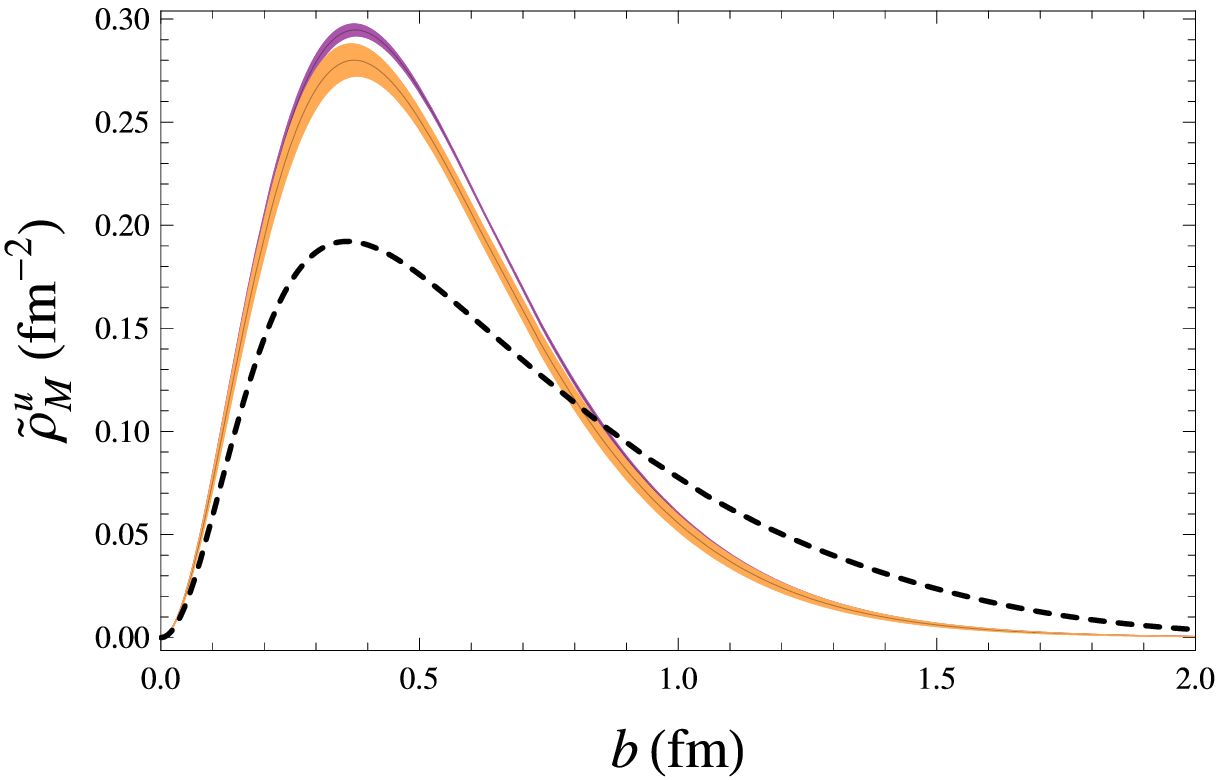}
\includegraphics[width=0.45\columnwidth]{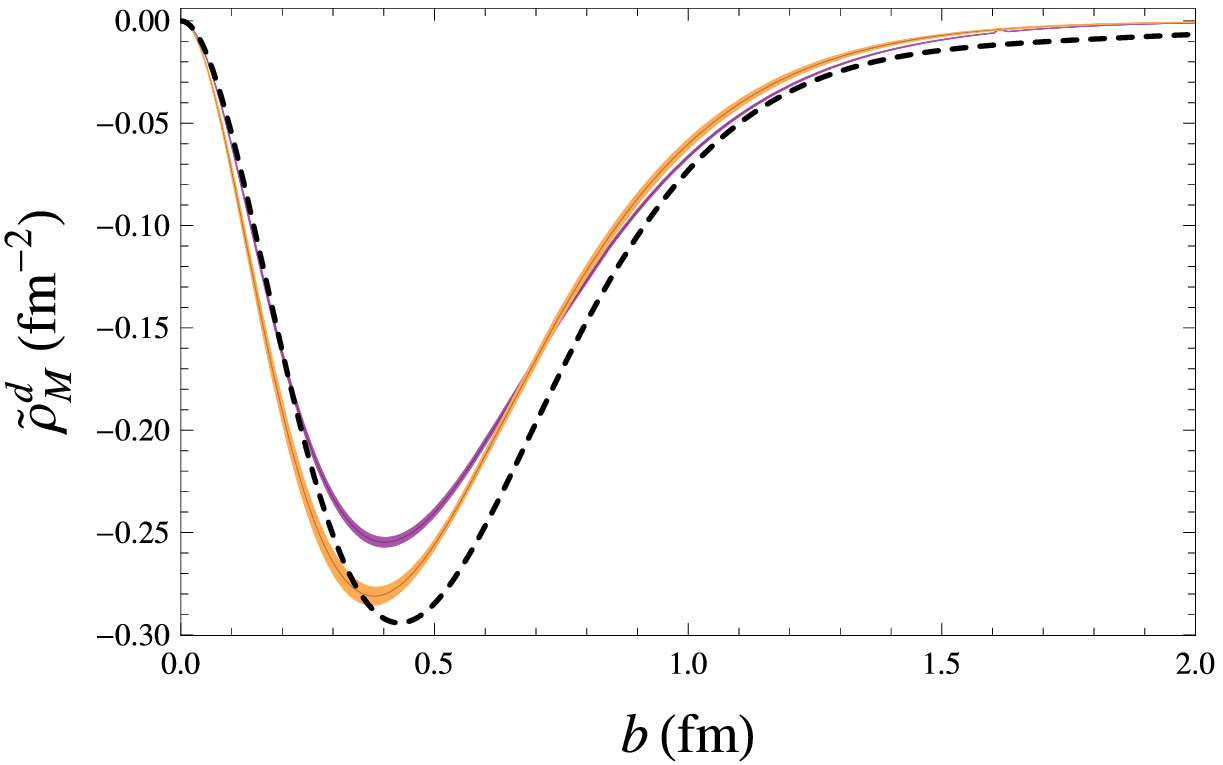}
\caption{The transverse magnetization density $\tilde{\rho}_M(b)$ of the proton (upper left), neutron (upper right), up (lower left) and down (lower right) quark
contributions, obtained by integrating the fitted $F_1$ form at the physical pion mass. Results from the dynamical 2+1-flavor ensembles are shown as orange bands. Results from quenched ensembles are shown as purple bands. The experimental result derived from the parametrization of Ref.~\cite{Arrington:2007ux,Kelly:2004hm} is shown as a dashed line.}
\label{fig:F2density}
\end{figure}

\begin{figure}
\includegraphics[width=0.45\columnwidth]{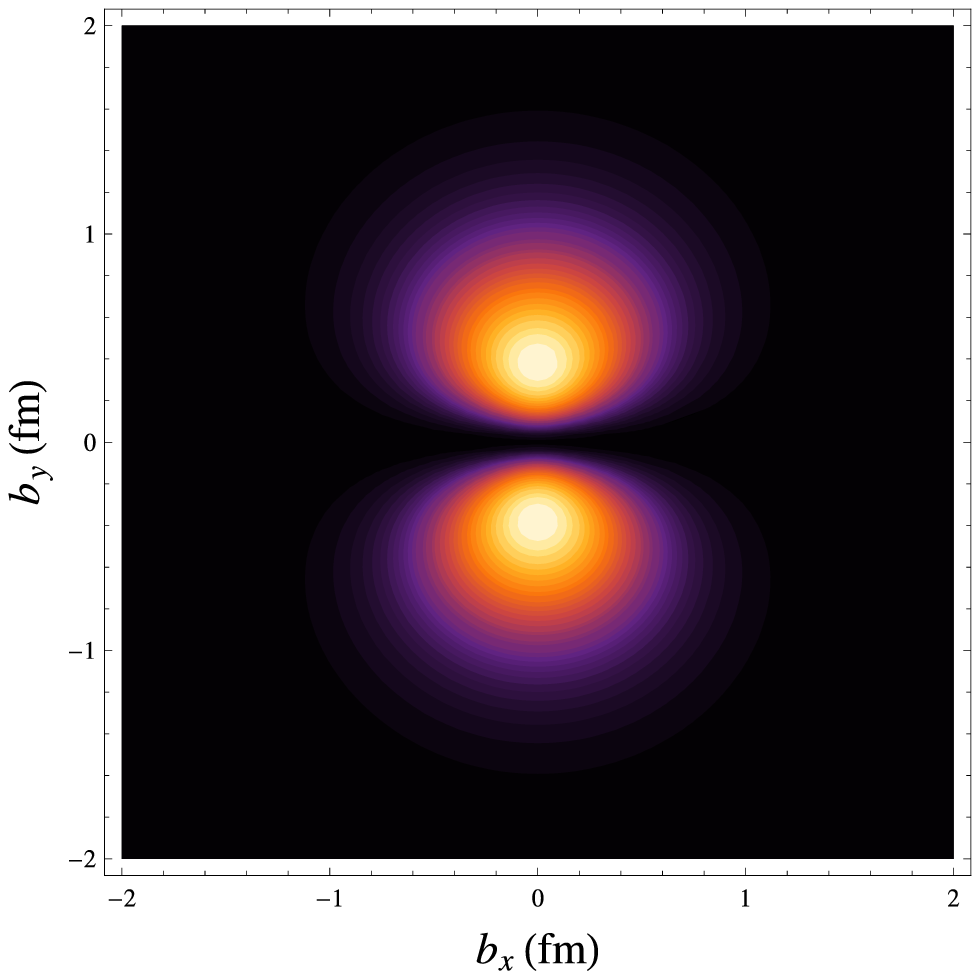}
\includegraphics[width=0.45\columnwidth]{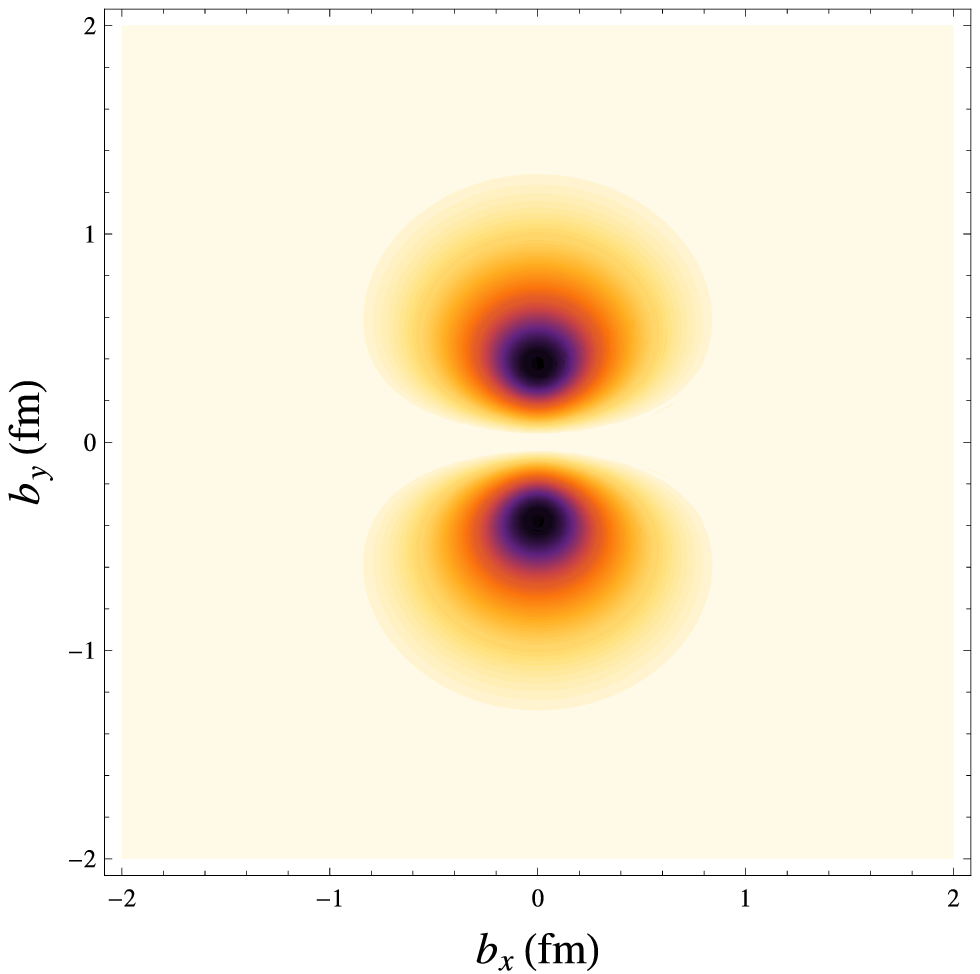}
\caption{The two-dimensional dependence of the transverse magnetization density $\tilde{\rho}_M(\mathbf{b})$ of the proton (left) and neutron (right). Black indicates the lowest values, followed by purple, orange and white in increasing order. For the proton, zero is black and values are positive; for the neutron zero is white and values are negative. Figure~\ref{fig:F2density} corresponds to slices through the vertical axis.}
\label{fig:F2pndensity-2d}
\end{figure}

Another kind of transverse density can be derived for the magnetic moment of the nucleons, the transverse magnetization density. The anomalous magnetic moment density is defined as a Fourier transformation of the $F_2$ form factor:
\begin{equation}
\rho_M(\mathbf{b}) \equiv \int\!\frac{d^2\mathbf{q}}{(2\pi)^2}F_2(\mathbf{q}^2)e^{-i\mathbf{q}\cdot\mathbf{b}}.
\end{equation}
However, Ref.~\cite{Miller:2007uy} finds it more convenient to consider this in terms of a magnetization density. Therefore, we take a derivative with respect to the $b_y$ direction orthogonal to the magnetic field. This yields
\begin{align}
\tilde{\rho}_M(b) &\equiv -b_y\frac{\partial\rho_M(b)}{\partial b_y}\\
                  &=b \sin^2\phi \int_0^\infty\!\frac{Q^2\,dQ}{2\pi}J_1(bQ)F_2(Q^2),
\end{align}
where $\phi$ is the angle between $\mathbf{b}$ and the magnetic field. Again, we calculate this quantity for the proton, neutron, up- and down-quark contributions to the proton and the isovector. We show one-dimensional cuts of the magnetization density along the $\phi=\pi/2$ axis of the transverse plane in Fig.~\ref{fig:F2density}. We also show the full dependence of the proton and neutron magnetization densities across the transverse plane in Fig.~\ref{fig:F2pndensity-2d}.

For the magnetization, we see poorer agreement with experiment than in the case of transverse charge density, but better agreement between the quenched and dynamical cases. The Pauli form factors from the dynamical and quenched ensembles are relatively similar (unlike the Dirac form factors), and this results in similar magnetization densities. The discrepancy in the up-quark channel is strongly reflected in the up-quark dominated proton, despite the down-quark contribution having relatively closer agreement with experiment; the converse effect is seen in the neutron.

In Subsec.~\ref{subsec:magnetic-moments}, we found most anomalous magnetic moments from the lattice calculation were roughly 2/3 the experimental values; this leads to a discrepancy in the overall scales of the densities. Also the fit forms for the Dirac form factors are better constrained in the low transfer-momentum region by conservation of charge, resulting in smaller discrepancies at large distances than the magnetization density. To get a better picture of the magnetization density, one would also need a better knowledge of the low transfer momentum form factors. Techniques such as twisted boundary conditions can provide smaller momenta for a particular combination of lattice spacing and box size, which should help to ameliorate these discrepancies.

\section{Conclusion}\label{sec:conclusion}

In this work, we attempted to determine the large-$Q^2$ form factors in lattice QCD. Using operators that have various degrees of overlap with boosted states in form-factor calculations significantly improves the signal. Here we demonstrated the method using a simple baryon operator with various smearing parameters. Since baryon systems typically have worse signal-to-noise ratio than mesons, the method should be applicable to other hadronic systems. 
We analyzed the nucleon ground-state form factor by explicitly including excited-state contributions. This allowed us to extract the ground-state signal more precisely, and combined with the use of a longer source-sink separation, reduces systematic uncertainty coming from excited-state contamination.

We checked the most commonly considered observables such as the Dirac and Pauli radii and anomalous magnetic moments on our dynamical ensemble, and compared with other $N_f=2+1$ calculations, finding reasonable agreement within the statistical errors. When we looked at the the dimensionless quantities $4(m_N^{\rm lat})^2\langle r_{1,2}^2\rangle$ in the later section, the chiral curvature was significantly reduced in our data; we further found the naive (linear) chiral extrapolation recovered the experimental values. Similar behavior was also seen in the RBC/UKQCD lighter pion-mass data points. Performing further investigation of such dimensionless quantities with higher signal-to-noise ratios in the lighter pion-mass region would be interesting. 
Nevertheless, the agreement in the low-$Q^2$ behavior among calculations with $N_f=2+1$ demonstrates a nice universality among various fermion actions.

We extended our study to the high-$Q^2$ region where no lattice-QCD calculations had reached in the past. We found significant differences in the Dirac form factor between quenched and dynamical studies, indicating non-negligible systematic error due to quenching, while Pauli form factors appear less sensitive to sea-fermion effects.
%We have not formulated a hypothesis to explain this behavior, and so will leave this to our phenomenological colleagues for further interpretation.
We compared our results with an interpolation to experimental results as a function of $Q^2$, although our knowledge of the $Q^2$-dependence of the $G_E^n$ form factor remains limited. These experimental lines may change after the collection of future precision and large-$Q^2$ neutron form factor data. Our lattice neutron form factors extend as far as 4--6~${\rm GeV}^2$ but due to the omission of ``disconnected'' diagrams (which are expected to contribute at $O(10^{-2})$), the $F_1^n$ and $G_E^n$ form factors in our calculations suffer from comparable systematic error. However, the other form factors (including those for up and down quarks) have relatively large magnitudes, making such $O(10^{-2})$ systematic uncertainty at the level of the statistical error; thus, these should be more reliable compared with experiment. All calculations may be further improved with lighter pions and more statistics.

We look at the transverse charge and magnetization densities using the infinite-momentum frame definition from Dirac and Pauli form factors. For these quantities, we see (possibly coincidental) exceptional agreement with experiment for the charge density of the proton. The level of agreement is particularly striking when compared to the quenched results. The neutron charge density agrees less well, as best seen in the isovector charge density, where the central density (inside 0.2~fm) greatly exceeds experiment. Since this region is most sensitive to high-$Q^2$ contributions, large systematics probably exist for both the lattice and experimental measurements. We showed that the magnetization densities are much less susceptible to sea-quark effects. In this case, the proton magnetization density has greater tension with experiment than the neutron density, but neither is in particularly good agreement. 

We have presented lattice calculations in the large momentum-transfer region up to 4 and 6~${\rm GeV}^2$. Our currently accessible momenta are limited by the available lattice spacing in the calculation, but this will soon be improved as finer lattices become available. The signals presented can be further improved by reducing the source-sink separation with the same analysis procedure; we plan to proceed in that direction and crosscheck with the calculation done in this work. 
Future generalizations to operators constructed in irreducible representations of the cubic group will probably allow us to analyze form factors for radially excited states of the nucleons.

To reach even higher $Q^2$ regions, we propose performing a numerical step-scaling calculation (for example, Ref.~\cite{Lin:2006ur}). On a small volume with very fine lattice spacing, we can easily reach high momentum. By calculating the step-scaling function at overlapping momentum points (or interpolating momentum function) we can reduce the systematic error due to finite-volume or lattice discretization artifacts. However, generating several volumes of dynamical $N_f=2+1$ lattices requires a large amount of computational resources; we hope such a proposal will become feasible as the petascale computing facilities become available in the near future.

\section*{Acknowledgments}
This work was done using the Chroma software suite\cite{Edwards:2004sx}; part of the propagator calculation used the EigCG solver\cite{Stathopoulos:2007zi}; and calculations were performed on clusters at Jefferson Laboratory using time awarded under the USQCD Initiative. 
We thank Gerald Miller for his helpful comments and feedback on transverse densities. HWL thanks Ian Cloet discussion about the pion-cloud model. 
SDC thanks the Institute of Nuclear Theory for their hospitality during the period working on this paper. 
SDC is supported by U.S. Dept. of Energy grants DE-FG02-91ER40676 and DE-FC02-06ER41440, and NSF grant OCI-0749300. 
RGE and DR are supported by U.S. DOE Contract No. DE-AC05-06OR23177. 
HWL is supported in part by the U.S. Dept. of Energy under grant No. DE-FG03-97ER4014. 
KO is supported in part by the U.S. Dept. of Energy contract No. DE-AC05-06OR23177 (JSA), DOE
grants DE-FG02-04ER41302 and DE-FG02-07ER41527 and NSF grant CCF-0728915. 
This work is coauthored by Jefferson Science Associates, LLC under U.S. DOE Contract No. DE-AC05-06OR23177.

\bibliographystyle{apsrev}

\end{document}